\documentclass[10pt]{article}
\usepackage{amsmath}
\usepackage{graphicx}
\usepackage{amsfonts}
\usepackage{amssymb}
\usepackage{epsf}
\usepackage{latexsym}

\def\80{\hspace{0.8in}}

\def\bq{\bf q}
\def\bM{\bf M}

\def\bR{\bf R}
\def\br{\bf r}
\def\bz{\bf z}

\def\brho{\mbox{\boldmath$\rho$}}
\newcommand{\be}{\begin{enumerate}}
\newcommand{\ee}{\end{enumerate}}
\newcommand{\bi}{\begin{itemize}}
\newcommand{\ei}{\end{itemize}}
\newcommand{\bd}{\begin{description}}
\newcommand{\ed}{\end{description}}

\def\beq{\begin{equation}}
\def\eeq{\end{equation}}
\def\bea{\begin{eqnarray}}
\def\eea{\end{eqnarray}}
\def\pa{\partial}
\def\d{\textrm{d}}

\def\ttA{\mbox{\tt A}}
\def\ttB{\mbox{\tt B}}
\def\ttC{\mbox{\tt C}}

\def\ttL{\mbox{\tt L}}
\def\ttD{\mbox{\tt D}}
\def\ttE{\mbox{\tt E}}
\def\ttF{\mbox{\tt F}}

\def\brho{\mbox{\boldmath$\rho$}}
\def\bPi{\mbox{\boldmath$\Pi$}}
\def\bpi{\mbox{\boldmath$\pi$}}
\def\cr{\mbox{\scriptsize{\bf $\mbox{ } \times \mbox{ }$}}}

\def\NSI{Na\"{\i}ve Schr\"{o}dinger Interpretation }

\def\NSII{Na\"{\i}ve Schr\"{o}dinger Interpretation}

\def\lt{\mbox{\Large $t$}}

\def\ma{\mbox{a}}
\def\mb{\mbox{b}}
\def\mc{\mbox{c}}
\def\md{\mbox{d}}

\def\mh{\mbox{h}}
\def\mi{\mbox{i}}
\def\mj{\mbox{j}} 

\def\ml{\mbox{l}}

\def\mn{\mbox{n}}

\def\mp{\mbox{p}} 
\def\np{\mbox{p}}

\def\mt{\mbox{t}}

\def\mz{\mbox{z}}

\def\mB{\mbox{B}} 
\def\mC{\mbox{C}}
\def\mD{\mbox{D}}
\def\mE{\mbox{E}}
\def\mF{\mbox{F}}

\def\mK{\mbox{K}}
\def\mL{\mbox{L}}
\def\mM{\mbox{M}}

\def\mP{\mbox{P}}

\def\mR{\mbox{R}}
\def\mS{\mbox{S}}
\def\mT{\mbox{T}}

\def\mZ{\mbox{Z}} 
\def\sa{\mbox{\scriptsize a}}
\def\sb{\mbox{\scriptsize b}}
\def\sc{\mbox{\scriptsize c}}
\def\sd{\mbox{\scriptsize d}}
\def\se{\mbox{\scriptsize e}}

\def\sh{\mbox{\scriptsize h}} 

\def\sj{\mbox{\scriptsize j}} 

\def\sll{\mbox{\scriptsize l}}  
\def\sm{\mbox{\scriptsize m}}
\def\sn{\mbox{\scriptsize n}} 
\def\so{\mbox{\scriptsize o}} 
\def\sp{\mbox{\scriptsize p}}
\def\sq{\mbox{\scriptsize q}}
\def\sr{\mbox{\scriptsize r}}
\def\sss{\mbox{\scriptsize s}}
\def\st{\mbox{\scriptsize t}}
\def\su{\mbox{\scriptsize u}}

\def\sA{\mbox{\scriptsize A}} 
\def\sB{\mbox{\scriptsize B}}
\def\sC{\mbox{\scriptsize C}}
\def\sD{\mbox{\scriptsize D}}
\def\sE{\mbox{\scriptsize E}}
\def\sF{\mbox{\scriptsize F}}

\def\sH{\mbox{\scriptsize H}}

\def\sJ{\mbox{\scriptsize J}}
\def\sK{\mbox{\scriptsize K}}
 
\def\sM{\mbox{\scriptsize M}}

\def\sR{\mbox{\scriptsize R}}
\def\sS{\mbox{\scriptsize S}}
\def\sT{\mbox{\scriptsize T}}

\def\sW{\mbox{\scriptsize W}}

\def\barp{\bar{p}}
\def\barq{\bar{q}}
\def\barr{\bar{r}}

\def\eph(B){\mbox{\scriptsize emergent(LMB)}}

\def\ta{\mbox{\tiny a}}

\def\te{\mbox{\tiny e}}

\def\tl{\mbox{\tiny l}}
\def\tm{\mbox{\tiny m}}

\def\to{\mbox{\tiny o}}

\def\tr{\mbox{\tiny r}}

\def\ttt{\mbox{\tiny t}}

\def\tD{\mbox{\tiny D}}

\def\tT{\mbox{\tiny T}}

\def\sbfM{\mbox{{\bf\scriptsize \sffamily M}}}

\def\sbM{\mbox{{\bf \scriptsize M}}}

\def\fa{\mbox{\sffamily a}}

\def\bQ{\mbox{\sffamily q}}

\def\fs{\mbox{\sffamily s}}

\def\fE{\mbox{\sffamily E}}

\def\fG{\mbox{\sffamily G}}
\def\fH{\mbox{\sffamily H}}

\def\fM{\mbox{\sffamily M}}
\def\fN{\mbox{\sffamily N}}

\def\fP{\mbox{\sffamily P}}
\def\fQ{\mbox{\sffamily Q}}
\def\fR{\mbox{\sffamily R}}
\def\fS{\mbox{\sffamily S}}
\def\fT{\mbox{\sffamily T}}

\def\fV{\mbox{\sffamily V}}
\def\fW{\mbox{\sffamily W}}

\def\sfa{\mbox{\sffamily{\scriptsize a}}}

\def\sfA{\mbox{\sffamily{\scriptsize A}}}
\def\sfB{\mbox{\sffamily{\scriptsize B}}}
\def\sfC{\mbox{\sffamily{\scriptsize C}}}

\def\sfZ{\mbox{\sffamily{\scriptsize Z}}}

\def\sumIN{\sum\mbox{}_{\mbox{}_{\mbox{\scriptsize $I$=1}}}^{N}}

\def\K{Kucha\v{r} }

\def\capFrG{\mbox{\boldmath$\mathfrak{G}$}}                    
\def\scapFrG{\mbox{\boldmath\scriptsize$\mathfrak{G}$}}        
\def\bupSigma{\mbox{\boldmath$\Sigma$}}      
\def\sbSigma{\mbox{\scriptsize\boldmath$\Sigma$}} 

\def\bn{\mbox{\bf n}}

\def\bB{\mbox{\bf B}}
\def\bP{\mbox{\bf P}}
\def\bQ{\mbox{\bf Q}}
\def\bS{\mbox{\bf S}}

\def\bZ{\mbox{{\bf Z}}}

\begin{document}

\begin{titlepage}

\begin{center}

{\LARGE\bf RELATIONAL QUADRILATERALLAND. I}

\vspace{.1in}

{\LARGE\bf THE CLASSICAL THEORY}

\vspace{.2in}

{\bf Edward Anderson}$^1$ 

\vspace{.2in}

{\em DAMTP, Centre for Mathematical Sciences, Wilberforce Road, Cambridge CB3 OWA} \normalsize

\end{center}

\begin{abstract}

Relational particle mechanics (RPM) models bolster the relational side of the absolute versus relational motion debate.  
They are additionally toy models for the dynamical formulation of General Relativity (GR) and its Problem of Time (PoT).  
They cover two aspects that the more commonly studied minisuperspace GR models do not: 
1) by having a nontrivial notion of structure and thus of cosmological structure formation and of localized records.
2) They have linear as well as quadratic constraints, which is crucial as regards modelling many PoT facets.

I previously solved relational triangleland classically, quantum mechanically and as regards a local resolution of the PoT. 
This rested on triangleland's shape space being $\mathbb{S}^2$ with isometry group $SO(3)$, allowing for use of widely-known Geometry, Methods and Atomic/Molecular Physics analogies.
I now extend this work to the relational quadrilateral, which is far more typical of the general $N$-a-gon, represents a `diagonal to nondiagonal Bianchi IX minisuperspace' 
step-up in complexity, and encodes further PoT subtleties.
The shape space now being $\mathbb{CP}^2$ with isometry group $SU(3)/\mathbb{Z}_3$, I now need to draw on Geometry, Shape Statistics and Particle Physics to solve this model; 
this is therefore an interdisciplinary paper.
This Paper treats quadrilateralland at the classical level, and then Paper II provides a quantum treatment.

\end{abstract}

\vspace{0.2in} 


\vspace{0.1in} 


\mbox{ } 

$^1$ ea212@cam.ac.uk

\end{titlepage}

\section{Introduction}\label{Intro}

This paper concerns a toy model that exhibits many of the facets of, and strategies for, a notorious foundational problem of Quantum Gravity: the Problem of Time (PoT).  
This toy model turns out to be mathematically tractable by application of a number of inter-disciplinarities.   

\mbox{ } 

\noindent The PoT, which dates back to the insights of John Archibald Wheeler \cite{W68} and Bryce DeWitt \cite{DeWitt67} and is reviewed by Karel Kucha\v{r}, Chris Isham and the 
Author in {\cite{Kuchar92, I93, APoT, APoT2, FileR}, is often regarded as an incompatibility by what is meant by `time' in General Relativity (GR) and in Quantum Theory. 
However, I have emphasized \cite{ARel, APoT2, FileR} that this problem is already present (albeit simpler to handle) in classical models if one takes seriously the demands of 
background-independence that go back to Einstein and Ernst Mach.
Indeed the present paper is a purely classical treatment, though Sophie Kneller and I provide the quantum counterpart in Paper II \cite{QuadIII}.
See Secs 3, 4 and 5 for further outline of the PoT.  

\mbox{ } 

\noindent The toy model in question is the mechanics of the relational quadrilateral \cite{QShape, QSub}.
The mathematical space of relational quadrilaterals -- quadrilateralland -- is the complex-projective space $\mathbb{CP}^2$ \cite{Kendall84, Kendall}.
This physical role of this space is as a configuration space (see Sec 2). 
It is, furthermore, a relational configuration space. 
(This means that its configurations include no reference to absolute structures, most usually Newton's absolute space. 
I.e. absolute position and absolute axes relative to which absolute rotation occurs in Newton's conceptual framework.)   
Even more specifically, it is a {\it shape space} -- it makes no reference to (absolute) size either.  
This notion of shape space is familiar from Shape Geometry \cite{Kendall84, Kendall89, Small, Kendall}, in which references it is applied further to the study of Shape Statistics.  
In the present paper I make particular use of the seminal approach to these subjects that was initiated by David Kendall in the early 1980's.
The present paper exposits the application of this to Mechanics instead, noting that this is the right way to remove Newton's absolute space and thus meaningfully contribute to 
the relational side of the famous `absolute versus relational motion debate' that dates back to Newton versus Leibniz.  
The relational side of this debate was further developed by Mach \cite{M}, Wheeler \cite{W68, Wheeler-Mach-1, Wheeler-Mach-2} and Julian Barbour\footnote{Wheeler and Barbour were each supported 
by collaborators in these endeavours: mathematical relativists James Isenberg and Niall \'{o} Muchadha, and the theoreticians Bruno Bertotti and Brendan Foster.}
\cite{BB82, B94I, RWR-1, RWR-3}; although Einstein's own formulation of GR was only an indirect implementation of Mach's ideas, \cite{RWR-1, FileR} 
demonstrated that GR was {\sl also} formulable directly in relational Machian terms.  
Mach, Wheeler and Barbour's work moreover represents a {\sl pivot} between the relational side of this debate and the PoT as the continuation of this debate to the modern-physics arena 
of Quantum Gravity.  
This paper's type of Relational Particle Mechanics (RPM, see Sec 3) was first conceived of by Barbour and Bertotti in 1982 \cite{BB82}, and, following Mach \cite{M}, 
succeeds in also freeing oneself \cite{B94I, FileR} from Newton's absolute time. 
This latter aspect is one reason why such models are highly suitable as arenas in which to consider the PoT.  
Further reasons for this eminent suitability (below and in Secs 3, 6 and 11) revolve around the large number of analogies between RPM's and General Relativity (GR) 
written in dynamical form.  
\cite{BB82} itself is a scaled RPM; it took 20 further years for Barbour's RPM of pure shape \cite{B03} to appear.
However, once this did appear, I quickly noted that its notion of shape is the same as Kendall's \cite{FORD}, and that \cite{BB82} itself can be considered in scale--shape split form. 
By this, Kendall's grasp of the requisite mathematics unlocks the classical and quantum mechanics of that model also \cite{FileR}.  

\mbox{ } 

\noindent There are also a number of ties between RPM's and each of Celestial Mechanics \cite{ArchRat, Montgomery2, ARel2} and Atomic and Molecular Physics 
\cite{LR97, Zick1, Zick2, Zick3, ACG86, Tessi}.  
These and the shape space analogy entirely sufficed to solve relational triangleland at all of the following levels: classical, quantum-mechanical, and a local PoT resolution
\cite{08I, 08II, +tri, 08III, SemiclIII, ARel2, ACos2, AHall, FileR}.  
However, the next most complex model -- the present paper's relational quadrilateralland -- requires more than just these analogies.  
This is for the two reasons given below.  

\mbox{ } 

\noindent I first comment that Celestial Mechanics, Atomic and Molecular Physics operate in 3-$d$ due to being {\sl direct} attempts at modelling physical reality. 
However, RPM's value as PoT models is via the {\sl configuration space level analogy} with GR in dynamical form, which does {\sl not} require a match in the {\sl space} dimensions 
of the two theories involved. 
It then turns out (Sec 7) that 1-$d$ RPM's are not complex enough to encapsulate all of the PoT-significant features of GR that RPM's can encapsulate, but 2-$d$ RPM's {\sl are}, 
whilst 3-$d$ RPM's do {\sl no better} in this respect.
Also, 3-$d$ RPM's turn out to have {\sl vastly} harder mathematics than 2-$d$ ones. 
(This is for many of the reasons that the 3-$d$ $N$-body problem is hard and has to be treated case by case even for the smallest values of $N$.  
On the other hand, Kendall's work can be interpreted as showing that the 2-$d$ $N$-body problem has simple $\mathbb{CP}^{N - 2}$ mathematics, and this is {\sl systematic} for all $N$.)
Moreover, the first nontrivial 2-$d$ RPM -- triangleland -- has uncharacteristically simpler mathematics than any of the larger-$N$ `$N$-a-gonlands' due to the well-known geometrical 
relation $\mathbb{CP}^1 = \mathbb{S}^2$.   
Thus the step-up to quadrilateralland is significant from the point of view of ensuring typicality/robustness of the toy model in use.

\mbox{ } 

\noindent As a general principle, it makes much more sense to toy-model a theory with hard mathematics (here the PoT for GR) with a toy model with simple mathematics (here 2-$d$ RPM's) 
rather than with a toy model which has hard mathematics {\sl in a different way from the original} (here 3-$d$ RPM's with their 3-$d$ $N$-body problem complexities).  
Moreover, one should not confuse the classical and quantum mechanics of a model possessing simple mathematics with these models remaining simple at higher levels of study 
-- here the PoT. 
PoT calculations can become very formidable even when one's configuration space is $\mathbb{S}^2$.  
However, the ways in which these do become formidable {\sl parallel} many of the ways in which the PoT manifests itself in GR.  
However, then humankind's store of prior knowledge of Geometry and Methods of Mathematical Physics for $\mathbb{S}^2$ often comes to the rescue of the toy model. 
This allows for its calculations, unlike GR's, to {\sl be completed}.  
Relational triangle is indeed a whole-universe toy model (suitable for investigating closed-system and Quantum Cosmology considerations) that manifests many of the facets of, and 
strategies for, the PoT, {\sl whilst possessing mathematics no harder than that of the rotor and the basic model of the hydrogen atom} due to its underlying $\mathbb{S}^2$ shape space.
It is important to note at this point that QM unfolds not on space but on configuration space. 
The trouble is that space and configuration space {\sl coincide} for the case of the 1-particle models that are meaningful in the absolutist context in which basic Physics texts are cast. 
Only rarely then do Physics texts recommended in {\sl Theoretical Physics} go on to consider the QM of $N$-body systems for which configuration space and space cease to be the same. 
This makes clear the QM significance of having a simple configuration space -- here $\mathbb{S}^2$ -- for those trained in other than the quantum $N$-body problems of Molecular Physics.

\mbox{ } 

\noindent Reason 1) As explained above, specifically spatially 2-$d$ $N$-particle models are very largely not studied in the Molecular Physics literature.  
Triangleland remained lucky in this sense, due to the 2- and 3-$d$ 3-body problems being identical.  
Due to this, Molecular Physics still provides suitable classical kinematics/coordinate systems. 
E.g. the work of Alex Dragt \cite{Dragt} or of Richard Littlejohn and Kevin Mitchell \cite{Tessi} (Sec \ref{TriStart}).
Richard Montgomery \cite{ArchRat, Montgomery2} developed an approach in Celestial Mechanics that is similar to the preceding and to some of David Kendall's Shape Geometry work.  
However, from quadrilateralland upward, one needs a new source of such kinematics/coordinate systems.  
What is required here is use of results on $\mathbb{CP}^k$ from more specialized pure Geometry and Geometrical Methods for Physics literature.
For this I draw upon the work of Nicolaas Kuiper \cite{Kuiper} in pure Geometry, and of Gary Gibbons and Chris Pope on $\mathbb{CP}^2$ case of gravitational instantons 
(see also the work of Andrzej Trautman \cite{Trautman77}).  

\mbox{ } 

\noindent Reason 2) $\mathbb{S}^2$'s isometry group $SO(3)$ [which shares Lie algebra with $SU(2)$] are obviously crucial and very widely known in basic Atomic and Molecular Physics. 
This is because they encapsulate the mathematics of the individual particle in absolute space, in which context the quantum rotor, the atomic orbitals and subsequently the Periodic Table 
and much of the theory of molecules are based.  
This enters in e.g. how perturbed models of triangleland and 4-stop metroland (4 particles in 1-$d$, which also has $\mathbb{S}^2$ configuration space) have the same mathematics 
as the Stark Effect \cite{Stark} and the Raman Effect \cite{Raman} of Atomic and Molecular Physics.
Upon this transitioning to $\mathbb{CP}^{N - 2}$'s isometry group $SU(N - 1)/\mathbb{Z}_{N - 1}$ (which shares Lie algebra with $SU(N - 1)$. 
Here $\mathbb{Z}_p$ is the cyclic group of order $p$), Atomic Physics ceases to be at least a direct source of shared mathematical results. 
It is now, rather, Particle Physics which is such a source.  
In particular, the $\mathbb{CP}^2$ case's $SU(3)$ is well-known twice over in the theory of the strong interactions.
Firstly, it is the approximate flavour group, and secondly it is the exact colour group (which, moreover, by the arbitrary-label nature of the `red', `green' and `blue' colours ascribed 
to the quarks, is {\sl precisely} $SU(3)/\mathbb{Z}_3$: Alan MacFarlane's analogy \cite{MF79, MacFarlane}).  
Quantum quadrilateralland is then a cross between (in the sense detailed in Paper II) the Periodic Table and Gell-Mann's `eightfold way' interpretation of hadrons bound together by 
the strong force.
  
\mbox{ }

\noindent Note how one has gone from triangleland's {\sl double} use of Atomic/Molecular Physics for its kinematics and the consequences of its isometry group to two 
new and separate analogies for each of these for the quadrilateral (Geometry and Particle Physics).  
I sum up triangleland's and quadrilateralland's net of interdisciplinarities in Fig \ref{Interdisc}.  
On such a multiple frontier, please excuse that some of the notation will be unfamiliar to each reader, 
and that concepts basic to one's own discipline may not be obvious to those from others, and so require explanation from scratch.
%
{            \begin{figure}[ht]
\centering
\includegraphics[width=1.05\textwidth]{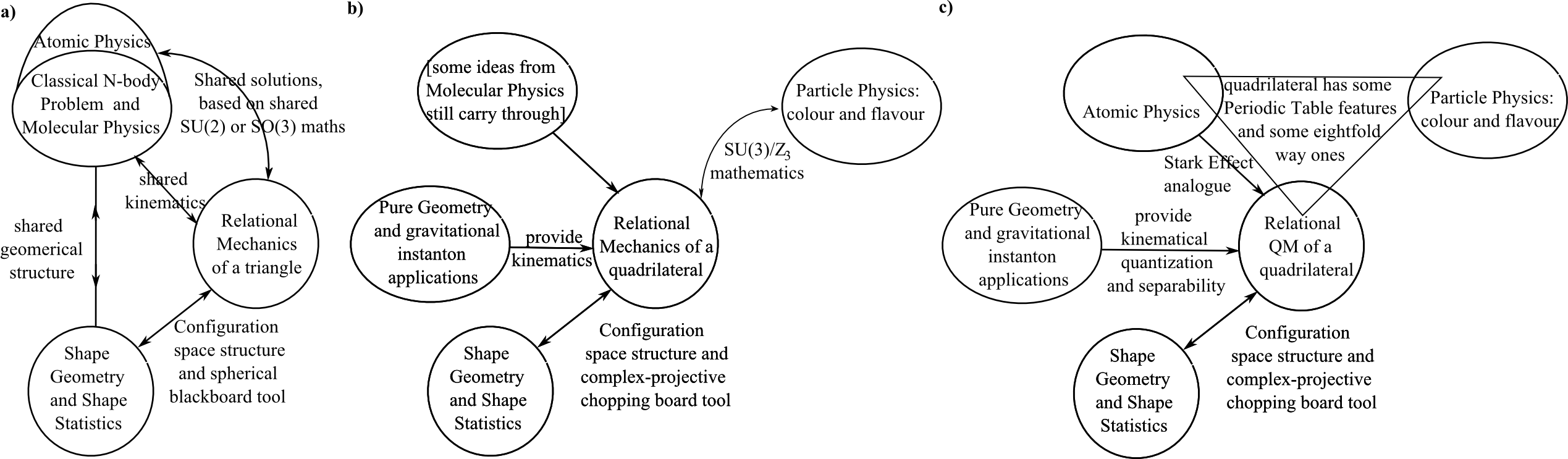}
\caption[Text der im Bilderverzeichnis auftaucht]{        \footnotesize{a) Triangleland's interdisciplinary connections. 
b) Quadrilateralland's more complicated net of interdisciplinary connections at the classical level.   
c) is how quadrilateralland's carry over to the quantum level.} }
\label{Interdisc} \end{figure}          }

\noindent This multitude of subjects that are drawn upon and cross-fertilized require a number of preliminary Sections. 
In Sec \ref{Config} I explain the important notion of configuration space in all of the classical and quantum $N$-body, RPM and GR contexts used in the present paper.
Then in Sec \ref{Relational}, I can make precise this program's meaning of the word `relational'. 
This permits introduction of the Principles of Dynamics actions for the RPM models and 
the dynamical formulation of GR that they are toy models of, the constraint equations that ensue from these and all of the facets of the PoT that are manifest at the classical level.  
I introduce the other 5 classical PoT facets in Sec \ref{Cl-PoT}.  
Secs \ref{Q-PoT} and \ref{PoT-Strat} then complete this with, respectively, outlines of the quantum versions of the PoT facets and of those strategies for facing the PoT that are used 
in the current program.
Secs \ref{RPM} and \ref{Quad-Mot} motivate, respectively, RPM models in general and quadrilateralland in particular.  
Sec \ref{2nd-Intro} then outlines the rest of this paper's material on quadrilateralland proper, as rendered comprehensible by the material in Secs \ref{Config}--\ref{Quad-Mot}.

\section{Configuration spaces}\label{Config}

It is useful to first consider the notion of configuration space $\fQ$.  
This is the set of possible values that the instantaneous configurations $Q^{\sfA}$ of one's theory can take, as illustrated by the following examples.  

\mbox{ } 

\noindent Example 1) The $q^{I\mu}$ particle position vectors of mechanics theories \cite{Lanczos} (for $I$ a particle index running from 1 to $N$: the number of particles, 
and $\mu$ a spatial index running from 1 to $d$ --- the dimension of space) straightforwardly form the configuration space $\mathbb{R}^{Nd}$.

\noindent Example 2) The relative Lagrange coordinates are whichever choice of basis of the relative inter-particle position vectors, $q^{I\mu} - q^{I\nu}$.  
An alternative representation for the information contained in these are the relative Jacobi vectors $\rho^{i\mu}$ (Sec \ref{Jacobi}).  
These generalize inter-particle position vectors to inter-particle {\sl cluster} vectors (between particles and/or centres of mass of clusters of particles).
They are commonly used in Celestial Mechanics \cite{Marchal} and in Molecular Physics \cite{LR97}.
Their advantage over the Lagrangian ones is that in terms of them the kinetic term remains diagonal.
In either case, the corresponding configuration space is $\mathbb{R}^{nd}$. 
Here, $i$ is a relative particle (cluster) index running from 1 to $n := N$ -- 1 for $N$.

\mbox{ }

\noindent What has been removed in passing from particle position vectors to relative vectors is the position of the centre of mass of the particles relative to an arbitrary 
absolute-space origin. 
Clearly this is a very straightforward operation mathematically; however, freeing oneself of the absolute axes with respect to which Newton's absolute rotation would occur is 
more problematic, especially in 3-$d$.  

\mbox{ } 

\noindent Example 3) Relational Particle Mechanics (RPM's) \cite{BB82, B94I, EOT, B03, FORD, Cones, ARel, FileR} are mechanics in which only relative times, relative angles and 
(ratios of) relative separations are physically meaningful.
A basis of relative angles and relative separations are jointly the configurations for scaled RPM, and a basis of relative angles and ratios of relative separations are the 
configurations for pure-shape RPM. 
These form, respectively, the relational space and shape space configuration spaces. 
See Sec \ref{TopMet} for an outline of the geometry and topology of these spaces.  

\noindent Example 4) Next, we consider full GR itself.  
While most commonly and historically encountered as a theory of 4-$d$ curved spacetime, this spacetime presentation can be decomposed by hypersurface geometry mathematics.  
This generalizes the very well-known Gauss's Outstanding Theorem that relates the intrinsic and extrinsic notions of curvature \cite{Frenkel}. 
Moreover, in the present application, this decomposition amounts to being a dynamical presentation in terms of evolving spatial 3-geometries that Wheeler therefore termed {\it geometrodynamics}. 
Here, the classical picture is that spacetime consists of evolving spatial 3-geometries, which play the role of configurations for GR. 
In fact, one usually works with redundant configurations -- the 3-metrics $h_{\mu\nu}$ of the spatial slices, which form the redundant configuration space Riem($\Sigma$) for 
$\Sigma$ the fixed topology that has to be shared by all the slices present in a spacetime in this formulation of GR.
As we shall see in the next Section in very general terms, such redundancy is nothing to worry about. 
[For now, note that the presence of the centre of mass motion in Example 1) makes it a more redundant description of ordinary Mechanics 
than in using Example 2)'s also-commonplace configurations.]  
The 3-geometries themselves are taken not to contain any coordinate information, that being physically empty through its being arbitrarily chosen out of convenience by 
people studying that system. 
By this they are indeed less redundant, the remaining information being a more complicated notion of (scaled) shape than the `$N$-a-gon in a plane' type notion of shape in Shape Geometry 
and Shape Statistics.\footnote{Note that this more complicated notion of shape has already received some geometrical study \cite{DeWitt67, Fischer70, Fischer86}, but has not, to the best 
of my knowledge, received a statistical study.  
That is (perhaps the hardest) suggestion for future interdisciplinary work based on the connections forged in this article.  
One should probably start considering such in the case of minisuperspace.}
Together, the 3-geometries form the less redundant configuration space Superspace($\Sigma$).  

\noindent Example 5) The previous example being in many ways hard, I end with consideration of a highly simplified version of it that General Relativist Charles Misner termed 
minisuperspace.  
One restricts attention here to spatial 3-metrics that are homogeneous, i.e. the same at each point in space. 
This example is simple enough that Riem and Superspace coincide as the configuration space minisuperspace($\Sigma$).  
Whereas it is immediately obvious that the scalefactor $a$ of Cosmology remains meaningful in minisuperspace, minisuperspace is also permitted to be anisotropic, 
with there existing a wide range of different Bianchi models of this nature.  
One of these that the present article makes useful comparison with is diagonal Bianchi IX, for which there are two independent anisotropy parameters $\beta_{\pm}$ \cite{MTW, Mini-1, 
Mini-2, Mini-3}.  

\mbox{ }

\noindent Finally, note that Mechanics configuration spaces have finite-dimensional positive-definite (Riemannian) metrics, $\bM$, whereas GR's has an infinite-dimensional indefinite 
(semi-Riemannian) metric upon it -- the so called inverse DeWitt supermetric \cite{DeWitt67}.  
Minisuperspace is finite-dimensional, whilst retaining this indefiniteness (provided that matter or anisotropy are included so as to have some of the infinity of plus signs 
to the scalefactor dynamics' highly-unusual minus sign).  
The indefinitenesses in this paragraph are not to be confused with the even more widely known indefiniteness of the spacetime metric itself.

\section{This program's meaning of `relational'} \label{Relational}

\noindent This simultaneously concerns the following two types of relationalism \cite{BB82, B94I, ARel, FileR}.

\mbox{ } 

\noindent 1) {\bf Temporal Relationalism}.  
Following Leibniz \cite{L}, one postulates that there is to be no time for the universe as a whole.  
In this sense, the Frozen Formalism Problem facet of the PoT, which is more usually only spoken of once it appears at the quantum level, is present right from the outset at 
the classical level. 
It arises from insisting on modelling closed universes from a background-independent perspective.  

\noindent Leibniz's postulate can be implemented mathematically by considering actions (in the Principles of Dynamics sense) that are 
a) parametrization-irrelevant and 
b) include no extraneous time-like variables such as absolute Newtonian time or the lapse \cite{MTW} (see Fig \ref{ADM}).  
Such actions are of Jacobi type \cite{Lanczos} for mechanics and a variant of Baierlein--Sharp--Wheeler (BSW) \cite{BSW} type actions for GR-as-geometrodynamics.  
For example, the Jacobi reformulation of mechanics itself involves the action 
\beq
S =  \sqrt{2}\int\sqrt{E - V(\bq^{I})}\,\d s \mbox{ } , \mbox{ } \mbox{ } \d s := \sqrt{\sumIN m_I \d \bq^{I\, 2}}
\label{Actio}
\eeq
for particle masses $m_I$, potential energy $V$ and constant total energy $E$.
Actions of this type were in use for many years prior to it being realized that these implement Temporal Relationalism in the work of Barbour and collaborators 
\cite{BB82, B94I, RWR-1, RWR-2, RWR-3, San, Phan, FEPI}; see \cite{B11, GrybTh, FileR, Pooley, ARel} for reviews.   
Such actions can additionally be viewed as associating a Mechanics to a given geometry, which then plays the role of configuration space for that theory: the Jacobi--Synge geometrical 
formulation of Mechanics \cite{Lanczos} (this is well-known in e.g. Celestial Mechanics).  

\mbox{ } 

\noindent Next, as envisaged in Dirac's work \cite{Dirac} on the Principles of Dynamics, parametrization-irrelevant actions oblige {\it primary constraints} to appear.  
These are constraints that arise solely due to the form of the Lagrangian rather than requiring any kind of variational procedure.  
For the above example, the structure of the Lagrangian is such that a single primary constraint arises that has quadratic and not linear dependence in the momenta: 
the energy constraint $T + V = E$. 
In the GR-as-geometrodynamics analogue below, we shall see that the corresponding constraint is the Hamiltonian constraint, 
whose form is well-known to lead to the {\sl quantum} Frozen Formalism Problem facet of the PoT. 
[It is, moreover, much less well known to arise by the complete chain of reasoning presented here.]

\mbox{ } 

\noindent Additionally, Leibniz's `there is no time for the universe as a whole' primary-level postulate can be countered at an emergent secondary level by Mach's Time Principle: 
`time is to be abstracted from change' \cite{M}.  
Then three varieties of such implementations concern `any change' (e.g. Carlo Rovelli's \cite{Rfqxi}), `all change' (e.g. Barbour's \cite{Bfqxi}) or my 
`sufficient totality of locally relevant change' (STLRC) \cite{ARel2}.  
A particular such emergent Machian time defined by the property of simplifying the momenta and equations of motion is the Jacobi--Barbour--Bertotti (JBB) emergent time 
\cite{B94I, SemiclI, FileR}, 
\beq
\lt^{\se\sm(\sJ\sB\sB)} =  \left.\int\d s \right/\sqrt{2\{E - V\}} \mbox{ }   
\eeq
for $\lt^{\se\sm(\sJ\sB\sB)} = t^{\se\sm(\sJ\sB\sB)} - t^{\se\sm(\sJ\sB\sB)}(0)$ i.e. allowing for an adjustment of `calendar year zero'.  
It attains this simplification by constituting a relational recovery of the Newtonian time of ordinary Mechanics.  
It implements Mach's time principle, at first sight in an `all change' manner since all changes are present in the $\d s$.  
However, the nontrivial and approximate way in which this time is in practise to be interpreted \cite{ARel2, FileR} takes one to a STLRC implementation.  

\mbox{ }

\noindent 2) {\bf Configurational Relationalism} \cite{BB82, RWR-1, RWR-2, RWR-3, Lan, FileR, ARel}.  
Here, a given configuration space $\fQ$ is subjected to some group $G$ of transformations that are held to be physically irrelevant.  
I use $g$ denotes the general element of $\fG$ and $\stackrel{\rightarrow}{\fG}$ to denote {\sl group action}.
For instance, for RPM's $\fG$ consists of translations, rotations and dilatations 
(or some subset of these due to whether scale itself is absolute or relative and the observation that removing the translations is mathematically trivial).
For geometrodynamics, $\fG$ consists of the spatial 3-diffeomorphisms,\footnote{These are, from a passive perspective, recoordinatizations embodying the 
`physics should not depend on choice of coordinates' maxim.  
However, the more modern active perspective considers these to be, rather, point-shuffling transformations corresponding to the deeper but superficially more 
counter-intuitive maxim that `no physical significance can be attached to points of the manifold in generally-relativistic physics'.
One can try to study Riem($\Sigma$) with larger or smaller groups of irrelevant transformations than Diff($\Sigma$).
One can include alongside these conformal transformations \cite{ABFO} or global volume preserving conformal transformations  \cite{ABFKO}.
However, one cannot use $\fG$ = id for GR \cite{Teitelboim73}, though this is consistent for strong gravity (i.e. the strong-coupled limit of GR) \cite{San}.}  
Diff($\Sigma$).
In fact, Physics is often presented in this way for mathematical convenience, 
since it is usually hard to work on the quotient space $\widetilde{\fQ} = \fQ/\fG$ for which such redundancy does not occur. 
Indeed, the term Configurational Relationalism covers both spatial relationalism and the internal groups of irrelevant transformations 
that are well-known from Particle Physics' own approach to Gauge Theory.
(It is rather less well-known that distinct gauge theories also occur in  Molecular Physics \cite{LR97}.)

\mbox{ } 

\noindent One particular indirect and widely useable implementation of Configurational Relationalism consists in building one's action not out of plain changes $\d{Q}^{\sfA}$ 
but out of $\fG$-corrected ones $\d_gQ^{\sfA} := \d{Q}^{\sfA} - \stackrel{\rightarrow}{\fG}_{\d{g}}Q^{\sfA}$.  
(For the examples given in the present paper, the group action on the $Q^{\sfA}$ themselves cancels out in the action). 
Combining this and Temporal Relationalism gives an action of the form (for the moment for finite theories as opposed to field theories)
\beq
S_{\sr\se\sll} = \sqrt{2}\int \sqrt{E - V(Q^{\sfC})}\d s \mbox{ } , \mbox{ } \mbox{ } 
\d s = ||\d_g \bQ||_{\sbM} := M_{AB}\d_g\d{Q}^{\sfA}\d_gQ^{\sfB} \mbox{ } . 
\eeq
\noindent For instance, for scaled RPM in relative Jacobi coordinates, the action implementing both Configurational and Temporal Relationalism is\footnote{ $\pi_{i\mu}$ are the momenta 
conjugate to $\rho^{i\mu}$ and $B^{\mu}$ is a rotational auxiliary variable.}
\beq
S = \sqrt{2}\int\sqrt{E - V(\rho^{i\mu})}\d s \mbox{ } , \mbox{ } \mbox{ } \d s := ||\dot{\brho} - \dot{\bB} \cr \brho|| \mbox{ } .
\label{Uuno}
\eeq
Then as for plain Mechanics, parametrization irrelevance implies as a primary constraint an energy constraint, 
\beq
{\cal E} := \delta^{IJ}\delta^{\mu\nu}\pi_{i\mu}\pi_{j\nu} \_  /2\mu_i + V(\bQ) = E \mbox{ } .    
\label{EnCon}
\eeq
However now also variation with respect to $B^{\mu}$ produces as a secondary constraint a zero total angular momentum constraint
\beq
{\cal L}_{\mu} := \sum_{i = 1}^n\{\rho^i \cr \pi_i\}_{\mu} = 0 \mbox{ } . 
\label{ZAM}
\eeq
This is how the indirect implementation of Configurational Relationalism works: although one has added $k$ auxiliary variables $g$, each of these produces a linear gauge constraint, 
and each of these uses up {\sl two} degrees of freedom.  
Thus indeed one is sent to a space with $k$ variables {\sl less} than $\fQ$: the quotient space $\fQ/\fG$ as required by Configurational Relationalism.
In the case of pure-shape RPM's, (\ref{ZAM}) is supplemented by the zero total dilational momentum constraint for the whole universe, 
\beq
{\cal D} := \sum\mbox{}_{\mbox{}_{\mbox{\scriptsize $i = 1$}}}^n \brho^i \cdot \bpi_i = 0 \mbox{ } .  
\eeq
This arises from variation of a dilational auxiliary $C$ that occurs as a further correction $\d C \rho^{i\mu}$ to the $\d \rho^{i\mu}$.
In the general case, I denote the linear constraints thus produced by Lin$_{\sfZ}$.
\noindent In the presence of Configurational Relationalism, the emergent JBB time becomes subjected to the above extremization becoming attainable, 
\beq
\lt^{\se\sm(\sJ\sB\sB)} = \stackrel{\mbox{\scriptsize extremum}}{\mbox{\scriptsize $g$ $\in$ $\fG$ \mbox{ }  of $S_{\tr\te\tl}$}  }                                                              
\left(                                                              
\int||\d_{g}\bQ||_{\sbfM}/\sqrt{2\{E - V(\bQ)\}} 
\right)  \mbox{ } .  
\label{G-tem}
\eeq
\noindent Note that in Atomic and Molecular Physics, one does not take out the absolute rotations. 
These subjects only treats these in a weaker sense, by splitting them off from the relational degrees of freedom. 
This is done in a manner that corresponds to an atom or a molecule {\sl in} a larger universe, rather than considering an `atom or molecule' that {\sl is} a whole-universe model.  
This parallel still permits RPM's to borrow classical kinematics and useful coordinate systems from Molecular Physics.   
However, the buck stops at the quantum level, at which level the presence of a separated-out absolute sector causes nontrivial differences with what happens in purely relational models 
\cite{Cones, FileR, QuadIII}. 

\mbox{ } 

\noindent Finally, let us turn to the GR case.
GR is usually and historically treated as a spacetime theory, the action for which is Einstein--Hilbert's:\footnote{$g_{\Gamma\Delta}$ is the spacetime metric on a manifold M,
with determinant $g$ and Ricci scalar curvature R.}
\beq
S_{\sE\sH} = \int_{\sM} d^4x\sqrt{|g|} \mR \mbox{ } . 
\eeq
{            \begin{figure}[ht]
\centering
\includegraphics[width=0.8\textwidth]{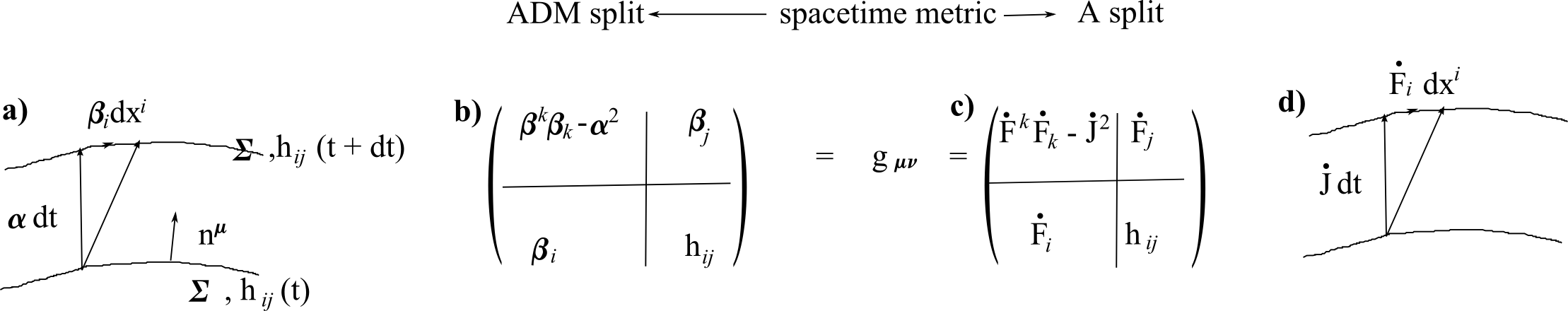}
\caption[Text der im Bilderverzeichnis auftaucht]{        \footnotesize{b) ADM split, with geometrical significance a). 
c) The A split, with geometrodynamical significance d).} }
\label{ADM} \end{figure}          }
%
A dynamical formulation for GR then follows from splitting $g_{\Gamma\Delta}$ in Arnowitt--Deser--Misner (ADM)'s way as per Fig \ref{ADM} a) and b) and applying a hypersurface geometry 
relation to R.  
This gives the ADM action for GR-as-geometrodynamics,\footnote{For geometrodynamics, the spatial topology $\bupSigma$ is fixed and taken to be compact without boundary  
for simplicity.  
$h_{\mu\nu}$ is then a spatial 3-metric thereupon, with determinant $h$, covariant derivative $D_{\mu}$, Ricci scalar $R$ and conjugate momentum $p^{\mu\nu}$.  
$\pounds_{\beta}$ is the Lie derivative with respect to the shift $\beta^{\mu}$.  
The GR configuration space metric is ${\fM}^{\mu\nu\rho\sigma} := h^{\mu\rho}h^{\nu\sigma} - h^{\mu\nu}h^{\rho\sigma}$, i.e. 
the undensitized inverse DeWitt supermetric \cite{DeWitt67} with determinant ${\fM}$ and inverse ${\fN}_{\mu\nu\rho\sigma}$ that is itself 
the undensitized DeWitt supermetric, $h_{\mu\rho}h_{\nu\sigma} - h_{\mu\nu}h_{\rho\sigma}/2$. 
The densitized versions are $\sqrt{h}$ times the former and $1/\sqrt{h}$ times the latter.}
\beq
S_{\sA\sD\sM} = \int \d\mt \int_{\Sigma}\d^3x\sqrt{h} \alpha \{||\dot{h}_{ab} - \pounds_{\beta}h_{ab}||_{\sbM}\mbox{}^2/\alpha^2 +  R\} \mbox{ } .
\label{S-ADM-L}  
\eeq
Following Wheeler's work with his students Baierlein and Sharp, one can then eliminate the lapse $\alpha$ from its own Lagrange multiplier variational equation to arrive at the BSW action, 
\beq
S_{\sB\sS\sW} = \int \d\lambda \int \d^3x \sqrt{h} \sqrt{R} ||\dot{h}_{ab} - \pounds_{\beta}h_{ab}||_{\sbM} \mbox{ } , \mbox{ } \mbox{ }  
\label{BSWaction}   
\eeq
This looks quite a lot like the above relational Jacobi action with meaningless label-time parameter $\lambda$'s inserted to match, 
except that the shift $\beta^{\mu}$ spoils the parametrization.  
This is remedied by using Fig \ref{ADM}'s other split, which replaces the shift by the velocity of a new variable, the `frame', $F^{\mu}$.  
Thus one arrives at the action the Author constructed with Barbour, Foster and \'{o} Murchadha, which casts GR in a fully relational form: 
\beq
S_{\sr\se\sll} = \int\d^3x\sqrt{h}\int_{\sbSigma}\sqrt{R}\,\d s \mbox{ } , \mbox{ } \mbox{ } \d s := ||\d {h}_{ab} - \pounds_{\d F}h_{ab}||_{\sbfM} \mbox{ } .  
\eeq
\noindent From parametrization irrelevance \cite{Dirac, FileR}, one gets as a primary constraint the quadratic GR Hamiltonian constraint, 
\beq
{\cal H} := \fN_{\mu\nu\rho\sigma}p^{\mu\nu}p^{\rho\sigma}/\sqrt{h} - \mbox{Ric$(x^{\mu}; h_{\mu\nu}]$}\sqrt{h} = 0 \mbox{ } . 
\eeq
Furthermore, from variation with respect to $F^{\mu}$, one gets the GR momentum constraint 
\beq
{\cal M}_{\mu} := - 2D_{\nu}{p^{\nu}}_{\mu} = 0 \mbox{ } .  
\eeq 
I also note that the GR case of $t^{\se\sm(\sJ\sB\sB)}$ amounts to the relational recovery of proper time and of cosmic time in suitable contexts.

Solving the GR momentum constraint at the Lagrangian level for the auxiliary $\beta$ or $\dot{F}^{\mu}$ is the {\it Thin Sandwich Problem} \cite{BSW, W63} of geometrodynamics. 
This is traditionally taken to be a second PoT facet \cite{Kuchar92, I93}. 
However, this facet generalizes firstly to the arbitrary-$\fG$ case of this extremization/Lagrangian-level reduction procedure.  
I.e. the {\it Best Matching Problem} \cite{BB82, RWR-1, RWR-2, RWR-3, Lan, B11, GrybTh, FileR, ARel} as first envisaged by Barbour.
Secondly, to Configurational Relationalism itself applied at {\sl whatever} level of the structure of the theory \cite{APoT2}.   
\noindent Note that RPM's in 1- and 2-$d$ do have this extremization resolved \cite{FileR}.  
This follows from the kinematics of the pure-shape problem producing parallel mathematics to that in Kendall's work \cite{Kendall84, Kendall}\footnote{This was 
applied to mechanics in \cite{FORD, FileR}.}
on Shape Statistics.   
The scaled case's kinematics coincides with the coning over the shape space geometry \cite{Cones, FileR}, which also has some links with Molecular Physics \cite{LR97}.  
These results can also be attained by reduction as per \cite{TriCl, FORD, FileR}.  
Thus this review serves as an example of how to resolve the PoT in approaches that eliminate Configurational Relationalism right at the outset. 
Overall, solving the Best Matching Problem facet of the PoT once and for all at the classical level gives one an explicit\footnote{This is as opposed to (\ref{G-tem}), in which 
one needs to carry out an extremization in the process of computing the emergent JBB time.} 
expression for the emergent JBB time that solves the classical Frozen Formalism Problem.

\section{The other five classical Problem of Time facets}\label{Cl-PoT}

\noindent In addition to 1) and 2) above being reformulations of two of the PoT facets at the classical level, there are five further facets of the PoT that are present at the 
classical level.  

\mbox{ } 
 
\noindent 3) {\bf The Classical Problem of Beables}.  
This is more commonly known as the Problem of Observables. 
However, the notion of beable (due to quantum physicist John Bell \cite{Bell}) is a quantity that just {\sl is}, as opposed to an observable which implies the role of an observation 
process with the difficult and unresolved Foundations of QM connotations that such processes carry.  
Furthermore, the notion of beables is particularly relevant to the study of the early universe. 
This did not contain any observers. 
It also goes hand in hand with QM wavefunction collapse due to decoherence by natural phenomena as opposed to by observation.
The observable/beable concept does have some classical roots, as a Poisson brackets precursor to the quantum-mechanical condition of such quantities having to commute with 
one's theory's constraints.  
There are three different conceptualizations of this, which differ in stringency (with how many constraints the observable/beable has to commute with). 
{\it Classical Dirac beables} \cite{OldDir} are functionals\footnote{I denote 
these with square brackets and functions with round brackets.} 
D = F[$Q^{\sfA}$, $P^{\sfA}$] that Poisson-bracket-commute with {\sl all} of a theory's constraints. 
{\it Classical \K beables} \cite{Kuchar93, Kuchar99, BF08} are functionals K = F[$Q^{\sfA}$, $P^{\sfA}$] that are just required to Poisson-bracket-commute with the theory's {\sl linear} 
constraints, Lin$_{\sfZ}$.
{\it Rovelli's partial observables} \cite{Rovellibook, Rfqxi} do not require commutation with any constraints.  
The Problem of Beables is then that in both of the first two of these conceptualizations (especially the first), 
it is hard to construct a sufficiently large set of these to describe physical theory, particularly for gravitational theory.  

\mbox{ }

\noindent Note 1) It is useful to remark that \K beables are an uncontroversial notion of gauge-invariant quantities for the gauge group $G$ that corresponds to the Lin$_{\sfZ}$ 
involved in their definition. 
It is rather less clear whether Quad --- the generalized quandratic constraint that includes both the GR Hamiltonian constraint ${\cal H}$ and the RPM energy constraint ${\cal E}$ --- 
itself can be interpreted as a gauge constraint, which would imply that evolution is gauge and therefore unphysical.  

\noindent Note 2) In addition to having an explicit JBB time, a second consequence of solving the Best Matching Problem is the possession of a full set of classical \K beables. 
Thus one has classical \K beables available for 1- and 2-$d$ RPM's.   
Additionally, quantum theoretician Jonathan Halliwell \cite{H03, H09-1, H09-2} has shown how to construct objects that Poisson-brackets-commute with Quad in $\fG$-free theories, 
and I have shown how to extend this construction to triangleland \cite{AHall}. 
Halliwell's construction does however involve Histories Theory, and as such is not addressed until Paper III.  

\mbox{ }   

\noindent 4) {\bf Constraint Closure Problem}.  
This involves the possibility that Dirac-type evolution procedures for handling a theory's constraints produces further constraints in addition to Quad and the Lin$_{\sfZ}$ that 
corresponds to the $\fG$ under investigation.  
One can establish by computation of the Poisson brackets between the constraints that this is not the case for classical RPM's or for GR. 
(In the latter case one obtains thus the `Dirac algebroid' of constraints \cite{Teitelboim73, HKT76, BojoBook}).

\noindent 5) {\bf Foliation Dependence Problem}. 
This concerns the desirable property that evolving from slice 1 to slice 2 does not depend on how spacetime is envisaged to be sliced in between these 
(This is via the dotted hypersurface or the dashed hypersurface in Figure \ref{PathIndep}.) 
This is a desirable result since how one foliates one's spacetime is closely akin to how one coordinatizes it, which is not to affect the physics of a generally-relativistic theory. 
Claudio Teitelboim demonstrated that this works out fine in classical GR \cite{Teitelboim73}, via this property being a geometrical interpretation of the Dirac algebroid commutator.  

{            \begin{figure}[ht]
\centering
\includegraphics[width=0.35\textwidth]{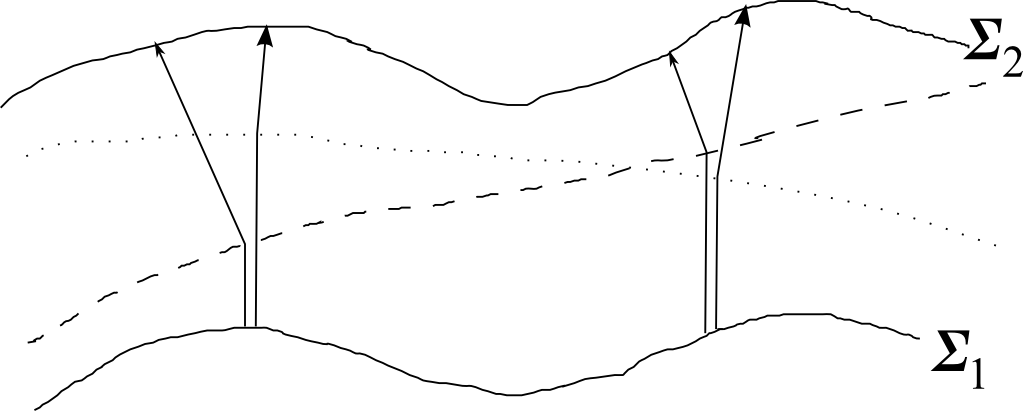}
\caption[Text der im Bilderverzeichnis auftaucht]{        \footnotesize{b) Path independence in GR as geometrodynamics: the only difference between evolving via the dashed hypersurface 
and via the dotted hypersurface is a coordinate transformation on $\Sigma_2$.} }
\label{PathIndep} \end{figure}          }

\noindent 6) {\bf Spacetime Reconstruction Problem} While we shall see further motivation for this at the quantum level in the next Section, 
it already has a counterpart at the classical level.  
This concerns taking space as primary and seeing if one can recover GR spacetime without presupposing features of it. 
This is a down-up counterpart to ADM's top-down approach, as performed by Barbour, Foster and \'o Murchadha \cite{RWR-1} and 
subsequently refined by the Author and Flavio Mercati \cite{Phan, Lan2, AM13}.  

\noindent 7) {\bf Global issues with the PoT} unfortunately lie outside the scope of the present program, which only investigates a {\sl local} resolution of the PoT.  
I also comment that 5) and 6) are trivially non-issues for RPM models themselves, since these have no notion of GR spacetime.

\section{Quantum Problem of Time facets}\label{Q-PoT}

\noindent At the quantum level, the PoT has further roots in the conflict of definitions between `time' in each of GR and ordinary Quantum Theory 
\cite{Kuchar92, I93, Kieferbook, Rovellibook, APoT, APoT2, FileR}.  
This incompatibility underscores a number of problems with trying to replace these two branches with a single framework in situations in which both apply, 
such as in black holes or the very early universe.  

\mbox{ } 

\noindent 1) {\bf Quantum Frozen Formalism Problem}.  
The purely quadratic Quad = 0 becomes 
\beq
\widehat{\mbox{Quad}}\Psi = 0
\label{FFP}
\eeq
(as opposed to e.g. 
\beq
\widehat{\mbox{Quad}}\Psi = i\pa_t\Psi
\eeq
for some notion of time $t$). 
In specific GR case, (\ref{FFP}) is the Wheeler--DeWitt equation \cite{DeWitt67, W68} $\widehat{H}\Psi = 0$.
This is the more well-known quantum-level Frozen Formalism Problem, and an apparent disaster strikes. 
I.e. using $t^{\se\sm(\sJ\sB\sB)}$ fails to unfreeze this equation, so that the classical resolution of this facet needs to be abandoned.  
Moreover, as we shall see in the next Sec, a striking resolution of this is that the emergent semiclassical time approach produces a new timestandard that is both approximately aligned 
with $t^{\se\sm(\sJ\sB\sB)}$ and itself as much of an `all change' or STLRC implementation of Mach's `time is to be abstracted from change' as $t^{\se\sm(\sJ\sB\sB)}$ is.  
One can think of this as a somewhat more `bottom up' recovery of the classical notion. 
Furthermore, semiclassicality suffices for e.g. many quantum-cosmological applications such as studying the possible quantum origin of galaxies and of cosmic microwave background 
inhomogeneities \cite{HallHaw}.

\mbox{ }

\noindent 2) For the present program for 1- and 2-$d$ RPM's, having solved the Configurational Relationalism facet of the PoT at the classical level, it remains solved.  

\mbox{ } 

\noindent 3) One can likewise write down quantum \K beables, modulo having to choose a suitable subalgebra of the classical ones to involve in the quantization (see Paper II).
Also, Halliwell has provided a {\sl separate} method for constructing semiclassical quantities that commute with $\widehat{\mbox{Quad}}$.
I have again combined the two methods for 1- and 2-$d$ RPM's \cite{AHall, QuadII, FileR} in order to produce quantum Dirac beables.  

\mbox{ } 

\noindent 4), 5), 6) all become unresolved problems for GR at the quantum level \cite{Kuchar92, I93}.
Moreover, the quantum level is much of the motivation for the Spacetime Reconstruction Problem. 
For, the dynamical object are 3-geometries, it is the dynamical objects that undergo QM fluctuations, and once 3-geometries fluctuate, these fluctuations no longer fit within a single 
spacetime \cite{W68}.   
Thus at the quantum level the notion of spacetime appears to have dissolved.
The problem is then how do we recover this notion in a suitable (semi)classical limit?
For RPM's, however, 5) and 6) have no meaning and 4) can be confirmed to work out by straightforward computation.  

\mbox{ } 

\noindent 7) The Global PoT is harder at the QM level.

\noindent 8) The Multiple Choice Problem also appears at the quantum level (that different choices of time can lead to inequivalent quantum theories).  

\noindent However, the current program for now focusses on the well-posed question of `a local resolution' of the PoT, which is independent of the further issues of addressing 7) and 8).

\section{Strategies for the Problem of Time}\label{PoT-Strat}

\noindent Some of the strategies toward resolving the Problem of Time in Quantum Gravity that can be modelled by RPM's are as follows (see \cite{Kuchar92, I93, FileR} 
for a more complete set of known PoT strategies).

\mbox{ } 

\noindent A) Perhaps one has slow heavy `$\mh$'  variables that provide an approximate timestandard with 
respect to which the other fast light `$\ml$' degrees of freedom evolve \cite{DeWitt67, Banks, HallHaw, Kuchar92, Kieferbook}.  
In the Halliwell--Hawking \cite{HallHaw} scheme for GR Quantum Cosmology, $\mh$ is scale (and homogeneous matter modes) and $\ml$ are small inhomogeneities.
Thus the scale--shape split of scaled RPM's afford a tighter parallel of this \cite{MGM, SemiclIII, Forth} than pure-shape RPM's.    
The Semiclassical Approach involves 

\noindent 1) making the Born--Oppenheimer ansatz $\Psi(\mh, \ml) = \psi(\mh)|\chi(\mh, \ml) \rangle$ 
This ansatz's name comes from its use in Molecular Physics, where the positions of the nuclei are h and the electrons are l.

\noindent 2) One then makes the WKB ansatz $\psi(\mh) = \mbox{exp}(iW(\mh)/\hbar)$.  
Such an ansatz is well-known throughout Quantum Physics (and even other kinds of Wave Physics). 
However, I must caution here that the reasons for its applicability elsewhere, such as careful laboratory preparation of the `in' state, themselves do not carry over to Quantum Cosmology. 
Thus there will be more to say below about the justifiability of this ansatz there.

\noindent 3) Next, one forms the $h$-equation ($\langle\chi| \widehat{\mbox{Quad}} \Psi = 0$ for RPM's), which, under a number of simplifications, 
yields a Hamilton--Jacobi\footnote{For simplicity, I 
present this in the case of 1 $\mh$ degree of freedom and with no linear constraints.} 
equation
\beq
\{\pa W/\pa \mh\}^2 = 2\{\fE - \fV(\mh)\} \mbox{ } 
\label{HamJac} 
\eeq
for $\fV(\mh)$ the $\mh$-part of the potential. 

\noindent 4) Then one way of solving this is for an approximate emergent semiclassical time,  $t^{\se\sm} = t^{\se\sm}(\mh)$. 

\noindent 5) One also forms the $\ml$-equation.  
This starts off as $\{1 - |\chi\rangle\langle\chi|\}\widehat{\mbox{Quad}}\Psi = 0$. 
This is manifestly a fluctuation equation. 
However, it can then be recast (modulo further approximations) as an emergent-time-dependent Schr\"{o}dinger equation for the $\ml$ degrees of freedom, 
\beq
i\hbar\pa|\chi\rangle/\pa t^{\te\tm((\sW\sK\sB)}  = \widehat{H}_{\sll}|\chi\rangle \mbox{ } .  
\label{TDSE2}
\eeq
(Here the left-hand side arises from the cross-term $\pa_{\sh}|\chi\rangle\pa_{\sh}\psi$ and $\widehat{H}_{\sll}$ is the remaining surviving piece of $\widehat{H}$).  
Note that the working leading to such a time-dependent wave equation ceases to work in the absense of making the WKB ansatz and approximation. 
Furthermore, in the quantum-cosmological context, is not known to be a particularly strongly supported stipulation.    
See Paper II for the quadrilateralland treatment of this approach. 

\mbox{ }

\noindent B) A number of approaches take timelessness at face value. 
One considers only questions about the universe `being', rather than `becoming', a certain way.  
This has at least some limitations as regards what questions it can cover in practise, but can, nevertheless, address some questions of interest. 

\mbox{ } 

\noindent Example 1) the {\it Na\"{\i}ve Schr\"{o}dinger Interpretation} \cite{HP86, UW89} concerns the  
probabilities of `being' for universe properties such as: what is the probability that the universe is large? 
Flat? 
Isotropic? 
Homogeneous?   
One obtains these by considering the probability that the universe belongs to region R of the configuration space that corresponds to a quantification of the property in question, 

\noindent
\beq 
\mbox{Prob(R)} \propto \int_{\sR}|\Psi|^2\d\Omega \mbox{ } , 
\label{NSI}
\eeq 
for $\d\Omega$ the configuration space volume element.
See Paper II also for the quadrilateralland treatment of this approach. 

\mbox{ } 

\noindent Example 2) {\it Records Theory} (\cite{PW83, GMH, EOT, H99, Records1, Records2} and Paper III) concerns localized subconfigurations of a single instant.  
More concretely, it concerns whether these contain useable information, are correlated to each other, and a semblance of dynamics or history arises from this.  
This requires notions of localization in space and in configuration space as well as notions of information.  
One can view Shape Statistics problems such as the standing stones alignment problem from Archaeology \cite{Kendall84} as a classical-level model of Records Theory \cite{FileR, AStats}. 
RPM's are superior to minisuperspace for a study of records since, firstly, they have a notion of localization in space. 
Secondly, they have more options for well-characterized localization in configuration space 
(i.e. of `distance between two shapes' \cite{FileR}) through their kinetic/configuration space metrics being positive-definite.  
See Paper III for the quadrilateralland treatment of this approach. 

\mbox{ }

\noindent C) Perhaps instead it is the histories that are primary: {\it Histories Theory} \cite{GMH, Hartle} and Paper III).    

\mbox{ }

\noindent Combining A) to C) (for which RPM's are well-suited) is a particularly interesting prospect \cite{H03}, along the following lines. 
(See also \cite{GMH, H99, H09-1, H09-2, AHall, FileR, CapeTown12} and Paper IV for further development of this.)   
There is a Records Theory within Histories Theory.  
Histories decohereing (`self-measuring') is one possible way of obtaining a semiclassical regime in the first place. 
I.e. finding an underlying reason for the crucial WKB assumption without which the Semiclassical Approach does not work.    
What the records are will answer the elusive question of which degrees of freedom decohere which others in Quantum Cosmology.

\section{Motivating RPM models in general}\label{RPM}

\noindent Many configuration space analogies between RPM's and GR are given in Fig \ref{Fig2}. 
\noindent Both are relational theories, with subsequent matching quadratic and linear constraints.  
\noindent Furthermore, RPM's exhibit 6 of the 8 facets of the Problem of Time (all bar Foliation Dependence and Spacetime Reconstruction), 
and many of the strategies for dealing with it, including all of those in the present program.  
\noindent \cite{Kuchar92, B94I, EOT, RWR-1, B03, Piombino-1, Piombino-2, Piombino-3, Piombino-4, Piombino-5, Piombino-6, Piombino-7, Piombino-8, Kieferbook, 06I, 06II, SemiclI, ScaleQM, MGM, 
Records1, Records2, FEPI, AF, Cones, AHall, ARel, ARel2, CapeTown12, Lostaglio, BKM13, ACos2} and especially \cite{FileR}) support this analogy. 

\mbox{ } 

\noindent Also, RPM's are useful for the qualitative study of the quantum-cosmological origin of structure formation/ inhomogeneity. 
(Scaled RPM's are a tightly analogous, simpler version of Halliwell and Hawking's \cite{HallHaw} model for this. 
Moreover, scalefree RPM's such as this paper's occurs as a subproblem within scaled RPM's, corresponding to the l-modes/inhomogeneities.) 
Relatedly, RPM's are likewise useful for the study of correlations between localized subsystems of a given instant. 
RPM's also allow for a qualitative study of notions of uniformity/of maximally or highly uniform states in Classical and Quantum Cosmology, 
which are held to be conceptually important notions in these subjects.

\section{Motivating Quadrilateralland in particular}\label{Quad-Mot}

\noindent 1) One needs $N$-a-gonland to have simultaneously scale-driven cosmology and nontrivial linear constraints.  

\noindent 2) One needs at least quadrilateralland for mathematical typicality (complex-projective mathematics that cannot be reduced to spherical mathematics).

\noindent 3) Passing from triangleland to quadrilateralland is comparable in step-up in complexity from diagonal to nondiagonal Bianchi IX minisuperspace models.  
Moreover, it is to be emphasized that these RPM's already possess two significant midisuperspace features each.
Namely 1) and nontrivial clustering/inhomogeneity/structure. 

\noindent 4) Moreover, one needs at least quadrilateralland in order to have 1) and the first nontrivialities in subsystem structure, as are appropriate for 
structure formation and localized records.
Namely, relationally nontrivial non-overlapping subsystems and hierarchies of nontrivial subsystems (see the Conclusion for further detail of these).

\mbox{ } 

\noindent Note 1) Modelling features 1)--4) ogether render this suitable as a qualitative toy model 
of Halliwell and Hawking's \cite{HallHaw} quantum cosmological origin of structure formation in the universe.

\noindent Note 2) Using multiple toy models to cover various different PoT facets and strategies is recommended in e.g. the reviews \cite{Kuchar92, I93, Kieferbook}. 

\mbox{ } 

\noindent Further motivations from Quadrilateralland having $\mathbb{CP}^2$ and $SU(3)$ mathematics are as follows.  
In the $\mathbb{CP}^2$ context, the $SU(3)$ group is nonlinearly realized; this has been studied by MacFarlane \cite{MacFarlane, MF03b, MF79} 
(some results of which are already in \cite{BGM71}; also see \cite{W82b}).  
This work of MacFarlane's represents the start of the present paper's quadrilateralland interpretation of $\mathbb{CP}^2$'s $SU(3)$ of conserved quantities.  
Other occurrences of $\mathbb{CP}^2$ mathematics in Physics include the following.  

\noindent A) qutrits in Quantum Information Theory \cite{QuTrit-1, QuTrit-2, QuTrit-3}; these can also be interpreted in terms of quadrilaterals and qu$n$its in terms of $N$-a-gons for $N = n + 1$.
These are motivated by there being much more information storage in a system based on qu$n$its rather than on qubits.  

\noindent B) $SU(3)$ corresponding to $\mathbb{CP}^2$ also occurs in the theory of skyrmions and in nonlinear sigma models. 
Each of these are theoretical concepts in Particle Physics along the following lines.  
Skyrmions are good toy models \cite{Skyrme} for some aspects of the Yang--Mills theory of the strong and electroweak forces (instantons, $\theta$-vacua and topological quantum numbers).
On the other hand: nonlinear sigma models \cite{+Sig-1, +Sig-2} usefully model a number of aspects of diffeomorphisms and of string-theoretic target spaces. 
$\mathbb{CP}^2$ is here present in its quotient form as the target space.  

\noindent C) Complex projective spaces play a prominent role in Penrose's twistor theory approach to GR (and the whole of Theoretical Physics more generally).   
For instance, planes in projective twistor space are given by $\mathbb{CP}^2$ spaces \cite{Huggett}.  

\noindent D) Finally, $\mathbb{CP}^2$ is well-known to be Einstein. 
(I.e. its Ricci tensor is constantly proportional to its metric tensor.)
Hence it is a solution of the Euclidean-signature version of GR \cite{GiPo1, GiPo2}, which is the defining condition for gravitational instantons \cite{GH79-1, GH79-2}. 
This is how Gibbons and Pope's work on gravitational instantons comes to be of relevance in the present paper.

\section{Outline of the subsequent quadrilateralland sections}\label{2nd-Intro}

I can provide this now that enough basic exposition has been accrued for interdisciplinary readers.  
\noindent Sec \ref{Jacobi} casts the general RPM in terms of the Jacobi coordinates of Molecular Physics/Celestial Mechanics.
\noindent Sec \ref{TopMet} describes the shape spaces of Kendall and the cones over these that are relevant to scaled RPM's.
\noindent Sec \ref{Q-1} covers the use of tessellations. 
This was particularly emphasized in Kendall's work \cite{Kendall84}: interpreting statistics of triangles.  
However, Montgomery \cite{ArchRat, Montgomery2} does also make use of a simple tessellation in Celestial Mechanics. 
Littlejohn, Mitchell and Aquilanti also make use of another \cite{Tessi} in Molecular Physics.
In the present paper's context, tessellations are used as a back-cloth for reading off the meaning of dynamical trajectories and of {\sl QM} probability distributions. 
This follows from Kendall's insightful use of the `spherical blackboard' presentation of the triangleland shape sphere for use in Shape Statistics \cite{Kendall84, Kendall89}.  
\noindent Sec \ref{Demo} applies the Molecular Physics theory of democracy transformations and democracy invariants to the case of the quadrilateral. 
This gives the Kuiper coordinates and their extension to what I term the Gell-Mann quadratic forms -- an early manifestation of the eventual isometry group of quadrilateralland being, 
at least locally, $SU$(3).  
\noindent Sec \ref{Q-2} covers the most useful coordinate systems for the simpler RPM's. 
I.e. the obvious interpretation of spherical coordinates for 4-stop metroland, and a less obvious interpretation of spherical coordinates for triangleland, 
for which parabolic-type coordinates also play a significant role.
Also homogeneous and inhomogeneous coordinates for $\mathbb{CP}^2$ do not suffice for some purposes, for which one needs the Kuiper and Gibbons--Pope type coordinates provided.    
This Sec ends with RPM-GR configuration space comparison and notions of collapsing Jacobi vectors to pass to smaller sets of such.
\noindent Sec \ref{Subman} considers submanifolds - as good an analogue of tessellations as one can get for a 4-$d$ shape space.  
With the two relative angle coordinates suppressed, I present the complex-projective chopping board analogue of the spherical blackboard, with the shapes thereupon described as axes. 
The axe-head is the triple of particles privileged by the coordinate system in question.  
\noindent Sec \ref{GM-GP} casts the Gell-Mann quadratic forms into Gibbons--Pope type coordinates, which turns out to be a useful preliminary for kinematical quantization in Sec II.2. 
\noindent Sec \ref{Unif} covers Uniformity and that it is a useful classical and quantum cosmological application.  

\mbox{ } 

\noindent Next, in Sec \ref{Mom-Int}, I provide and interpret the shape momenta conjugate to the shape coordinates for  the various RPM theories up to quadrilateralland.
In Sec \ref{Hams}, I consider the corresponding Hamiltonians.  
Secs \ref{SSec: Cons} and \ref{Quad-Case} consider the isometries for RPM's up to quadrilateralland.  
\noindent I previously studied \cite{08I, +tri} how triangleland has $SO(3) = SU(2)/\mathbb{Z}_2$ as its isometry group.  
On the other hand, $N$-stop metroland has $SO(N - 1)$ as its isometry group, among which 4-stop metroland's is $SO$(3) again.  
This led to spherical polar mathematics and various further analogies \cite{AF, +tri} with Molecular Physics. 
I.e. rigid rotors, the Stark Effect, Pauling's study of the rotations of molecules in crystals \cite{Pauling} and the theoretical underpinnings for Raman spectroscopy.  
\noindent In the present paper, I consider the conserved quantities for quadrilateralland. 
I.e. the quadrilateralland interpretation of $\mathbb{CP}^2$'s isometry group, $SU(3)/\mathbb{Z}_3$ \cite{MacFarlane}.  
Whilst this is no longer analogous to Molecular Physics, it is now analogous to Particle Physics.
I give analogues of what are hypercharge and isospin \cite{PS} in Particle Physics in terms of the Gibbons--Pope type coordinates \cite{GiPo1, GiPo2} from the study of gravitational 
instantons and then interpret these conserved quantities in quadrilateralland terms.
Sec \ref{Cl-PoT-Q} covers the classical-level Best Matching Problem, Temporal Relationalism, \K beables and Constraint-Closure resolutions of the Problem of Time for quadrilateralland.  

\mbox{ } 

\noindent N.B. that the classical dynamics and QM are required prior to further PoT and Quantum Cosmology applications.  
Thus in Sec \ref{dyn}, I give the classical equations of motion for quadrilateralland along with particular potentials with some symmetries. 
In Sec \ref{HO-PoT}, I interpret the geodesics on $\mathbb{CP}^2$ in quadrilateralland terms (i.e. as a sequence of quadrilaterals).  
In Sec \ref{cl-soln} I consider HO dynamics on $\mathbb{CP}^2$ from a qualitative perspective in quadrilateralland terms.  
These free problem and HO problem cases are then considered at the quantum level in Paper II (see the Conclusion for an outline). 
The above classical solution work feeds into the Semiclassical Approach to the PoT and Quantum Cosmology.  
These classical dynamics sections include obtaining inputs for the Classical and Semiclassical Machian Approaches, as well as being interesting in their own right (c.f. study of Bianchi 
IX) and a necessary precursor for understanding the corresponding quantum theory in Paper II.  
I conclude in Sec \ref{Concl} by listing the counterparts to $N$-stop  metroland and triangleland's key Steps for Quadrilateralland.  
These can no longer rely on Molecular Physics and its ready generalization from the 2-sphere to the $N$-sphere \cite{Norway}. 
I provide also some comments on the higher-$N$ extension of the present paper.
I also provide a discussion of regions of configuration space, which are to be applied to the \NSII, semiclassical and Halliwell approaches to Problem of Time. 

\mbox{ } 

\noindent The crucial observations and results to solve quadrilateralland are labelled as keys in the text. 
$N$-stop metroland and triangleland cracked using similar but simpler keys with strong support from the analogy with Atomic and Molecular Physics.    
The present article then gives the corresponding classical-level keys for quadrilateralland. 
For this, Molecular and Atomic Physics are longer an appropriate analogy.  
One is to use instead the Geometry and Particle Physics analogies summarized in Fig \ref{Fig1}.

\section{Positions, relative positions and Jacobi vectors}\label{Jacobi}  

Consider $N$ particle-position vectors ${\bq}^{I}$.  
We are really dealing with constellations and not figures, i.e. with the particle `dots' and not with how one `joins the dots'. 
I emphasize this distinction since it only becomes nontrivial for $N$-a-gonlands that are larger than the previously-studied triangleland.  
For quadrilateralland upward, `joining the dots' is no longer unique and the way in which they are joined is not particularly meaningful.  
Recollect from Sec 2 that $N$ particles in dimension $d$ has an incipient Cartesian {\it configuration space} $\fQ(N, d) = \mathbb{R}^{Nd}$.  
There is then for now absolute position, absolute rotation and absolute scale.
Then rendering absolute position irrelevant (e.g. by passing from ${\bq}^I$ to any sort of relative coordinates) leaves 
one on the configuration space {\it relative space}, $\fR = \mathbb{R}^{2n}$, for $n$ = $N$ -- 1.
Relative Lagrangian coordinates are the most obvious for this: a basis set of $n$ of the ${\br}^{IJ} = {\br}^J - {\br}^I$.  
The kinetic line element is unfortunately not diagonal in these.  

\noindent \underline{Key 1} A simple first key point for understanding RPM's of any $N$ or $d$ is as follows. 
(Though it was missed from 1982 right through \cite{Kuchar92, BS89-1, BS89-2, BS89-3, BS89-4, BS89-5, B03} until 2005 \cite{Paris}).  
One to resolve this non-diagonality by using instead $N$ -- 1 relative Jacobi vectors (inter-particle {\sl cluster} vectors), ${\bR}^i$.\footnote{Lower-case Latin letters as relative 
position variables indices running from 1 to $n = N - 1$, barred lower-case Latin letters as indices running from 1 to $n - 1$.}   
%
This is clear from basic Theoretical Molecular Physics \cite{LR97}. 
(Though it is also present in Celestial Mechanics, see e.g. \cite{Marchal} and is indeed first due to Jacobi around 150 years ago.)  
Relative Jacobi coordinates have associated particle cluster masses $\mu_i$. 
In fact, it is tidier to use {\it mass-weighted} relative Jacobi coordinates $\brho^i = \sqrt{\mu_i} {\bR}^i$ (Fig \ref{Fig1}). 
The squares of the magnitudes of these are the partial moments of inertia $I^i = \mu_i|{\bR}^i|^2$.  
I also denote $|\brho^i|$ by $\rho^i$, $I^i/I$ by $N^i$ for $I$ the moment of inertia of the system, and $\brho^i/\rho$ by $\bn^i$ for $\rho = \sqrt{I}$ the {\it configuration space 
radius} (termed the {\it hyperradius} in the Molecular Physics literature.  
Moreover, I do not use this name since it is not conceptually descriptive. 
I rather call it by its `true name' \cite{WheelerInt, Kvothe}).
I next consider the various further levels of structure that one can envisage for such clusters of particles. 

{           \begin{figure}[ht]
\centering
\includegraphics[width=1.0\textwidth]{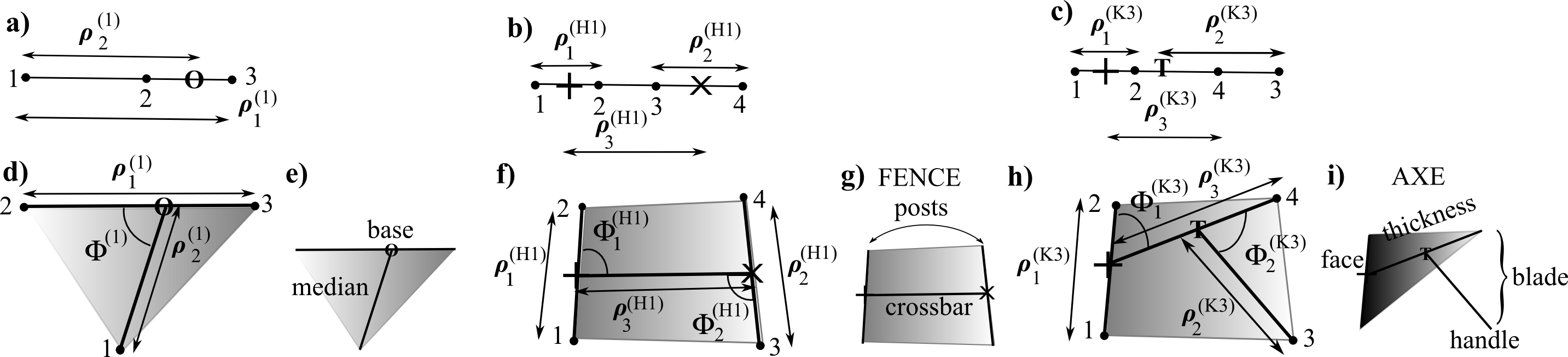}
\caption[Text der im Bilderverzeichnis auftaucht]{  \footnotesize{
a) 3-stop metroland's Jacobi coordinates.
b) 4-stop metroland's Jacobi H-coordinates. 
c) 4-stop metroland's Jacobi K-coordinates.
These are `squashed' versions of d), f) and h) respectively, for which the tree names T, H and K are self-evident.  
d) Triangleland's Jacobi coordinates. 
I term the two relative separations of the Jacobi triangle as the {\it base} and the {\it median} [e)].
f) Quadrilateralland in Jacobi H-coordinates.
g) I term the three separations of the Jacobi H tree as the two {\it posts} separated by the {\it crossbar} (viewing the configuration as a fence).
h) Quadrilateralland in Jacobi K-coordinates.
i) term the three separations of the Jacobi K tree as the {\it face of the blade}, the {\it thickness of the blade} and the {\it handle} (viewing the configuration as an axe).  
O, +, $\times$ and T denote COM(23), COM(12), COM(34) and COM(124) respectively, where COM(ab) is the centre of mass of particles a and b.
I write specific coordinate components' indices downstairs for ease of presentation.

For overall compactness of presentation, I also include here the definitions of the following relative angles that are used in subsequent Sections.  
For the triangle, $\Phi^{(\sa)}$ is the `{\it Swiss army knife}' angle $\mbox{arccos}\big( \brho_1^{(\sa)} \cdot \brho_3^{(\sa)} / \rho_1^{(\sa)} \rho_3^{(\sa)} \big)$.
For Jacobi H-coordinates [with $\Phi_1^{(\sH\sb)}$ and $\Phi_2^{(\sH\sb)}$ are the `Swiss army knife' angles 
$\mbox{arccos}\big(\brho_1^{(\sH\sb)}\cdot\brho_3^{(\sH\sb)}/\rho_1^{(\sH\sb)}\rho_3^{(\sH\sb)}\big)$ 
and 
$\mbox{arccos}\big(\brho_2^{(\sH\sb)}\cdot\brho_3^{(\sH\sb)}/\rho_2^{(\sH\sb)}\rho_3^{(\sH\sb)}\big)$, whilst for 
Jacobi K-coordinates, $\Phi_1^{(\sK\sa)}$ and $\Phi_2^{(\sK\sa)}$ are the `Swiss army knife' angles  
$\mbox{arccos}\big(\brho_1^{(\sK\sa)}\cdot\brho_3^{(\sK\sa)}/\rho_1^{(\sK\sa)}\rho_3^{(\sK\sa)} \big)$  and  
arccos$\big( \brho_2^{(\sK\sa)}\cdot\brho_3^{(\sK\sa)}/\rho_2^{(\sK\sa)}\rho_3^{(\sK\sa)} \big)$.
I note that there is a third relative angle for each tree's representation of quadrilateralland -- that between the inclinations of the posts or of the face of the blade and the handle. 
Whilst these are in each case not independent, it is sometimes useful to talk about them, and so I term these {\it post angles}. 
These carry the implication of involving extrapolating a Jacobi separation in order to be defined locally.}        } 
\label{Fig1} \end{figure}         } 


\noindent {\bf Jacobi trees} (from Molecular Physics \cite{ACG86}).  
For 3 particles, one particular choice of mass-weighted relative Jacobi coordinates are as indicated in Fig \ref{Fig1}.d). 
This is a T-shaped {\it tree} on the Jacobi clustering structure of the constellation.   
For 4 particles, there are then (up to permutation) two possible trees on the Jacobi clustering structure: the H and K of Fig \ref{Fig1}.e) and f).  
See e.g. \cite{ACG86} for further trees.
[Graph-theoretically, {\it trees} are 'loopless and connected', as is familiar in Particle Physics from the expression `tree amplitudes'. 
Moreover, these trees are also {\it irreducible}, meaning that there are no vertices of degree 2.  
N.B. the constellation is not in general constituted of the vertex particles due to the benefits of diagonality, attained from the Jacobi clustering structure.  
This means that some of the vertices are centres of mass of clusters rather than point particles, hence `Jacobi tree'.]  

\mbox{ } 

\noindent {\bf Cluster notation}  Each of the T's is then labelled according to its clustering structure. 
I.e. I use $\{\ma, \mb\mc\}$ read left-to-right in 1-$d$ and anticlockwise in $\geq$ 2 $d$, which I abbreviate by (\ma).
I use \{a...c\} to denote a cluster composed of particles a, ... c, ordered left to right in 1-$d$ and anticlockwise in 2-$d$.  
I take these to be distinct from their right to left and clockwise counterparts i.e. I consider plain configurations.  
I insert commas and brackets to indicate a clustering, i.e. a partition into clusters.
These notations also cover collisions, in which constituent clusters collapse to a point.
I use (Hb) as shorthand  

\noindent for \{ab, cd\} i.e. the clustering (partition into subclusters) into two pairs \{ab\} and \{cd\}, and (Ka) as shorthand for 

\noindent \{\{cd, b\}, a\} i.e. the clustering into a single particle a and a triple \{cd, b\} which is itself partitioned into a pair cd and a single particle b.
In each case, a, b, c, d form a cycle.
unless indicated otherwise, I take clockwise and anticlockwise labelled triangles and quadrilaterals to be distinct. 
I.e. I make the plain rather than mirror-image-identified choice of set of shapes.  
I also consider just distinguishable particles.
See \cite{QSub, FileR} for otherwise on these two counts.  
I also assume equal masses for simplicity.  
Later references to H and K coordinates refer explicitly to the (H1) and (K3) cases depicted above; I drop these labels to simplify the notation. 

\mbox{ } 

\noindent N.B. it is the constellation and not the tree, cluster-label or the shape made by joining the dots that is the genuinely relational concept.  
Contrast with Bookstein's distinct shape geometry \cite{Bookstein}. 
This gives material significance to the plane figures themselves -- a kind of non-total crushability -- 
whereas relational mechanics considers the constellation of points to be the primary entity.  
\noindent  And yet, the choice of tree and of cluster label can have some particular {\sl contextual} meaning as regards the regime of study or the propositions being addressed.
E.g. if the three particles are the Sun, Earth and Moon of Celestial Mechanics, then there is particular significance to having the apex particle be 1) the Sun.
This is because the Earth and Moon base pair are far more localized. 2) The Moon, because it is considerably lighter and therefore amenable to treatment as a perturbation.  
E.g. H-coordinates are particularly suited to the study of two pairs of binaries/diatomics. 
On the other hand, K-coordinates are suited to a single binary/diatomic alongside two single bodies, emphasizing a triple particle subsystem within. 
(`Binary' and `diatomic' are, respectively, Celestial Mechanics and Molecular Physics terminology).

\section{Topological and metric structure of configuration space}\label{TopMet}  

If one quotients out the rotations also, one's configuration space is {\it relational space} ${\cal R}(N, d)$.    
If one furthermore quotients out the scalings, one is on {\it shape space} $\fS(N, d)$.\footnote{I use $\sfA$ indices for shape coordinates 
(running from 1 to 6 for quadrilateralland and from 1 to 3 for triangleland), and $\fa$ indices for a subset of these running from 1 to 5 for quadrilateralland.}     
%
If one quotients out the dilations but not the rotations, one is on {\it preshape space} \cite{Kendall} $\fP(N, d)$.
These last two names come from David Kendall's geometrical study of the shape of spaces prior to his well-known study of the statistics of shape \cite{Kendall}.  
On the other hand, the first two names come from Julian Barbour's study of of the relational mechanics of shapes with \cite{BB82} and without \cite{B03} scale.
It is highly satisfying that these two authors in entirely different fields came across exactly the same concept of, and name for, shape space \cite{FORD}. 
It turns out to be advantageous to treat the pure-shape case first, hence \cite{B03} is valuable in advancing the study of RPM's.

\underline{Key 2} Topology is only simple for 3 series, and only two of these remain metrically simple.  
This is due to Kendall and Casson's work on Shape Geometry (see e.g. \cite{Kendall}).
The two metrically simple series are the $N$-stop metrolands (configuration space $\mathbb{S}^{N- 2}$) and the $N$-a-gonlands (configuration space $\mathbb{CP}^{N - 2}$). 
$\mathbb{CP}^{n - 1}$ involves $n$ lines, whilst $n$ lines can be used to form whichever Jacobi tree for an $N$-a-gon. 
This provides a lucid insight in to why $\mathbb{CP}^{n - 1}$ is representable as the space of all $N$-a-gons.
The 3-particle case of this is, moreover, special, by $\mathbb{CP}^1 = \mathbb{S}^2$.
Shape space geometry is usually a prequel to shape statistics but here it is a prequel to classical dynamics and QM.  

\noindent Topological part of \underline{Key 3}.     
Additionally, relational space ${\cal R}(N, d)$ is mathematically equal to \cite{Cones} {\it the cone} over shape space, denoted by $\mC(\fS(N, d))$. 
[This is a topological and metric generalization of viewing the usual cone as $\mathbb{S}^1 \times$ the real half-line with a special `cone point' at its apex.
At the topological level, then, for C(X) to be a {\it cone} over some topological manifold X, 
\beq
\mbox{C(X) = X $\times$ [0, $\infty$)/\mbox{ }$\widetilde{\mbox{ }}$} \mbox{ } , 
\eeq
where $\widetilde{\mbox{ }}$ means that all points of the form \{p $\in$ X, 0 $\in [0, \infty)$\} are `squashed' or identified to a single point termed the {\it cone point}, 
and is denoted by 0. 
For what a cone further signifies at the level of Riemannian geometry, see below.]  
In particular, ${\cal R}(N, 1) = \mC(\mathbb{S}^{N - 2} )= \mathbb{R}^{n}$ and ${\cal R}(N, 2)$ = C$(\mathbb{CP}^{N - 2})$ [among which additionally $\mC(\mathbb{CP}^1) 
= \mC(\mathbb{S}^2) = \mathbb{R}^3$ at the topological level].  
Thus, this paper's quadrilateralland case is the first case with {\sl nontrivial} complex-projective mathematics.  

{           \begin{figure}[ht]
\centering
\includegraphics[width=0.95\textwidth]{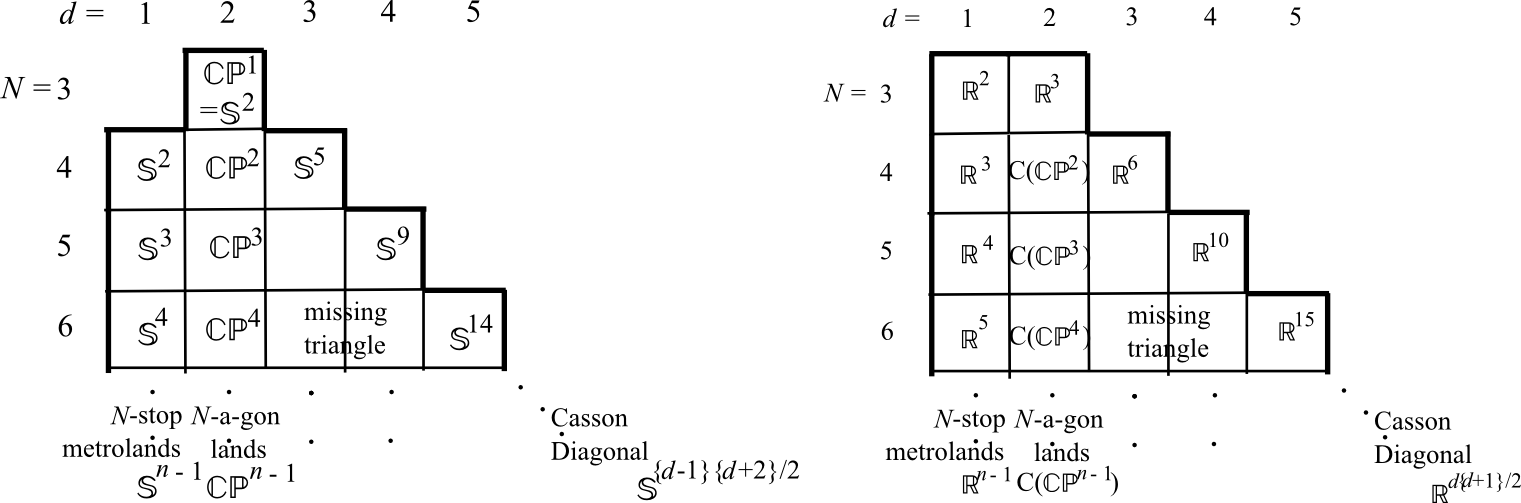}
\caption[Text der im Bilderverzeichnis auftaucht]{  \footnotesize{`Partial Periodic Table' of the discernible a) shape spaces and b) relational spaces at the topological level.  
In each case, only the first two columns (rather than the `Casson diagonal' \cite{Kendall}) continue to be straightforward at the metric level.
These constitute \underline{Key 2} to unlocking RPM's.}        } 
\label{DiscernSS-R} \end{figure}         } 

\noindent \underline{Key 4} For 1-$d$, one has the usual \{$N$ -- 2\}-sphere metric on shape space \cite{Kendall, 06II}, $\d^2s_{\sss\sp\sh\se}$. 
Then the Euclidean metric on the corresponding relational space is $\d^2s_{\sE\su\sc\sll} = \d\rho^2 + \rho^2\d^2s_{\sss\sp\sh\se}$ \cite{Cones}.   
Associated simplifications include that these coordinates cast each metric in diagonal form.  
For 2-$d$, the kinetic metric on shape space is indeed  \cite{Kendall} the metric that is natural on these complex projective spaces: the Fubini--Study metric 
\beq
\d s_{\sF\sS}^2 = \big\{\{1 + |{\mZ}|_{\sc}^2\}|\d {\mZ}|_{\sc}^2 - |({\mZ} ,\d\overline{{\mZ}}\}_{\sc}|^2\big\}/\{1 + |{\mZ}|_{\sc}^2\}^2  \mbox{ } .
\label{FS} 
\eeq
The coordinates in use here have the following significance. 
Firstly, \{${\mz}^i$\} are, mathematically, complex homogeneous coordinates for $\mathbb{CP}^2$. 
I denote their polar form by $\mz^i = \rho^i\mbox{exp}({i\phi^{i}})$.\footnote{All of this is well-known from the basic 
Geometry of complex projective spaces, see e.g. \cite{Nakahara}. \label{Peus}} 
Moreover, in the present physical application in terms of relational quadrilaterals, these contain 2 redundancies.  
Their moduli are the magnitudes of the Jacobi vectors and their arguments are angles between the Jacobi vectors and an absolute axis.  
Next, \{${\mZ}^{\barp}$\} are, mathematically, complex inhomogeneous coordinates for $\mathbb{CP}^2$. 
I denote their polar form by $\mZ^{\barp} = {\cal R}^{\barp}\mbox{exp}({i\Phi^{\barp}})$.$^{\ref{Peus}}$.    
These are independent ratios of the $\mz^{i}$.  
Thus in the present physical application in terms of relational quadrilaterals, their magnitudes ${\cal R}^{{\barp}}$ are ratios of magnitudes of Jacobi vectors.  
Additionally their arguments $\Phi^{\barp}$ are angles between Jacobi vectors, which are entirely relational quantities.   
Also, I use $|{\mZ}|_{\sc}^2 := \sum_{\barp}|{\mZ}^{\barp}|^2$, $( \mbox{ } , \mbox{ } )_{\sc}$ for the corresponding inner product, 
overline to denote complex conjugate and $|\mbox{ }|$ to denote complex modulus.
Note that using the polar form for the $\mZ^{\barp}$, the line element and the corresponding kinetic term can be cast in a real form. 
Moreover, in this 2-$d$ case, this fails to cover that the metric is not block-minimal form in these coordinates (see Key 4b later as regards how to attain this).  

\noindent Metric part of \underline{Key 3}.   
The cone structure carries over to the metric level of structure: the kinetic metric on relational space in scale-shape split coordinates is then of the cone form 
\beq
\d s_{\sC(\sF\sS)}^2 = \d \rho^2 + \rho^2\d s_{\sF\sS}^2 \mbox{ }  
\eeq
[for $\d s^2$ is the line element of X itself and $\rho$ a suitable `radial variable'\footnote{In the spherical 
presentation of the triangleland case, coordinate ranges dictate that the radial variable is, rather, $I$.  
Also note that, whilst this cone is topologically $\mathbb{R}^3$, the metric 
it comes equipped with is {\sl no}t the flat metric (though it is exploitably {\it conformally flat} \cite{08I, 08III}).}
that parametrizes the [0, $\infty$), which is the distance from the cone point; such `cone metrics' are smooth everywhere except (possibly) at the troublesome cone point].

The action for RPM's in relational form is then (\ref{Actio}), with, in 1- or 2-$d$, the line element is $\d\fs^2_{\sss\sp\sh\se}$ or $\d s^2_{\sE\su\sc\sll}$, 
$\d\fs^2 = \d\fs^2_{\sF\sS}$ or $\d s^2_{\sC(\sF\sS)}$ built from the above selection of metrics (thus directly implementing Configurational Relationalism).  
\noindent Note that in the spherical presentation of the triangleland case, coordinate ranges dictate that the radial variable is, rather, $I$.  
Also note that, whilst this cone is topologically $\mathbb{R}^3$, the metric it comes equipped with is {\sl no}t the flat metric (though it is exploitably {\it conformally flat} 
\cite{08I, 08III}).
\noindent Finally, RPM's in 3-$d$ are much more intractable \cite{Kendall}, whilst not enhancing much the geometrodynamical analogy.   
Thus these are not considered to be useful toy models to develop in the present context.  

\mbox{ }

\noindent I use `r-formulation' to mean reduction of the previous 2 Secs' presentations of RPM's. 
Or, equivalently by \cite{FORD, Cones, FileR}, to mean the direct construction of a Mechanics from the relational configuration space's geometry by the already--mentioned Jacobi--Synge 
procedure.

\section{Configuration spaces and coordinate systems for smaller RPM's}\label{Q-1}

This is a useful recap both as simpler cases of the present paper's construct and as structures that recurr as submanifolds within quadrilateralland itself.

\subsection{3-stop metroland }

{            \begin{figure}[ht]
\centering
\includegraphics[width=0.95\textwidth]{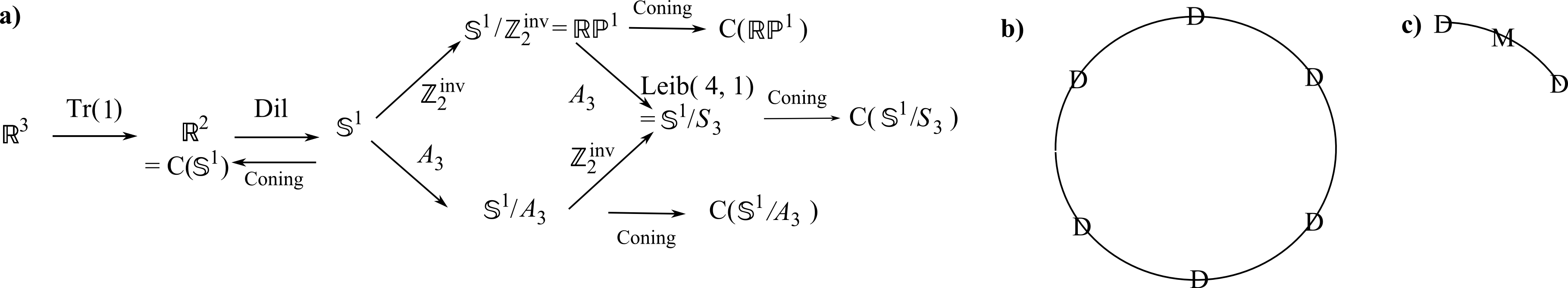}
\caption[Text der im Bilderverzeichnis auftaucht]{        \footnotesize{a) The various 3-stop metroland configuration spaces. 
$\mathbb{Z}_2$ is the 2-element group. 
This is superscripted by the nature of its action (inv for inverse here, and conj for conjugate in Fig \ref{Fig4}), $S_p$ is the permutation group on p objects and $A_p$ 
is the alternating group of even permutations of $p$ objects.
Leib is the `Leibniz space' that maximally encapsulates Leibniz's `identity of indiscernibles': the relational configuration space for indistinguishible particles.
b) are the tessellations for the distinguishable particles plain shape case; here, D denotes `double collision'. 
I.e. the collision of particles (exact coincidence of point particles, which is already meaningful at the topological level).
c) At the metric level, each piece of the preceding contains a distinguished mid-point: the {\it merger}, M, which physically corresponds to one particle lying upon the centre of 
mass of the other two.}  }
\label{Fig-3-Stop} \end{figure}          }
%
3-stop metroland's shape space is the circle $\mathbb{S}^1$, decorated as per Fig \ref{Fig-3-Stop}b). 
As a dynamics of pure shape, this is trivial by degree of freedom count, but it is nontrivial as the shape part of the scaled 3-stop metroland theory.  
A useful coordinatization is by $\varphi$ running from 0 to 2$\pi$, whose physical meaning is 
\beq
\varphi = \mbox{arctan}(\rho_2/\rho_1) \mbox{ } .
\eeq  
The corresponding RPM kinetic line element is 
\beq
\d \fs^2 = \d{\varphi}^2/2 \mbox{ } .
\label{3s}
\eeq

\subsection{4-stop metroland}

{            \begin{figure}[ht]
\centering
\includegraphics[width=0.97\textwidth]{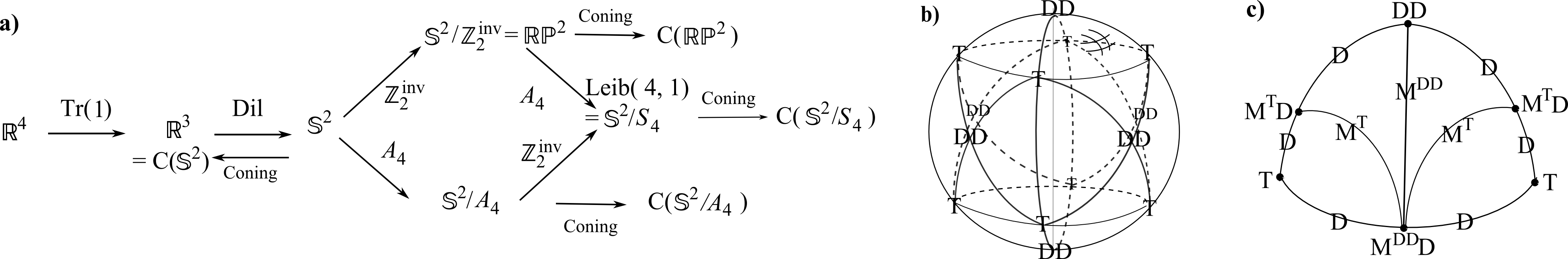}
\caption[Text der im Bilderverzeichnis auftaucht]{        \footnotesize{a) The sequence of configuration spaces for 4-stop metroland.
b) is the tessellation for the distinguishable particles plain shape case 
c) illustrates the notions of merger new to 4-stop metroland within a single triangular face.}}
\label{Fig-4-Stop} \end{figure}          }

\noindent 4-stop metroland's shape space is the sphere $\mathbb{S}^2$, decorated as in Fig \ref{Fig-4-Stop}b). 
The configuration space here for distinguishable particles and in the plain shape case is the sphere, decorated with the physical interpretation of Figs \ref{Fig-4-Stop}b) and c).
All the lines in Fig \ref{Fig-4-Stop}b)) are lines of double collisions, D.
The DD points are double double collisions (i.e. configurations containing two double collisions, each at a non-coincident place) whilst the T points are triple collisions 
(three of the point particles coincide in a single place).
D, DD and T are all already meaningful at the topological level.   
At the metric level, each `unit cell' of this configuration space is decorated as per Fig \ref{Fig-4-Stop} with the net of more complicated notions of merger that having 4 particles 
affords.
There are $\mM^{\tT}$ arcs (for which the fourth particle is at the COM of the other three) and $\mM^{\tD\tD}$ arcs. 
(For this the COM's of 2 pairs of particles coincide), as well as $\mM^{\tT} \bigcap \mD = \mM^{\tT}$D and $\mM^{\tD\tD} \bigcap \mM^{\tT} \bigcap \mD = \mM^*\mD$ points.    
These are also the types of merger present in quadrilateralland, though there of course their geometry as configuration space regions is more complicated.

A useful coordinatization is in terms of $\theta$ running from 0 to $\pi$ and $\phi$ running from 0 to 2$\pi$, whose physical meanings are, in Jacobi H-coordinates, 
\beq
\theta = \mbox{arctan}\big(\sqrt{\rho_1\mbox{}^2 + \rho_2\mbox{}^2}/\rho_3\big) \mbox{ } , \mbox{ } \mbox{ } 
\phi = \mbox{arctan}(\rho_2/\rho_1) \mbox{ } .
\label{4StopPolars}
\eeq  
These are, respectively, 1) a measure of the size of the universe's contents relative to the size of the whole model universe. 
2) A measure of inhomogeneity among the contents of the universe (whether one of the constituent clusters is larger than the other one.)  
On the other hand, for Jacobi K-coordinates 
\beq
\theta = \mbox{arctan}\big(\sqrt{\rho_1\mbox{}^2 + \rho_2\mbox{}^2}/\rho_3\big) \mbox{ } , \mbox{ } \mbox{ } 
\phi = \mbox{arctan}(\rho_1/\rho_2) \mbox{ } . 
\label{4StopPolars2}
\eeq
These are, respectively, 1) a measure the sizes of the \{12\} and \{T3\} clusters relative to the whole model universe. 
2)  A measure of the sizes of the \{12\} and \{T3\} clusters relative to each other. 
The corresponding RPM kinetic line element is 
\beq
\d \fs^2 = \{\d{\theta}^2 + \mbox{sin}^2\theta\,\d{\phi}^2\}/2 \mbox{ } .
\label{4s}
\eeq
(\ref{3s}), (\ref{4s}) and their $\mathbb{S}^{n - 1}$ generalization to ultraspherical coordinates is \underline{Key 4[$N$-stop]}. 
 
\noindent\underline{Key 5 [$N$-stop]} is then a useful set of redundant coordinates for a surrounding flat Euclidean space.  
(This is here equal to relational space).  
For 4-stop metroland, these simply are the three relative Jacobi coordinate magnitudes, $\mn^i$.  
 
\noindent \underline{Key 6[$N$-stop]} is that these also then serve as subsystem-describing coordinates.

\subsection{Triangleland}\label{TriStart}

{            \begin{figure}[ht]
\centering
\includegraphics[width=0.96\textwidth]{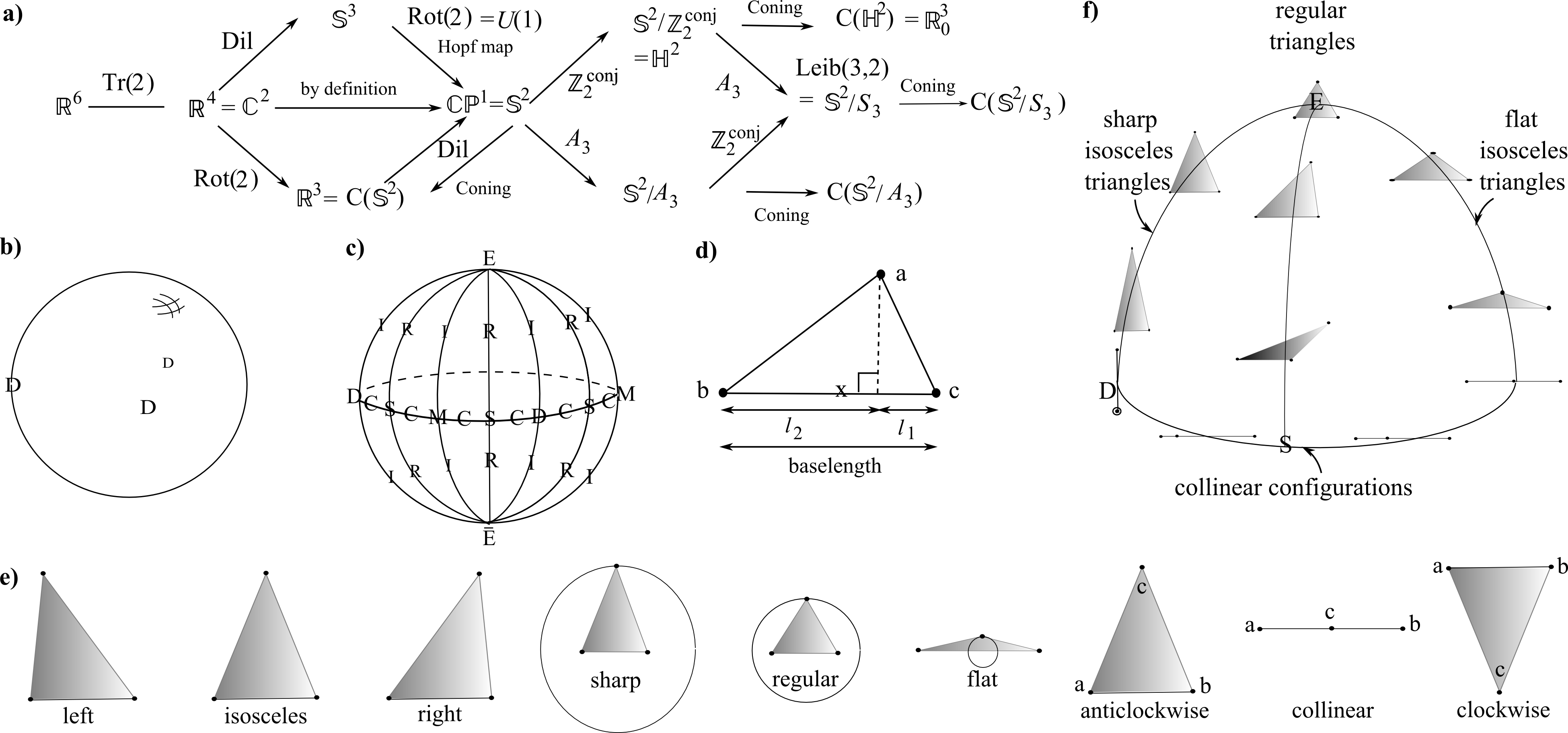}
\caption[Text der im Bilderverzeichnis auftaucht]{        \footnotesize{ a) The sequence of configuration spaces for triangleland; $\mathbb{H}^2$ is here the hemisphere with edge included.

\noindent b) and c) the configuration spaces for the distinguishable particle plain shape case at the topological and metric levels respectively.   
These pictures constitute \underline{Key 7[$\triangle$]}.
d) Is the figure used in the definition of anisoscelesness: the anisoscelesness per unit base length is the amount by which the perpendicular to the base fails to bisect it. 
I.e. $l_1 - l_2$ for $l_1$ and $l_2$ as indicated.
e) The most physically meaningful great circles on the triangleland shape correspond to the isosceles, regular and collinear triangles.
These respectively divide the shape sphere into hemispheres of right and left triangles, sharp and flat triangles, and anticlockwise and clockwise triangles. 
f) {\bf Kendall's spherical blackboard} from Shape Statistics \cite{Kendall84}.}   }
\label{Fig4} \end{figure}          }

Here, we have $\mathbb{S}^2$, decorated as in Fig \ref{Fig4}b), c), f). 

\noindent \underline{Key 4[$\triangle$]} is that useful intrinsic diagonal coordinates for triangleland are $\Theta$ and $\Phi$.   
These are again spherical polar coordinates on the configuration space sphere, but their meaning in terms of the relative particle cluster separations is rather different.  
The interpretation of the azimuthal angle is now 
\beq
\Theta = 2\,\mbox{arctan}(\rho_2/\rho_1)
\label{TriAzi} \mbox{ } ,  
\eeq 
and that of the polar angle $\Phi$ is as in Fig \ref{Fig1}.d).  
In terms of these, the triangleland kinetic line element is 
\beq
\d \fs^2 = \{\d{\Theta}^2 + \mbox{sin}^2\Theta\,\d{\Phi}^2\}/2 \mbox{ } .
\eeq
We also make use of the complex coordinate form of this which is a particular collapse of the kinetic line element built out of (\ref{FS}): 
\beq
\d \fs^2 = \d{\mZ}\,\d{\overline{\mZ}}/2\{1 + |\mZ|^2\}^2 \mbox{ } .
\eeq
The configuration space here for distinguishable particles and in the plain shape case is the sphere, decorated as in Figs \ref{Fig4}b) and \ref{Fig4}f).
The labelled points and edges have the following geometrical/mechanical interpretations.  
E and $\bar{\mE}$ are the two mirror images of labelled equilateral triangles.  
C are arcs of the equator that is made up of collinear configurations. 
This splits the triangleland shape sphere into two hemispheres of opposite orientation (clockwise and anticlockwise labelled triangles, as per Fig \ref{Fig4}c). 
Each of these separates the triangleland shape sphere into hemispheres of right and left slanting triangles with respect to that choice of clustering [Fig \ref{Fig4}b)].
The R meridians of the shape sphere correspond to {\sl regular} configurations (those in which the base and meridian partial moments of inertial are equal).  
Each of these separated the triangleland shape sphere into hemispheres of sharp and flat triangles with respect to that choice of clustering [Fig \ref{Fig4}a)].  
The M are merger points: where one particle lies at the COM of the other two. 
S denotes spurious points, which lie at the intersection of R and C but have no further notable properties (unlike the D, M or E points that lie on the other intersections).  

\mbox{ }

\noindent Interpretation of various great circles and of the hemispheres they divide triangleland into are provided in Fig \ref{Fig4}.  

\noindent \underline{Key 5[$\triangle$]} provides useful redundant coordinates covering a surrounding Euclidean space. 
(This is here also the relational space.) 
These coordinates are now the complicated combinations of the two Jacobi vectors: the {\it Dragt coordinates} \cite{Dragt} of Molecular Physics, 
\beq
\mbox{dra}_1 = 2\,{\bn}_1 \cdot {\bn}_2 \mbox{ } \mbox{ } , \mbox{ } 
\mbox{dra}_2 = 2\{{\bn}_1 \cr {\bn}_2\}_3   \mbox{ } \mbox{ } , \mbox{ }
\mbox{dra}_3 = \mn_2^2 - \mn_1^2 \mbox{ } .
\eeq
\noindent Note that as compared to the 4-stop metroland Cartesian components, there are rather less straightforwardly realized.  
This is because there are now four rather than just three obvious Jacobi vector quantities to use. 
However, this turns out to be sorted out by the fairly well-known mathematics of the Hopf Map, giving the above three quadratic combinations.

\noindent Also note that these have a lucid interpretation in terms of the triangle, respectively as \cite{+tri}.  
Namely, the {\it anisoscelesness} aniso of the labelled triangle, four times the mass-weighted area of the triangle per unit moment of inertia and the {\it ellipticity} ellip of the two 
partial moments of inertia.

\mbox{ } 

\noindent The on-sphere condition is then $\sum_{\sfA = 1}^3\mbox{dra}^{\sfA\,2} = 1$.
These combinations appearing as surrounding Cartesian coordinates is much less obvious than the $\mn^e$ appearing in the same role for 4-stop metroland.  
These combinations arise from the sequence 
\beq
\stackrel{\mbox{$\fR(3,2)$}}{= \mathbb{R}^4} \mbox{ }   \mbox{ }    
\stackrel{\mbox{\scriptsize obvious on-sphere condition}}{\longrightarrow}   \mbox{ }   \mbox{ }
\stackrel{\mbox{$\fP(3,2)$}}{= \mathbb{S}^3} \mbox{ }   \mbox{ }      
\stackrel{\mbox{\scriptsize Hopf map}}{\longrightarrow} \mbox{ }   \mbox{ }
\stackrel{\mbox{$\fP(3,2)$}}{= \mathbb{S}^2}  \mbox{ }   \mbox{ }   
\stackrel{\mbox{\scriptsize coning}}{\longrightarrow}  \mbox{ }   \mbox{ } 
\stackrel{\mbox{${\cal R}(3,2)$}}{= \mathbb{R}^3} \mbox{ } .  
\label{TriSeq}
\eeq

\noindent \underline{Key 5b[$\triangle$]} One sometimes also swaps dra$_{2}$ 
%
%
for the scale variable $I$ in the non-normalized version of the coordinates to obtain the \{$I$, aniso, ellip\} system.  

\noindent \underline{Key 6[$\triangle$]} Then a simple linear recombination of this is \{$I_1$, $I_2$, aniso\}, i.e. the 
two partial moments of inertia and the dot product of the two Jacobi vectors.  
This is in turn closely related \cite{08I} to the parabolic coordinates on the flat $\mathbb{R}^3$ conformal to the triangleland relational space, which are $\{I_1, I_2, \Phi\}$.  
In the present physical situation, these are playing the subsystem role: base and meridian subsystems' partial MOI's alongside the relative angles between them. 

\mbox{ }

\noindent Then of course, the sphere has a lot of standard geometry and Methods of Mathematical Physics available for it \cite{08I, +tri}; there is almost as much for the $k$-sphere.

\subsection{Overview of configuration spaces for quadrilateralland}

The shape space is $\mathbb{CP}^2$ or some quotient as per Fig \ref{Fig5}.  
\noindent N.B. this is harder to visualise than the previous subsections' shape spaces, due to greater dimensionality 
as well as greater geometrical complexity and a larger hierarchy of special regions of the various possible codimensions. 
Some geometrical detail of this space is elucidated in Secs \ref{Q-2} and \ref{Subman}.  

{            \begin{figure}[ht]
\centering
\includegraphics[width=0.8\textwidth]{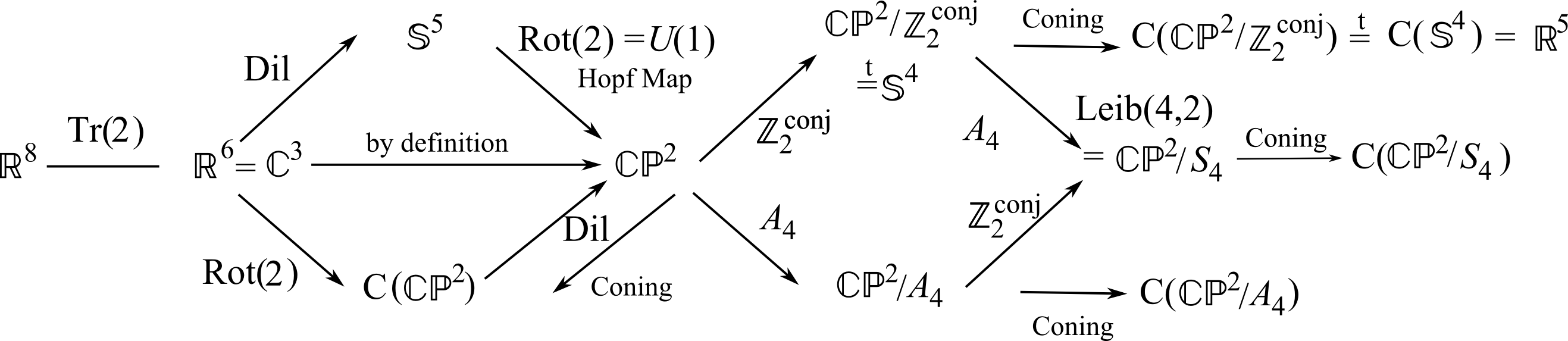}
\caption[Text der im Bilderverzeichnis auftaucht]{        \footnotesize{The sequence of configuration spaces 
for quadrilateralland.  
$\stackrel{t}{=}$ denotes equality at the topological level.  } }
\label{Fig5} \end{figure}          }

\noindent   \underline{Key 4[$\Box$]} \noindent The $N$-a-gons all possess complex projective space mathematics.
(Triangleland atypically simplifies via $\mathbb{CP}^1 = \mathbb{S}^2$, whereas quadrilateralland is much more mathematically typical for an $N$-a-gonland, the first typical such.)  
Quadrilateralland extends triangleland for some purposes.  
This is in contrast with how Celestial Mechanics, Molecular Physics, Nuclear Physics, and simplex constructions/spin foams in Quantum Gravity \cite{Perez} pass instead to the tetrahaedron. 
$\mathbb{CP}^2$ geometry is in this paper.  
The associated Methods of Mathematical Physics is considered in Paper II (this is still within standard: Jacobi polynomials \cite{JP} and Wigner D-functions \cite{WignerD}).  
Fubini--Study metric in inhomogeneous coordinates for 2-$d$ RPM's \cite{FORD, Kendall}.  
%

\noindent It makes sense to pose question about analogues of the remaining Keys; much of the rest of this paper is dedicated to answering that. 

{\begin{figure}[ht]
\centering
\includegraphics[width=0.7\textwidth]{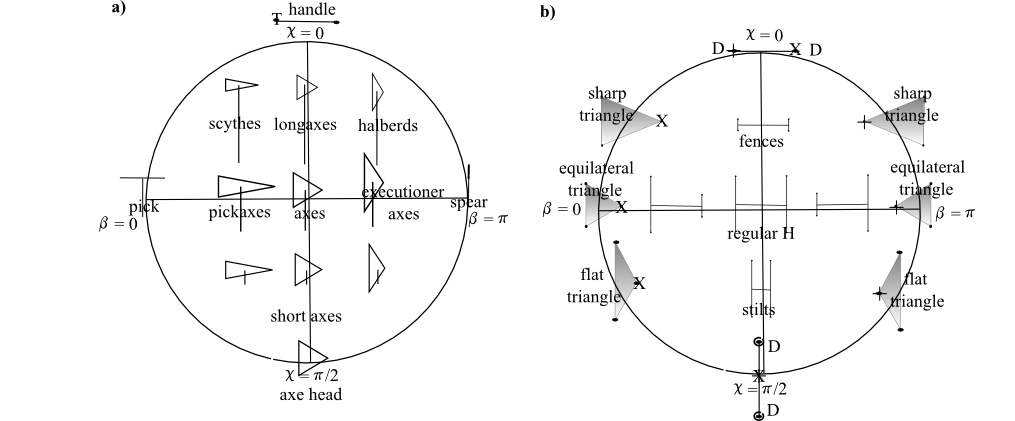}
\caption[Text der im Bilderverzeichnis auftaucht]{        \footnotesize{a) My `$\mathbb{CP}^2$ chopping board' counterpart of Kendall's spherical blackboard of Fig \ref{Fig4}f). 
This is displayed in pure-relative-ratio coordinates. 
(I.e. with relative angle coordinates $\phi$ and $\psi$ suppressed. It is drawn in the $\psi = \pi/2 = \phi$ case). 
This is for the Jacobi-K tree/triple clustering, depicted as an `axe' whose blade is the coordinate-privileged 3-cluster and whose handle runs from the COM T of that to the final particle.  
\noindent b)  The corresponding space of `fences' for the Jacobi-H case.  
\noindent N.B. how Figs \ref{Fig4} and \ref{FigX} supply the limiting cases.   }        }
\label{EndGame}\end{figure}            }

\subsection{Resum\'{e} of configuration spaces}\label{Resu}

This includes the GR--RPM analogies at the level of configuration spaces (Fig \ref{Fig2}). 

\vspace{10in}

{            \begin{figure}[ht]
\centering
\includegraphics[width=0.75\textwidth]{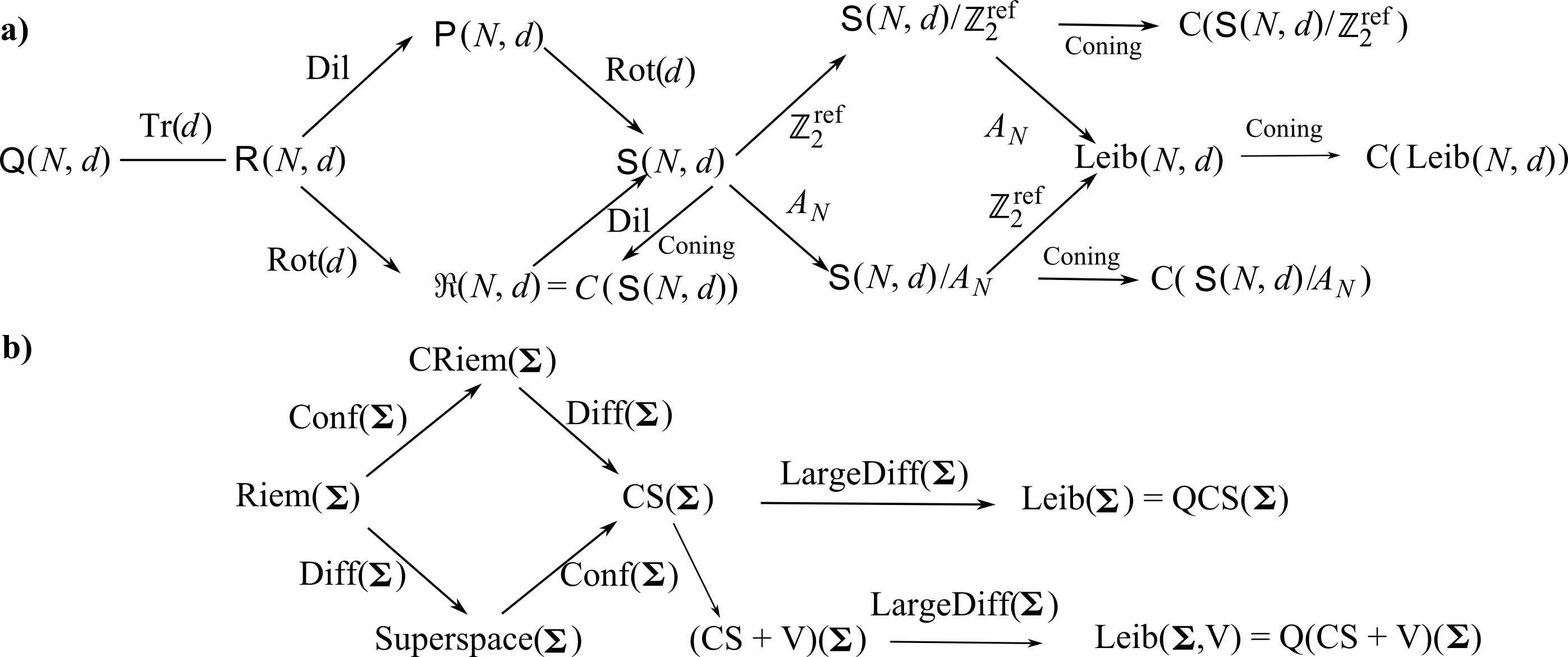}
\caption[Text der im Bilderverzeichnis auftaucht]{        \footnotesize{GR--shape dynamics configuration space analogies.  
a) The sequence of configuration spaces for RPM's. 
b) The corresponding sequence of configuration spaces for GR.       
Riem($\bupSigma$) is the space of spatial (i.e. positive-definite) 3-metrics on a fixed topology $\bupSigma$ that is taken to be a compact without boundary one for simplicity. 
It is named in honour of mathematician Bernhard Riemann, inventor of metric geometry.
This space corresponds most naturally to RPM's relative space, $\fR$($N, d$).  
Diff($\bupSigma$) and Conf($\bupSigma$) are the corresponding groups of 3-diffeomorphisms and conformal transformations. 
These most naturally correspond to the rotations Rot($d$) and dilations Dil respectively.  
Superspace($\bupSigma$) is meant in Wheeler's sense \cite{W68}.  
CRiem($\bupSigma$) is pointwise superspace and  CS($\bupSigma$) is conformal superspace.
Superspace($\bupSigma$), CRiem($\bupSigma$) and CS($\bupSigma$) most naturally correspond to relational space ${\cal R}(N, d)$, 
preshape space P($N, d$) and shape space S($N, d$) respectively.  
(RPM's also admit analogies with conformal/York-type \cite{York73-1, York73-2, York74, ABFO, ABFKO} initial value problem 
formulations of GR, with the conformal 3-geometries playing here an analogous role to the pure shapes.)
(CS + V)($\bupSigma$) is conformal superspace to which has been adjoined a single global degree of freedom: the spatial volume of the universe. 
CS($\bupSigma$) and (CS + V)($\bupSigma$) have on a number of occasions been claimed to be the space of true dynamical degrees of freedom of GR \cite{York73-1, York73-2, York74, ABFO, ABFKO, NewBO}.
The quotienting out of large diffeomorphisms gives the notion of quantum superspace and quantum conformal superspace, QCS($\bupSigma$), as per \cite{FM96}.  
This corresponds most naturally to identifying mirror image shapes and enforcing particle indistinguishability.
Leib($N, d) = \fS(N, d)/S_N$ is then the analogue of QCS($\bupSigma$) and C(Leib($N,d$)) is then in some ways the analogue of quantum CS + V.  
It is named thus as the most very Leibnizian of the possible configuration spaces for mechanics with equal particle masses. } }
\label{Fig2} \end{figure}          }

\subsection{Collapses and ratio choices}
%
{            \begin{figure}[ht]
\centering
\includegraphics[width=0.92  \textwidth]{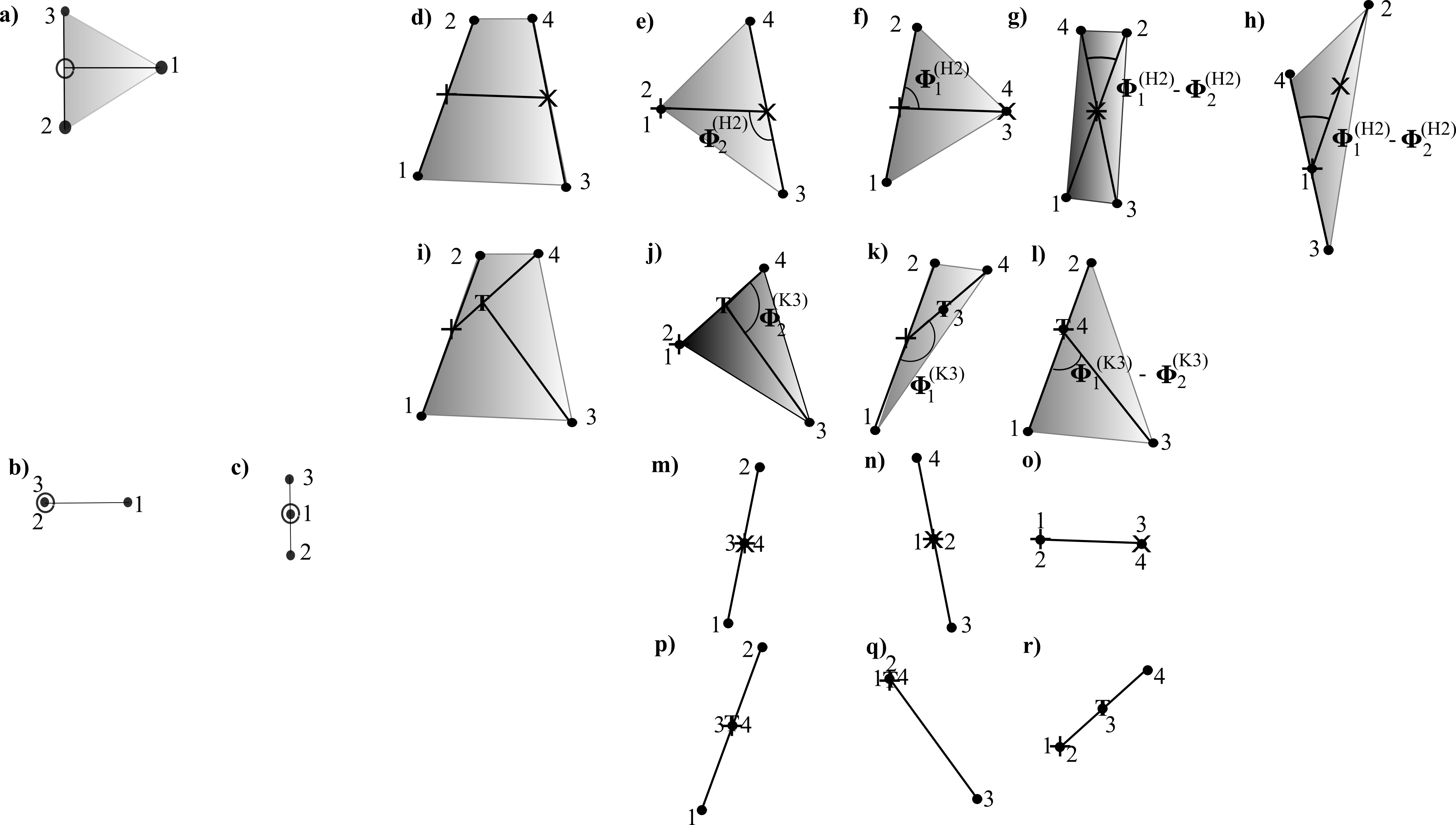}
\caption[Text der im Bilderverzeichnis auftaucht]{        \footnotesize{By collapse, I mean that 1 or more Jacobi vectors go to zero magnitude.  
The triangle a) can be collapsed in two ways within the indicated cluster: base to 0 b) and median to 0 c).
The quadrilateral in H-coordinates d) can have a single Jacobi vector collapse in 3 ways: the coarse-graining triangles e) and f) and the coarse-graining rhombus. 
[One can use h) to interpret the associated anisoscelesness.] 
The quadrilateral in K-coordinates i) can have 1 Jacobi vector collapse in 3 ways, which are now all coarse-graining triangles j), k), l). 
e), f) and g) contain D's, g) a quadrilateralland $\mM^{\tD\tD}$, k) a quadrilateralland M$^{\tT}$ and l) a coincidence of a particle, a 
triple COM and one double COM (a quadrilateralland weakening of M$^{*}$).   
There are also now double Jacobi vector collapses complementary to each single Jacobi vector: m),n),o),p),q),r) respectively.  
These have the following special mergers.  m),n) and p) are of type M$^*$D, o) is of type DD, q) is of type T and r) is of type M$^{\tT}$D    
 For the triangle, b) and c) complement each other in this way. 
This complementarity plays a role in Sec 17's analysis of quadrilateralland's geodesics.}         }
\label{FigX}\end{figure}          }

These collapses are useful through their describing physically-significant asymptotic configurations for the study of dynamics in Secs \ref{dyn} and \ref{cl-soln}. 
They are also useful for pinning various interpretations on coordinate systems and potentials.    
The physical significance of collapses depends both on the underlying tree (H or K) and on the finer detail of the choice of ratios used.
[Given a tree, which two ratios with a common denominator to make in passing to complex coordinates and then further coordinates whose interpretation depends on this prior choice.]  
The choice of ratios is furthermore related to collapses of the Jacobi vectors, which have geometrical significance as special, degenerate quadrilaterals.  
See Figure \ref{FigX} for these degenerate quadrilaterals; a characterization I developed with Serna for the significance of the various choices of ratio is given below.   

\mbox{ }

\noindent The basic H -- denoted H(DD) -- involves ratios $\rho_1/\rho_3$ and $\rho_2/\rho_3$.  

\noindent The other type of ratio choice for the H -- denoted H(M$^*$D) -- involves ratios $\rho_2/\rho_1$ and $\rho_3/\rho_1$, 
or the 1 $\leftrightarrow$ 2 permutation of this amounting to interchanging the two posts. 

\noindent The basic K -- denoted K(T) -- involves ratios $\rho_1/\rho_2$ and $\rho_3/\rho_2$.  
This is convenient for addressing questions about the 3-particle subsystem picked out by the K-construction.   
This does then involve a less intuitive `post' relative angle.  

\noindent Another K -- denoted K(M$^*$D) -- involves ratios $\rho_3/\rho_1$ and $\rho_2/\rho_1$.  
This one succeeds in involving two Swiss army knife relative angles at the expense of no longer focusing on the 3-particle subsystem. 
 
\noindent The final type of ratio choice for the K -- denoted K(M$^{\sT}$D) -- involves ratios $\rho_1/\rho_3$ and $\rho_2/\rho_3$.   
This one is `as H-like as possible', via using the thickness of the blade much like the H uses the crossbar.  

\mbox{ } 

\noindent In terms of which Jacobi distances these involve (enumerated), the following transpositions hold.  
$$
\mbox{\scriptsize H(M$^*$D) is the 1 $\leftrightarrow$ of H and K(M$^*$D) is the 1 $\leftrightarrow$ 3 of K(M$^{\tT}$D) } ,   
$$
$$
\mbox{\scriptsize H and K(M$^{\tT}$D) use the same ratios: K(M$^{\tT}$D) is the closest realization of treating the axe as if it had 2 subsystem posts (the blade face and handle) },
$$ 
\beq
\mbox{\scriptsize K(T) is the 2 $\leftrightarrow$ 3 of the preceding } .
\eeq 
The rest of this series of papers uses only the basic H and K choices of ratios.  

\vspace{10in}
 
\section{Democracy transformations for 1- and 2-d RPM's}\label{Demo}  

The {\it democracy transformations} \cite{Zick1, Zick2, Zick3, ACG86, LR95, LR97} 
\beq
\brho_{i} \longrightarrow \brho_{i}^{\prime} = 
D_{ij}\brho_{j}
\eeq
are interpolations between relative Jacobi vectors for different choices of clustering.
These constitute a very natural line of enquiry from the Molecular Physics perspective.
Moreover, the quadrilateral is not a case specifically done in Molecular Physics due to its focus on 3-$d$ models.

These transformations form the {\it democracy group} Demo($n$), which for this example is, mathematically, $O(n)$, but can be taken without much loss of generality to be $SO(n)$. 
Note that this is a symmetry group of the unreduced kinetic term.  
Also note that the democracy group is indeed independent of spatial dimension $d$.  
This is since democracy transformations only mix up whole Jacobi vectors rather than separately mixing up each vector's individual components.  
This means, firstly, that much about the theory of democracy transformations in 2-$d$ can be gleaned from their more widely studied counterpart in 3-$d$ (as cited above).
Secondly, it means that for spatial dimension more than 1, the democracy group is but a subgroup of the full 
symmetry group of the unreduced kinetic term, which itself is the dimension-dependent $O(nd)$.

\mbox{ }  

\noindent {\bf Lemma 1}. The configuration space radius $\rho$ is itself a democracy invariant.  
It is however a pure-scale quantity.

\mbox{ } 

\noindent {\bf Lemma 2}. 
$\mbox{dim}({\cal R}(n, d)) - \mbox{dim}(\mbox{Demo}(n)) = \{n + d - \{n - d\}^2\}/2$ gives a a minimum number of independent democracy invariants \cite{LR95}.
This bound is 2 both for triangleland and for quadrilateralland, revealing in each case the presence of at least one further democracy invariant.  

\mbox{ } 

\noindent It is intuitively clear that one such for triangleland should be the area of the triangle. 
It is not, however, clear what happens for quadrilateralland, due to RPM's in truth involving constellations rather than figures formed from `joining up the dots'. 
Since for $N > 3$ there is an ambiguity in how one would `join up the dots', 
the area ascribed to the constellation via drawing a quadrilateral between its points is not itself a democracy invariant.

To  find democracy invariants, consider the independent invariants that arise from the so-called $Q$-matrix \cite{LR95}
\beq 
Q_{ij} = \brho_i\brho_j^{\st} \mbox{ } ,  
\eeq  
where the t-superscript stands for transpose.  
The invariant trace($Q$) always gives the configuration space radius, hence proving Lemma 1.  
For $N$-stop metroland, $Q_{ij}$ is just $\rho_i\rho_j$, which is sufficiently degenerate that it and all of its nontrivial submatrices have zero determinants, 
so that there are no more democracy invariants.  
For triangleland, however, det($Q$) = 4 Area$^2$ (for Area the area of the mass-weighted triangle per unit $I$), while $Q_{ij}$ is but 2 $\times$ 2, so that there are no more invariants.  
For quadrilateralland, det($Q$) = 0.  
This can be understood in terms of volume forms being zero for planar figures.  
Also, being 3 $\times$ 3, there is one further invariant, $Q_{11}Q_{22} - Q_{12}\mbox{}^2$ + cycles = 
$\rho_1\mbox{}^2\rho_2\mbox{}^2 - \{\brho_1\cdot\brho_2\}^2$ + cycles = $|\brho_1\cr\brho_2|_3\mbox{}^2$ + cycles, 
i.e. proportional to the sum of squares of areas of various coarse-graining triangles (see Fig \ref{FigX}), $\sum_{i = 1}^3a_i^2$ 
%
%
Here the 3-suffix denotes that it is mathematically the component in a fictitious third dimension of the cross product.  
E.g. in the \{12, 34\} H-clustering, these are the coarse-grainings by taking 1, 2 and COM(34); 3, 4 and COM(12).  
12 and 34 with one shifted so that its COM coincides with the other's.  
%

\subsection{Shape quantities and democracy invariants for triangleland}\label{Demo-3}

\noindent I present here a different method of finding the useful Dragt-type shape coordinates for triangleland. 
I find that this method extends to quadrilateralland, as well as providing a useful prequel for this as regards notation and interpretation.   
$\underline{\underline{Q}}$ may be written as $\frac{1}{2}\{w \underline{\underline{1}} + 
w_1\underline{\underline{\sigma}}_1 + w_3\underline{\underline{\sigma}}_3\}$ (for Pauli matrices 
$\underline{\underline{\sigma}}_1$ and $\underline{\underline{\sigma}}_3$; the other Pauli matrix 
$\underline{\underline{\sigma}}_2$ does not feature since $\underline{\underline{Q}}$ is symmetric). 
This takes such an $SU(2)$ form due to the 3-particle case being exceptional through the various tensor fields of interest possessing higher symmetry in this case \cite{LR95}.
Then, reading off from the definition of $\underline{\underline{Q}}$ [and dropping (1)-labels] and dividing though by I to make pure-shape quantities, 
\beq
s_1 := q_1 := \mn_1^2 - \mn_2^2         := \mbox{ellip} 
\mbox{ } , \mbox{ } 
\eeq
\beq
s_2 = 2\,\bn_1\cdot\bn_2  := \mbox{aniso} 
\mbox{ } . \mbox{ } 
\eeq
A third significant quantity is 
\beq
s_3 = 2|\bn_1 \cr \bn_2|_3 = 4 \times \mbox{Area}/I := 4 \times \mbox{area} := 
\mbox{demo}(N = 3) 
\mbox{ } . \mbox{ }
\label{demo3}
\eeq
area is clearly clustering-invariant and thus a democratic invariant (\underline{Key 5b}); $s$ denotes shape, $q$ denotes $Q$-quantity and $a$ denotes area.  
Note that now $\sigma_2$ features as well, via (\ref{demo3}).  
These are all Dragt-type quantities as per Sec \ref{TriStart}.  
These are in $\mathbb{R}^3 = C(\mathbb{S}^2)$ and subject to the on-$\mathbb{S}^2$ restriction $\sum\mbox{}_{\mbox{}_{\mbox{\scriptsize $\sfA$ = 1}}}^3s^{\sfA\,2}  = 1$.

\subsection{\underline{Key 6[$\Box$]} Kuiper coordinates} 

These consist of all the possible inner products between pairs of Jacobi vectors. 
I.e. 3 magnitudes of Jacobi vectors per unit MOI $N^i$, alongside 3  $\bn^i \cdot \bn^j$ ($i \neq j$), which are very closely related to the three relative angles.  
I denote this sextet by $k^{\sfA}$.  
As such, they are, firstly, very much an extension of the parabolic coordinates for the conformally-related flat $\mathbb{R}^3$ of triangleland \cite{08I}. 
These physically correspond to the subsystem split underlying the choice of clustering in use.  
Secondly, in the quadrilateralland setting, they are a clean split into 1) 3 pure relative angles. 
(Any 2 of these are independent and can be interpreted as the anioscelesnesses of the coarse-graining triangles and one coarse-graining rhombus.) 
2) 3 magnitudes, supporting 2 independent non-angular ratios.
I therefore denote this coordinate system by \{$N^i$, aniso($i$)\}. 
Here, aniso($i$) = aniso($jk$) is the same notion of anisoscelesness as before but now applied to the triangle made out of the 1 and 2 relative Jacobi vectors, and so on.   
Thirdly, whilst dimension-counting clearly indicates that the Kuiper coordinates contain 2 redundancies, they are fully democratic in relation to the constituent 
Jacobi vectors and coarse-graining triangles/parallelogram made from pairs of them.

\subsection{Shape quantities and democratic invariants for quadrilateralland}\label{Demo-4}

Repeating the penultimate SSec's working for quadrilateralland closely parallels the analysis for the tetrahaedron in \cite{LR95}. 
This is because the democracy group does not care about dimension. 
Now, 
\beq
\underline{\underline{Q}} = 
\left.
\left\{
w \,\underline{\underline{1}} + \sum\mbox{}_{\mbox{}_{\sfa = 1}}^5 w_{\sfa}\, \underline{\underline{B}}\mbox{}_{\sfa}\right\} \right/2 \mbox{ } , 
\eeq
where the $\underline{\underline{B}}\mbox{}_{\sfa}$ are a priori \cite{LR95} $\mj = 1$  representation matrices of the $SO(3)$ democracy group,  
\beq
\stackrel{\mbox{$\underline{\underline{B}}_1 = \sqrt{3}$}}{} \mbox{\Huge (}
\stackrel{                \mbox{\scriptsize{1 \, 0 \, 0}} \hspace{0.03in}   }
         {    \stackrel{  \mbox{\scriptsize 0   -1 \, 0   }    }
                       {  \mbox{\scriptsize 0 \, 0 \, 0      }    }    } \mbox{\Huge )} 
\mbox{ } , \mbox{ } 
\stackrel{\mbox{$\underline{\underline{B}}_2 = \sqrt{3}$}}{}   \mbox{\Huge (}
\stackrel{                \mbox{\scriptsize{0 \,  1 \, 0}}    }
         {    \stackrel{  \mbox{\scriptsize 1 \, 0 \,  0   }    }
                       {  \mbox{\scriptsize 0 \, 0 \,  0  }    }    } \mbox{\Huge )}\mbox{ } , \mbox{ }
\stackrel{\mbox{$\underline{\underline{B}}_3 = \sqrt{3}$}}{}   \mbox{\Huge (}
\stackrel{    \mbox{\scriptsize{0 \, 0 \, 0}}    }
         {    \stackrel{  \mbox{\scriptsize 0 \, 0 \, 1   }    }
                       {  \mbox{\scriptsize 0 \, 1 \, 0  }    }    } \mbox{\Huge )}\mbox{ } , \mbox{ }
\stackrel{\mbox{$\underline{\underline{B}}_4 = \sqrt{3}$}}{}   \mbox{\Huge (}
\stackrel{    \mbox{\scriptsize{0 \, 0 \, 1}}    }
         {    \stackrel{  \mbox{\scriptsize 0 \, 0 \, 0   }    }
                       {  \mbox{\scriptsize 1 \, 0 \, 0   }    }    } \mbox{\Huge )}\mbox{ } , \mbox{ }
\stackrel{\mbox{$\underline{\underline{B}}_5 =$}}{}     \mbox{\Huge (}
\stackrel{    \mbox{\scriptsize{-1 \, 0 \, 0}}    }
         {    \stackrel{  \mbox{\scriptsize \, 0 -1  \, 0 }    }
                       {  \mbox{\scriptsize \, 0 \, 0 \,  2  }    }    } \mbox{\Huge )}\mbox{ } . \mbox{ }
\eeq
Thus, reading off from the definition of $\underline{\underline{Q}}$ and dropping (H2) labels, and dividing through by I to make shape quantities,


\noindent
\beq
q_1 := \sqrt{3}\{\mn_1\mbox{}^2 - \mn_2\mbox{}^2\}/2 = \sqrt{3}\,\mbox{ellip}(12)/{2} \mbox{ } ,  
\eeq
\beq
q_2 := \sqrt{3}\,{\bn_1}\cdot{\bn_2} = \sqrt{3}\,\mbox{aniso}(12)/2 \mbox{ } ,
\eeq
\beq
q_3 := \sqrt{3}\,{\bn_2}\cdot{\bn_3} = \sqrt{3}\,\mbox{aniso}(23)/2 \mbox{ } ,
\eeq
\beq
q_4 := \sqrt{3}\,{\bn_3}\cdot{\bn_1} = \sqrt{3}\,\mbox{aniso}(31)/2 \mbox{ } ,
\eeq
\beq
q_5 := \{- \mn_1\mbox{}^2 - \mn_2\mbox{}^2 + 2\,\mn_3\mbox{}^2\}/2 = \{\mbox{ellip}(31) + \mbox{ellip}(32)\}/2 \mbox{ } .
\eeq
Note that these quantities, alongside I, are either Kuiper coordinates or linear combinations of Kuiper coordinates.

\noindent Three further quantities of relevance are proportional to the individual triangles' areas,  
\beq
a_1 := \sqrt{3} \, \{\bn_1 \cr \bn_2\}_3 = 2\sqrt{3}\,\mbox{area}(12) \mbox{ } ,
\eeq
\beq
a_2 := \sqrt{3} \, \{\bn_2 \cr \bn_3\}_3  = 2\sqrt{3}\,\mbox{area}(23) \mbox{ } ,
\eeq
\beq
a_3 := \sqrt{3} \, \{\bn_3 \cr \bn_1\}_3  = 2\sqrt{3}\,\mbox{area}(31) \mbox{ } .
\eeq
One can then recognize the whole set of 8 as quadratic forms based on each of the Gell-Mann $\lambda$-matrices.  
Thus I call the 5 Kuiper quantities and the 3 areas together the {\sl Gell-Mann quadratic forms}.    
I label these shape quantities by $s^{\Gamma}$, with components ordered as follows. 
$(q_2, a_1, q_1, q_4, a_3, q_3, a_2, q_5)$ so to match up with the choice of basis employed in the standard representation of Gell-Mann's $\lambda$-matrices.
The triangleland counterpart of this characterization (Sec \ref{TriStart}) should then be called, from a conceptual perspective, {\sl Pauli quadratic forms}.
(For all that they are named after Hopf in the Geometry literature and after Dragt in the Molecular Physics literature.)


\noindent Finally, one has a ninth shape quantity,
\beq
s_9 := q_6 = \sqrt{3}\{|\bn_1 \cr \bn_2|_3\mbox{}^2 + \mbox{cycles}\}^{1/2} = \sqrt{12}\left\{\sum\mbox{ }_{\mbox{}_{\mbox{\scriptsize i = 1}}}^3 a_i^2\right\}^{1/2} =: \mbox{demo}(N = 4) \mbox{ } .  
\eeq
that is quadrilateraland's nontrivial democracy invariant, demo(4).   
Note the interpretation 
\beq
\mbox{demo($N$ = 4) = $\sqrt{\mbox{12 $\times$ sum of squares of mass-weighted areas of coarse-graining triangles per unit 
MOI}}$} \mbox{ }    .  
\eeq
\mbox{ } \mbox{ }
Then 
$1 = \sum\mbox{}_{\mbox{}_{\mbox{\scriptsize $\sfA$ = 1}}}^6q^{\sfA\,2}$: the on-$\mathbb{S}^5$ condition. 
That an $\mathbb{S}^5$ makes an appearance is not unexpected, from Fig \ref{Fig5} and using the generalization of the Hopf map (in the geometrical sense of $U(1) = SO(2)$ fibration),
\beq
\stackrel{\mbox{$\fR(4,2)$}}{= \mathbb{R}^6} \mbox{ }   \mbox{ }    
\stackrel{\mbox{\scriptsize obvious on-sphere condition}}{\longrightarrow}   \mbox{ }   \mbox{ }
\stackrel{\mbox{$\fP(3,2)$}}{= \mathbb{S}^5} \mbox{ }   \mbox{ }      
\stackrel{\mbox{\scriptsize Hopf map}}{\longrightarrow} \mbox{ }   \mbox{ }
\stackrel{\mbox{$\fP(3,2)$}}{= \mathbb{CP}^2}  \mbox{ }   \mbox{ }   
\stackrel{\mbox{\scriptsize coning}}{\longrightarrow}  \mbox{ }   \mbox{ } 
 \stackrel{\mbox{${\cal R}(4,2)$}}{= C(\mathbb{CP}^2)} \mbox{ } .  
\eeq
Moreover, since $\mathbb{CP}^2$ is 4-$d$ and $\mathbb{S}^5$ is 5-$d$, there is a further condition on the $q^{\sfA}$ for quadrilateralland.

Quadrilateralland's relational space is locally $\mathbb{R}^5$.  
One can envisage this from $\mathbb{CP}^2/\mathbb{Z}_2$ = $\mathbb{S}^4$ \cite{Kuiper} and then taking the cone.  
As we are treating $\mathbb{CP}^2$, we have, rather, two copies of $\mathbb{S}^4$.  
Such a doubling does not appear in the coplanar boundary of \cite{LR95}'s Molecular Physics study of 4 particles in 3-$d$.  
This is since 3-$d$ dictates that the space of shapes on its coplanar boundary is the mirror-image-identified one.
Another distinctive feature of 3-$d$ is that it has a third volume-type democracy invariant, $\brho_1 \cdot \brho_2 \cr \brho_3$ \cite{LR95}.
We summarize as follows.

\mbox{ }

\noindent \underline{Key 5[$\Box$]} \{$s^{\sfA}$\} are a redundant set of eight shape coordinates, which can be extended by a ninth quantity 
that is quadrilateralland's democracy invariant.  
These are the analogue of triangleland's Dragt coordinates according to the construction in \cite{LR95, QShape}.  

\mbox{ } 

\noindent This paper is an important prerequisite for the subsequent study of the classical and quantum mechanics of quadrilateralland \cite{QuadIII}. 
The eventual application to the Problem of Time in Quantum Gravity is mostly in Papers II to IV.

\section{\underline{Key 4b[$\Box$]}: Gibbons--Pope coordinates}\label{Q-2}

\noindent The Gibbons--Pope type coordinates (arising in their study of gravitational instantons \cite{GiPo1, GiPo2}) \{$\chi$, $\beta$, $\phi$, $\psi$\} are useful intrinsic coordinates for 
quadrilateralland in a number of senses that extend the rule of the spherical coordinates on the triangleland and 4-stop metroland shape spheres.       
These are a subcase of Euler-angle-adapted coordinates for a privileged $SU(2)$ subgroup of the $SU(3)$. 
See also \cite{Trautman77} and adapted coordinate systems along these lines were probably known previously to that.  
In fact they pick out one of the three constituent overlapping $SU(2) \times U(1)$'s worth of conserved quantities in simple and clear form.  
These coordinates additionally place the metric in `block-minimal' form (i.e. as diagonal as possible, Fig \ref{Matrix}).

{            \begin{figure}[ht]
\centering
\includegraphics[width=0.2\textwidth]{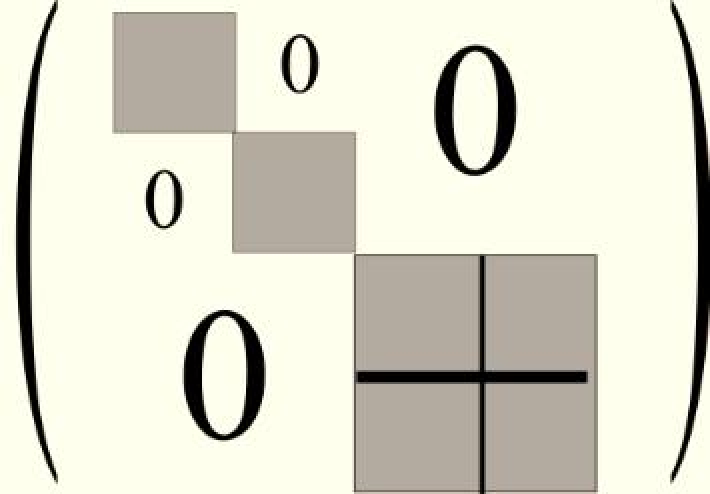}
\caption[Text der im Bilderverzeichnis auftaucht]{        \footnotesize{The block structure type of the Gibbons--Pope type coordinates.}  }
\label{Matrix} \end{figure}          }

\noindent This makes the form of the remaining coordinate less important. 
(I.e. is there is a wider class of $SU(2)$ Euler-angle-adapted coordinate systems with this same type of block structure for the different choices of the last coordinate).
The particular Gibbons--Pope type choice of the $\chi$ further tidies the constituent blocks.  

The coordinate ranges are $0 \leq \chi \leq \pi/2$, 
                 $0 \leq \beta \leq \pi$, 
                 $0 \leq \phi \leq 2\pi$ (a reasonable range redefinition since it is the third relative angle), and
                 $0 \leq \psi \leq 4\pi$.
These are related to the bipolar form of the Fubini--Study coordinates by
\beq
\psi^{\prime} = - \{\Phi_1 + \Phi_2\} \mbox{ } , \mbox{ } \mbox{ } \phi^{\prime} = \Phi_2 -\Phi_1, 
\eeq
with then $\psi = - \psi^{\prime}$ (measured in the opposite direction to match Gibbons--Pope's convention) and $\phi$ is taken to cover the coordinate range $0$ to $2\pi$, 
which is comeasurate with it itself being the third relative angle between the Jacobi vectors involved. 
\beq
\beta = 2\,\mbox{arctan}\,({\cal R}_2/{\cal R}_1)  \mbox{ } , \mbox{ } \mbox{ } \chi = \mbox{arctan}\big(\sqrt{{\cal R}_1\mbox{}^2 + 
{\cal R}_2\mbox{}^2}\big) \mbox{ } .  
\eeq
From here on, a different interpretation is to be attached to these last two formulae in terms of the $\rho_i$, according to the choice of H or K and of ratios.  

\mbox{ } 

\noindent{\bf General geometrical interpretation}.  By their ranges, $\beta$ and $\phi$ parallel azimuthal and polar coordinates on the sphere.
[Furthermore, $\beta$, $\phi$ and $\psi$ take the form of Euler angles on $SU(2)$, with the remaining coordinate $\chi$ playing the 
role of a compactified radius.]   
Now, in the quadrilateralland interpretation, $\beta$ has the same mathematical form as triangleland's azimuthal coordinate $\Theta$ (\ref{TriAzi}).
Additionally, $\chi$ parallels 4-stop metroland's azimuthal coordinate $\theta$ [the first equation in (\ref{4StopPolars})], except that it has only half the range.
(This halving is because the collinear 1234 and 4321 orientations have to be the same due to the existence of the second dimension via which they can be rotated into each other.)  

\mbox{ }

\noindent{\bf Specific interpretation by tree and ratio choice}. 
The basic H's 
$\beta$ is then contents inhomogeneity variable for the ratio of the two posts, whilst its $\chi$ is what proportion of the universe is occupied by its post contents. 
Its $\phi$ and $\psi$ are a sum and a difference of Swiss army knife angles. 
These may be less intuitive than the big $\Phi$'s, but their conjugate momenta are be revealed to have a more lucid interpretation in Paper II.  
\beq
\beta = 2\,\mbox{arctan}\,(\rho_2/\rho_1)  \mbox{ } , \mbox{ } \mbox{ } \chi = \mbox{arctan}\big(\sqrt{\rho_1\mbox{}^2 + \rho_2\mbox{}^2}/\rho_3\big) \mbox{ } .
\eeq 
Its $\phi$ and $\psi$ are a sum and a difference of a Swiss army knife angle and a post angle.  

\noindent The basic K's 
$\beta$ is then the ratio of the face to the thickness of the blade (i.e. a sharpness/flatness shape quantity for the obvious triangle subsystem),             
$\chi$ is the ratio of the face-and-thickness of the blade (which in $d > 1$ forms the obvious triangle subsystem) to the handle.
In this case,
\beq
\beta = 2\,\mbox{arctan}\,(\rho_1/\rho_3)  \mbox{ } , \mbox{ } \mbox{ } \chi = \mbox{arctan}\big(\sqrt{\rho_1\mbox{}^2 + 
\rho_3\mbox{}^2}/\rho_2\big) \mbox{ } .
\eeq
Its $\phi$ and $\psi$ are a sum and a difference of Swiss army knife angles. 

\mbox{ }  

\noindent Finally, in (whichever, with labels suppressed) Gibbons--Pope type coordinates, the Fubini--Study line element is 
\beq
\d \fs^2 = \d\chi^2 + 
  \mbox{sin}^2\chi\big\{\d \beta^2 + \mbox{cos}^2\chi\{\d\phi^2 + \d\psi^2 + 
2\,\mbox{cos}\,\beta\, \d\phi\d\psi\} + \mbox{sin}^2\chi\,\mbox{sin}^2\beta \,\d\phi^2 
\big\}/4 \mbox{ } . 
\eeq
As promised, this is of the form of Fig \ref{Matrix}.

\section{Inclusion of trianglelands, 4-stop metroland within quadrilateralland}\label{Subman}

\noindent \underline{Key 7[$\Box$]} (using tesselations as interpretative back-cloth) is less available due to 4-$d$ manifolds being harder to visualize, 
but one can still ask about the meaningful submanifolds within.  


\noindent In Gibbons--Pope type coordinates based on Jacobi H coordinates, when $\rho_3 = 0$, $\chi = \pi/2$ and the metric becomes 
\beq
\d s^2 = \{1/2\}^2\{\d \beta^2 + \mbox{sin}^2\beta\,\d\phi^2\}
\eeq
i.e. a sphere of radius 1/2, which corresponds to the conformally-untransformed $\mathbb{CP}^1$.  
When $\rho_2$ = 0, $\beta = 0$ and the metric becomes 
\beq
\d s^2 = \{1/2\}^2\{\d \Theta_1^2 + \mbox{sin}^2\Theta_1\d\Phi_1^2\}
\eeq
for $\Theta_1 = 2\chi$ having the correct coordinate range for an azimuthal angle.  
Finally, when $\rho_1$ = 0, $\beta = 0$ and the metric becomes 
\beq
\d s^2 = \{1/2\}^2\{\d \Theta_2^2 + \mbox{sin}^2\Theta_2\d\Phi_2^2\}
\eeq
for $\Theta_2 = 2\chi$ again having the correct coordinate range for an azimuthal angle.   
The first two of these spheres are a triangleland shape sphere included within quadrilateralland.    
The first and second of these are, respectively, collapses to the spaces of the triangles of Fig \ref{FigX}e) and \ref{FigX}f), whilst 
the third of these is collapse to the space of the rhombi of Fig \ref{FigX}g).

In Gibbons--Pope type coordinates based on Jacobi K coordinates, when $\rho_3 = 0$, $\chi = \pi/2$ and the metric becomes 
\beq
\d s^2 = \{1/2\}^2\{\d \beta^2 + \mbox{sin}^2\beta\,\d\phi^2\}
\eeq
i.e. a sphere of radius 1/2, which corresponds to the conformally-untransformed $\mathbb{CP}^1$.  
When $\rho_2$ = 0, $\beta = 0$ and the metric becomes 
\beq
\d s^2 = \{1/2\}^2\{\d \Theta_1^2 + \mbox{sin}^2\Theta_1\d\Phi_1^2\}
\eeq
for $\Theta_1 = 2\chi$ again having the correct coordinate range for an azimuthal angle.  
Finally, when $\rho_1$ = 0, $\beta = 0$ and the metric becomes 
\beq
\d s^2 = \{1/2\}^2\{\d \Theta_2^2 + \mbox{sin}^2\Theta_2\d\Phi_2^2\}
\eeq
for $\Theta_2 = 2\chi$ yet again having the correct coordinate range for an azimuthal angle.   
The first, second and third of these are, respectively, the collapses to the spaces of the triangles of Fig \ref{FigX}l), \ref{FigX}j) and \ref{FigX}k).

In Kuiper coordinates, each of the three on-$\mathbb{S}^2$ conditions, for whichever of H or K coordinates, involves losing one magnitude and two inner products.  
Thus the survivors are two magnitudes and one inner product (closely related to parabolic coordinates and  
linearly combineable to form the \{$I$, aniso, ellip\} system, as per Sec \ref{Q-1}).  
If we recombine the Kuiper coordinates to form the \{$I, s^{\sfa}$\} system, and swap the $I$ for the $k_6$ = demo(4), 
then the survivors of the procedure are the Dragt coordinates (since the procedure kills two of the three area contributions to the $k_6$).  
This gives another sense in which the \{$k^{\sfA}$\} system is a natural extension of the Dragt system.  
  
\noindent Quadrilateralland is also decorated by a net of 6 $\mathbb{S}^2$ trianglelands (in each case with one vertex being a double collision).  

\noindent While it is also easy to write conditions in these coordinates for rectangles, kites, trapezia... , these are less meaningful 
1) mathematically (e.g. they are not topologically defined and are less related to the underlying constellation) and 2) physically.

\subsection{The split into hemi-$\mathbb{CP}^2$'s of oriented quadrilaterals}

{\bf Definition}: The Veronese surface $V$ \cite{Veronese} is the space of conics through a point.
This is a quadratic mathematics parallel to the better-known linear mathematics of how a projective space is a set of lines through a point.
(Geometers and Particle Physicists are familiar with similar generalizations called Grassmann spaces \cite{Nakahara}).

\mbox{ }

\noindent {\bf Kuiper's Theorem} i) The map  

\beq
\eta: \stackrel{\mbox{$\mathbb{CP}^2$}}{ (\mz_1, \mz_2, \mz_3) }                  
\mbox{ } \mbox{ } \mbox{ }
      \stackrel{\mbox{$\longrightarrow$}}{   \stackrel{}{\mbox{$\longmapsto$}}   }  
\mbox{ } \mbox{ } \mbox{ }
      \stackrel{\mbox{$\mathbb{E}^5$}}{(|\mz_1|^2, |\mz_2|^2,|\mz_3|^2, 
\{\mz_2\bar{\mz}_3 + \mz_3\bar{\mz}_2\}/2, \{\mz_3\bar{\mz}_1 + \mz_1\bar{\mz}_3\}/2, \{\mz_1\bar{\mz}_2 + \mz_2\bar{\mz}_1\}/2)}
\eeq  
induces a piecewise smooth embedding of $\mathbb{CP}^2/\mathbb{Z}_2^{\sc\so\sn\sj}$ onto the boundary of the so-called convex hull of the Veronese surface $V$ in $\mathbb{E}^5$, 
which moreover has the right properties to be the usual smooth 4-sphere \cite{Kuiper}. 


\noindent N.B. this is at the topological level; it clearly cannot extend to the metric level by a mismatch in numbers of Killing vectors ($\mathbb{S}^4$ 
with the standard spherical metric has 10 whilst $\mathbb{CP}^2$ equipped with the Fubini--Study metric has 8).    

\mbox{ } 

\noindent Restricting $\{{\mz}_1, {\mz}_2, {\mz}_3\}$ to the real line corresponds in the quadrilateralland interpretation 
to considering the collinear configurations, which constitute a $\mathbb{RP}^2$ space as per Sec \ref{Q-1}.
Moreover the above embedding sends this onto the Veronese surface $V$ itself.  
Proving this proceeds via establishing that, as well as the on-$\mathbb{S}^5$ condition 
\beq
\sum\mbox{}_{\mbox{}_{\mbox{\scriptsize $i$ = 1}}}^3 N^{i} = \sum\mbox{}_{\mbox{}_{\mbox{\scriptsize $i$ = 1}}}^3n^{i\, 2} = 1 \mbox{ } ,
\label{JustS5} 
\eeq   
a second restriction holds, which in our quadrilateralland interpretation, reads 
\beq
\sum\mbox{}_{\mbox{}_{\mbox{\scriptsize $i$ = 1}}}^3 \mbox{aniso}(i)^2N^{i} - 4N_1N_2N_3 + \mbox{aniso}(1)\,\mbox{aniso}(2)\, \mbox{aniso}(3) = 0 \mbox{ }  
\label{Cubi}
\eeq 
(Knowledge of this restriction is also useful in kinematical quantization \cite{I84}, and it is clearer in the \{$N^i$, aniso($i$)\} system, 
which is both the quadrilateralland interpretation of Kuiper's redundant coordinates and a simple linear recombination of the shape coordinates \{$I$,$s^{\sfa}$\} coordinates.)

\mbox{ }

\noindent {\bf Another form for Kuiper's theorem} \cite{Kuiper}.
Moreover, $\mathbb{CP}^2$ itself is topologically a double covering of $\mathbb{S}^4$ branched along the $\mathbb{RP}^2$ 
of collinearities which itself embeds onto $\mathbb{E}^5$ to give the Veronese surface $V$. 

\mbox{ } 

\noindent Here, branching is meant in the sense familiar from the theory of Riemann surfaces \cite{Riemann}. 
Moreover,  the $\mathbb{RP}^2$ itself embeds 

\noindent {\sl non-smoothly} into the Veronese surface $V$.

\mbox{ }

\noindent Then the quadrilateralland interpretation of these results is in direct analogy with the plain shapes case of triangleland consisting of two hemispheres of opposite orientation 
bounded by an equator circle of collinearity. 
Thus quadrilateralland's distinction between clockwise- and anticlockwise-oriented figures is strongly anchored to this geometrical split, 
with the collinear configurations lying at the boundary of this split.

\section{Gell-Mann quadratic forms in terms of Gibbons--Pope coordinates}\label{GM-GP}

This is motivated by the Kuiper quantities occuring in the HO potentials (see Paper II) as well as 
furnishing interesting `shape operators' at the quantum level (e.g. it is part of Paper II's kinematical quantization scheme).   
In each case, however, the kinetic part of the Hamiltonian is far more conveniently expressed in terms of 
Gibbons--Pope type coordinates, thus necessitating the below conversions for the Gell-Mann quadratic forms:  

\noindent 
\beq
s_1 = \sqrt{3} \bn_1\cdot \bn_2 = \sqrt{3} \mbox{sin}^2\chi\,\mbox{sin}\beta\mbox{cos}\phi =: \sqrt{3} \mbox{sin}^2\chi\,\mbox{sin}\beta\mbox{cos}\,f_3 \mbox{ } ,
\label{n1n2}
\eeq
\beq
s_2 = \sqrt{3} \{\bn_1\cr \bn_2\}_3 = \sqrt{3} \mbox{sin}^2\chi\,\mbox{sin}\beta\mbox{sin}\phi =: \sqrt{3} \mbox{sin}^2\chi\,\mbox{sin}\beta\mbox{sin}\,f_3 \mbox{ } ,
\label{A12}
\eeq
\beq
s_3 = \sqrt{3} \mbox{sin}^2\chi\,\mbox{cos}\,\beta \mbox{ } ,
\label{3}
\eeq
\beq
s_4 = \sqrt{3} \bn_3\cdot \bn_1 = \sqrt{3} \mbox{sin}\,2\chi\,\mbox{cos}\mbox{$\frac{\beta}{2}$}\mbox{cos}\mbox{$\frac{\psi - \phi}{2}$} 
=: \sqrt{3} \mbox{sin}\,2\chi\,\mbox{cos}\mbox{$\frac{\beta}{2}$}\mbox{cos}\,f_2 \mbox{ } .  
\label{n3n1}
\eeq
\beq
s_5 = \sqrt{3} \{\bn_3\cr \bn_1\}_3 =  \sqrt{3} \mbox{sin}\,2\chi\,\mbox{cos}\mbox{$\frac{\beta}{2}$}\mbox{sin}\mbox{$\frac{\psi - \phi}{2}$} 
=: \sqrt{3} \mbox{sin}\,2\chi\,\mbox{cos}\mbox{$\frac{\beta}{2}$}\mbox{sin}\,f_2 \mbox{ } .  
\label{A31}
\eeq
\beq
s_6 = \sqrt{3} \bn_2\cdot \bn_3 = \sqrt{3} \mbox{sin}\,2\chi\,\mbox{sin}\mbox{$\frac{\beta}{2}$}\mbox{cos}\mbox{$\frac{\psi + \phi}{2}$} 
=: \sqrt{3} \mbox{sin}\,2\chi\,\mbox{sin}\mbox{$\frac{\beta}{2}$}\mbox{cos}\,f_1 \mbox{ } ,
\label{n2n3}
\eeq
\beq
s_7 = \sqrt{3} \{\bn_2\cr \bn_3\}_3 = \sqrt{3} \mbox{sin}\,2\chi\,\mbox{sin}\mbox{$\frac{\beta}{2}$}\mbox{sin}\mbox{$\frac{\psi + \phi}{2}$} 
=: \sqrt{3} \mbox{sin}\,2\chi\,\mbox{sin}\mbox{$\frac{\beta}{2}$}\mbox{sin}\,f_1 \mbox{ } ,
\label{A23}
\eeq
\beq
s_8 = \sqrt{3}\{3\mbox{cos}^2\chi  -  1 \} \mbox{ } ,
\label{8}
\eeq

\noindent [The $f_i$ notation is useful as regards transposition symmetry arguments interrelating interpretations in terms of different ratio choices.]
I also note that these are a nice rendition of quantities that auto-obey the two on-$\mathbb{CP}^2$ conditions as trigonometric {\sl identities}, 
which plays a role in Paper II's quantum scheme.
(\ref{JustS5}) obviously holds, whilst (\ref{Cubi}) becomes
\beq
\sum\mbox{}_{\mbox{}_{\mbox{\scriptsize$i$ = 1}}}^3\mbox{cos}^2f_i + 2\mbox{cos}f_1\mbox{cos}f_2\mbox{cos}f_3 =  1  \label{OrigId} \mbox{ } , 
\eeq
which trigonometric identity indeed holds for angles such that $f_1  + f_2 + f_3 = 0$, which is true since we are free to change 
signs of some of the $f_i$ here since they occur solely inside cos functions for which signs do not matter.  
Going from (\ref{Cubi}) to (\ref{OrigId}) amounts to reversing the working by which Kuiper obtained the non-quadrilateralland-interpreted version of (\ref{Cubi}) in the first place.  

\mbox{ }  

%
\noindent The other significant quantity, $s_9$, is more conveniently written in terms of Kuiper coordinates; its square is 
\beq
\mbox{demo}(N = 4)^2 = 3\big\{N_1N_2 + N_3\{1 - N_3\} - \sum\mbox{}_{\mbox{}_{\mbox{\scriptsize $i$ = 1}}}^3\mbox{aniso}(i)/16  \big\}    \mbox{ }  .
\eeq
\noindent Finally, the {\it irreducible homogeneous polynomials} IHP($\mathbb{C}^{k}$, p) involve $\mathbb{C}^k$ variables and are of degree p; 
these spaces occur under another name in \cite{BGM71}.
In particular, IHP($\mathbb{C}^{2}$, p) is $\mathbb{R}^3$ as a consequence of the $SU(2) = SO(3)$ `accidental relation' between the Pauli matrices and the $\mathbb{R}^3$ basis vectors, 
and IHP($\mathbb{C}^{k}$, p) is indeed then the correct $N$-a-gon generalization of some uses of $\mathbb{R}^3$ for triangleland.

\section{\underline{Key 8}: Notions of uniformity for quadrilateralland}\label{Unif}

Notions of uniformity are of classical and quantum cosmological significance due to the commonly-held stance that the very early universe/initial state was very highly uniform.  
Some particularly uniform configurations are, for quadrilateralland, three squares per hemi-$\mathbb{CP}^2$ of orientation 
in analogy to the single equilateral triangle per hemisphere of orientation of triangleland. 
This reflects the presence of a further threefold symmetry in quadrilateralland; in this case, choosing to use indistinguishable particles quotients this out.  
As explained in detail in \cite{+tri}, the nontrivial democratic invariant demo(3) for triangleland is the mass-weighted area of the triangle per unit moment of inertia.  
In analogy, for quadrilateralland the nontrivial democratic invariant demo(4), is proportional to the square root of the sum of the mass-weighted areas of the 
coarse-graining triangles/rhombi in Fig \ref{FigX}g).

Triangleland's area Dragt coordinate = demo(3) can be used as a measure of uniformity \cite{QShape}, its modulus running from maximal value at the most uniform configuration 
(the equilateral triangle) to minimal value for the collinear configurations. 
Quadrilateralland's square root of sums of squares of constituent subsystems' areas demo(4) likewise gives a quantifier of uniformity for model-universe configurations.  
Its minimal value likewise picks out quadrilateralland's collinear configurations, 
but extremizing demo(4) does not uniquely pick out the six labelled squares -- one gets one extremal curve per hemi-$\mathbb{CP}^2$, each containing three labelled squares.  
The present paper's choices of coordinate systems clarify the interpretation to be given to these two extremal curves.  
In $\{N^i$, aniso($i$)\} coordinates, these are given by aniso(1) = 0 = aniso(2) and $\left| \mbox{aniso(3)}\right|$ = 1,  i.e. (1)-isosceles, (2)-isosceles and maximally (3)-right 
or (3)-left i.e. (3)-collinear, with $N_3$ = 1/2 and $N_1$, $N_2$ varying (but such that the on-$\mathbb{S}^5$ condition $N_1 + N_2 + N_3 = 1$ holds).  
If one works instead in Gibbons--Pope type coordinates, the uniformity condition is $\phi = 0$, $\psi = 4\pi$, $\chi = \pi/4$ and $\beta$ free. 
I.e., in the H-coordinates case, freedom in the contents inhomogeneity i.e. size of subsystem 1 relative to the size of subsystem 2.
On the other hand, in the K-coordinates case, $\beta$-free signifies freedom in how tall one makes the selected \{12, 3\} cluster's triangle.

\section{Physical interpretation of shape momenta}\label{Mom-Int}

I use relative angular momenta and relative distance momenta as names for the conjugates of relative angles and of ratios of relative separations respectively.

\subsection{3- and 4-stop metroland}

For 3-stop metroland in polar coordinates, the momenta are [dropping (a) labels and recycling the notation $\mp_i$ to mean the conjugate of 
$\mn^i$],
\beq
{\cal D} := p_{\varphi} = \varphi^* = \mn_1\mp_2 - \mn_2\mp_1 = {\ttD}_2\mn_1/\mn_2 - {\ttD}_1\mn_2/\mn_1 
\eeq
for $\ttD_i$ the {\it partial dilations} (in parallel to the $I_i$ being partial moments of inertia) $\mn^i\np_i$ (no sum).  
The second form of this is manifestly a shape-weighted {\it relative dilational quantity} corresponding to a particular exchange of dilational 
momentum between the \{bc\} and \{a\} clusters.
It is indeed conceptually clear that the conjugate to the non-angular length ratio $\varphi^{(\sfa)}$ is a relative distance momentum.
I generally use the notation ${\cal D}$ for whichever type of relative distance momenta.  

\mbox{ }  

For 4-stop metroland in spherical coordinates, the momenta are [dropping (Hb) or (Ka) labels] 
\beq
{\cal D}_{\phi}   := p_{\phi} = \phi^* = \mn_1\mp_2 - \mn_2\mp_1  = {\ttD}_2\mn_1/\mn_2 - {\ttD}_1\mn_2/\mn_1 \mbox{ } ,   
\label{WillBeA3}
\eeq
i.e. a a weighted relative dilational quantity corresponding to a particular exchange of dilational momentum between the \{ab\} and \{cd\} 
clusters in the H-case or the \{bc\} and \{Ta\} clusters in the K-case, and
\beq
{\cal D}_{\theta} := p_{\theta} \mbox{ } . 
\eeq
%

\subsection{Triangleland}

The triangleland momenta and their interpretation are as follows. 
\beq
{\cal D}_{\triangle} =: p_{\Theta} := \Theta^* = \mbox{dra}_1\Pi^{\sd\sr\sa}_2 -  \mbox{dra}_2\Pi^{\sd\sr\sa}_1  \mbox{ } , 
\eeq
\beq
{\cal J} =: p_{\Phi} := \mbox{sin}^2\Theta\,\Phi^* \mbox{ } . 
\eeq
Here, and more generally, I use ${\cal J}$ to denote angular momenta.  
This ${\cal J}$, moreover, clearly cannot be an overall angular momentum since $\ttL = 0$ applies.  
It is indeed a relative angular momentum \cite{08I}: 
\beq
{\cal J} = I_1I_2\Phi^{*}/I = I_1I_2\{\theta_2^* - \theta_1^*\}/I = \{I_1\mL_2 - I_2\mL_1\}/I = 
\mL_2 = - \mL_1 = \{\mL_2 - \mL_1\}/2 \mbox{ } 
\eeq
[the fourth equality uses the zero total angular momentum constraint (\ref{ZAM})].   
Thus this can be interpreted as the angular momentum of one of the two constituent subsystems, minus the 
angular momentum of the other, or half of the difference between the two subsystems' angular momenta.  
That is indeed a relative angular momentum also ought to also be clear from it being the conjugate of a relative angle.
The $\Theta$ and $\Phi$  coordinates represent a clean split into pure non-angle ratios and pure angle ratios, 
by which they produce one relative dilational momentum and one relative angular momentum as their conjugates.

\mbox{ } 

\noindent Franzen and I \cite{AF} termed the collective set of quantities of this nature relative {\it rational momenta} since they correspond 
to the general-ratio generalization of angle-ratio's angular momenta.
Franzen and I already noted that the rational momentum concept also naturally extends to include mixed dilational momentum 
and angular momentum objects in addition to the above examples of purely dilational and purely angular objects.
Rational momenta was previously called {\it generalized angular momenta} by Felix Smith \cite{Smith60} in the Molecular Physics context.  
Serna and I do not use this name since, once again, it is not conceptually descriptive; 
we rather unravel exactly what it means physically and thus call it by its `true name' \cite{WheelerInt, Kvothe}.
The rather conceptually-cleaner introduction of this at the level of the momenta rather than \cite{AF}'s at the level of the conserved quantities is new to the present paper).
Our final proposal is to call them {\it shape momenta}, since what are mathematically ratio variables can also be seen to be dimensionless shape variables, 
and the quantity in question is the momentum conjugate to such a quantity.  
We celebrate this by passing from the notation ${\cal R}$ for `rational' to ${\cal S}$ for `shape'.

This `true naming' becomes clear in moving, away from the previous idea of interpreting in physical space the $SO(n)$ mathematics of 
the first few RPM models studied, along the following line of thought.  

\noindent 1) Scale--shape splits are well-defined.  
Then there are corresponding splits into scale momenta and shape momenta.  

\noindent 2) The shape momenta are conjugate to dimensionless variables, i.e. (functions of) ratios, accounting for why the previously encountered objects were termed rational momenta. 

\noindent 3) Then in some cases, shape momentum mathematics coincides with (arbitrary-dimensional) angular momentum mathematics, and also 
some shapes/ratios happen to be physically angles in space, so the interpretation in space {\sl indeed is} as angular momentum.  
  
\noindent 4) However, in other cases, shapes can correspond physically to ratios other than those that go into angles in space.
E.g. ratios of two lengths (then one's momentum is a pure relative distance momentum) or a mixture of angle and non-angle in space ratios (in which case one has a general shape momentum).
Moreover, there is no a priori association between shape momenta and $SO(n)$ groups. 
This {\sl happens} to be the case for the first few examples encountered 
($N$-stop metroland, triangleland) but ceases to be the situation for quadrilateralland (and $N$-a-gonlands beyond that).

\subsection{Quadrilateralland}

The Gibbons--Pope type coordinates for quadrilateralland extend the above triangleland spherical polar coordinates 
in constituting a clean split into pure non-angle ratios and pure angle ratios (two of each).
Thus their conjugates are again cleanly-split pure relative angular momenta and relative dilational momenta as their conjugates (two of each): 
\beq
{\cal J}_{\psi}  =: p_{\psi}  = \mbox{sin}^2\chi\mbox{cos}^2\chi\{\psi^* + \mbox{cos}\,\beta\,\phi^*\}/4 \mbox{ } , \hspace{1.15in}
\eeq
\beq
{\cal J}_{\phi}  =: p_{\phi}  = \mbox{sin}^2\chi\{\mbox{cos}^2\chi\{\phi^* + \mbox{cos}\,\beta\,\psi^*\} + \mbox{sin}^2\chi\,\mbox{sin}^2\beta\, \phi^*\}/4 \mbox{ } , \hspace{0.1in} 
\eeq
\beq
{\cal D}_{\beta} =: p_{\beta}  =  \mbox{sin}^2\chi\,\beta^*/4 \mbox{ } , \hspace{2.25in}
\eeq
\beq
{\cal D}_{\chi}  =: p_{\chi}  = \chi^* \mbox{ } . \hspace{2.75in} 
\eeq
\mbox{ } \mbox{ } The interpretation of these in terms of quadrilaterals are as follows (using Fig \ref{Fig1}'s and Sec \ref{Q-2}'s nomenclatures).  
\noindent The basic H's 
$p_{\beta}$ is then the relative dilation of the two posts -- universe contents, 
                     whilst its $p_{\chi}$ is the relative dilation of the posts contents relative to their `universe separation'.
\noindent The basic K's 
$p_{\beta}$ is then the relative dilation of the blade face to the blade thickness 
(i.e. a change of sharpness/flatness shape momentum for the obvious triangle subsystem in $d > 1$). 
$p_{\chi}$ is the relative dilation of the face-and-thickness of the blade (for the obvious triangle subsystem) to the remaining handle particle 
I.e. a relative dilation of the whole triangle relative to the separation between it and the remaining particle.
\noindent 
In each case, $p_{\phi}$ and $p_{\psi}$ are a co-rotation and a counter-rotation of the two selected objects without any discernible pattern.

\section{Forms of the shape Hamiltonians}\label{Hams}

\beq
\mbox{The corresponding Hamiltonians are} \hspace{1in}  \fH = p_{\varphi}\mbox{}^2/2 + \fV = {\cal D}^2/2 + \fV \mbox{ (3-stop metroland) ,} \hspace{2in}
\eeq
\beq
\fH = p_{\theta}\mbox{}^2/2 + p_{\phi}\mbox{}^2/2\,\mbox{sin}^2\theta + \fV = {\cal D}_{\theta}\mbox{}^2/2 + {\cal D}_{\phi}\mbox{}^2/2\,\mbox{sin}^2\theta + \fV 
\mbox{ } \mbox{ (4-stop metroland) } , 
\eeq
\beq
\fH = p_{\Theta}\mbox{}^2/2             +   p_{\Phi}\mbox{}^2/2\,\mbox{sin}^2\Theta + \fV
     = {\cal D}_{\triangle}\mbox{}^2/2   +   {\cal J}^2/2\,\mbox{sin}^2\Theta + \fV \mbox{ } \mbox{ (triangleland) and } 
\eeq
$$
\fH = \frac{p_{\chi}^2}{2} + \frac{2}{\mbox{sin}^2\chi}\left\{p_{\beta}^2 + \frac{1}{\mbox{sin}^2\beta}
\{p_{\phi}^2  \mbox{ } +   \mbox{ } p_{\psi}^2  \mbox{ } -  \mbox{ } 2\, p_{\phi}p_{\psi} \mbox{cos}\,\beta \}\right\} + \frac{2}{\mbox{cos}^2\chi}p_{\psi}^2  + \fV \hspace{0.4in}
$$
\beq
=\frac{{\cal D}_{\chi}^2}{2} + \frac{2}{\mbox{sin}^2\chi}\left\{{\cal D}_{\beta}^2 + \frac{1}{\mbox{sin}^2\beta}
\{{\cal J}_{\phi}^2 + {\cal J}_{\psi}^2 - 2\,{\cal J}_{\phi}{\cal J}_{\psi}\mbox{cos}\,\beta \}\right\} + \frac{2}{\mbox{cos}^2\chi}
{\cal J}_{\psi}^2  + \fV \mbox{ } \mbox{ (quadrilateralland) } . 
\eeq

\section{Physical interpretation of RPM isometries/conserved quantities}\label{SSec: Cons}

\noindent For a dynamical system, conserved quantities correspond to isometries of the kinetic metric that are also respected by the potential.
These are crucial for the classical and quantum understanding of a system.  
This Section and the next deal with isometries; see Sec 15 for which of these survive as conserved quantities for various potentials.
\beq
\mbox{Isom($\fS(N, 1))$ = Isom($\mathbb{S}^{n - 1}$) = $PSO(n)$ (the $n$-dimensional projective special orthogonal group) = $SO(n)$ } .
\eeq
\beq
\mbox{Isom($\fS(N, 2))$ = Isom($\mathbb{CP}^{n - 1}$) = $PSU(n)$ (the $n$-dimensional projective special unitary group) = $SU(n)/\mathbb{Z}_n$ } . 
\eeq
The $SU(n)$ versus  $SU(n)/\mathbb{Z}_n$ distinction does not affect the algebra involved, though it does matter as regards 
some further subtleties (along the lines of the much better-known $SU(2)$ versus $SU(2)/\mathbb{Z}_2 = SO(3)$ distinction).  
The above is fairly standard mathematics; moreover, the {\sl physical interpretation} of the generators of these 
(which for various classes of potentials are to be interpreted as conserved quantities) is somewhat unusual, as I shall build up case-by-case below.

\subsection{$N$-stop metroland cases}\label{Sprintina}

The pure-shape case of 3-stop metroland is relationally trivial as per \cite{FileR}, but it is part of dynamically nontrivial scaled 

\noindent3-stop metroland problem.  
Here the generator of Isom$(\fS(3, 1))$ = Isom$(\mathbb{S}^1) =  SO(2) = U(1)$ is just the above-described 
\noindent ${\cal D}\mi\ml$.   
This is mathematically the `component out of the plane' of  `angular momentum', albeit in {\sl configuration space}, 
there clearly being no meaningful physical concept of angular momentum in 1-$d$ space itself.

For 4-stop metroland, the three generators of Isom($\fS(4, 1)$) = Isom($\mathbb{S}^2$) = $SO(3)$ are 
\beq
{\cal D}_i = {{\epsilon_{ij}}^k}\mn^j\mp_{k} \mbox{ } .
\label{DDD}
\eeq 
(\ref{DDD}) are mathematically the three components of `angular momentum' albeit again in configuration space.  
Their physical interpretation (for the moment in the setting of H-coordinates) in space is an immediate extension of that of the already-encountered 3-component of this object 
(\ref{WillBeA3}):
\beq
{\cal D}_i = {\ttD}_k\mn^j/\mn^k - {\ttD}_j\mn^k/\mn^j \mbox{ } .  
\eeq
Moreover, this example's interpretation relies, somewhat innocuously, on the three conserved quantities ${\cal D}\mi\ml_i$ corresponding to three mutually perpendicular directions 
(the three DD axes picked out by using H-coordinates), as is brought out more clearly by the next example.

For 4-stop metroland in K-coordinates the above formulae recurr [dropping (Ka) labels instead of (Hb) ones]. 
They are clearly still all relative distance momenta, albeit corresponding to a different set of ratios.  
Then e.g. ${\cal D}_3$ is a (weighted) {\it relative dilational quantity} corresponding to a particular exchange of dilational momentum between the \{12\} and \{T3\} clusters. 
Here, one needs to use an axis system containing only one T-axis, e.g. a \{$\mT, \mM^*\mD, \mM^*\mD$\} axis system (c.f. Fig \ref{Fig-4-Stop}).

For 4-stop metroland the {\it total shape momentum} counterpart of the total angular momentum is 

\noindent ${\cal D}_{\sT\so\st} = \sum_{i = 1}^{3}{\cal D}_i\mbox{}^2 = {\cal D}_{\theta}\mbox{}^2 + \mbox{sin}^{-2}\theta\,
{\cal D}_{\phi}\mbox{}^2 = 2\fT$ in terms of momenta. 
For 3-stop metroland, this is just ${\cal D}_{\sT\so\st} = {\cal D}^2$. 
In each case, finally, $\fH = {\cal D}_{\sT\so\st}/2 + \fV$.

The above pattern repeats itself, giving,  for $N$-stop metroland, a) $n$ -- 1 hyperspherical coordinates that can be interpreted as a sequence of ratios of relative inter-particle 
cluster separations. b) Shape space isometry group $SO(n)$. c) A set of $n\{n - 1\}/2$ isometry generators which are, mathematically, components of `angular momentum' in configuration space.

\subsection{Triangleland case}\label{Sprague}

Here, the three Isom $(\mathbb{S}^2) = \mbox{Isom}\fS(3, 2) = SO(3) = SU(2)/\mathbb{Z}_2$ generators are given by 
\beq
{\cal S}_{\sfA} = {{\epsilon_{\sfA\sfB}}^{\sfC}}\mbox{dra}^{\sfB} \Pi^{\sd\sr\sa}_{\sfC} \mbox{ } ,
\label{SSS}
\eeq
which are mathematically the three components of `angular momentum' albeit yet again in configuration space rather than in space.  
Now on this occasion, there is a notion of relative angular momentum in space.
There are even three natural such, one per clustering: ${\cal J}_{(\sa)}$.    
Are these the three components of ${\cal S}_{\sfA}$?
No! 
These  three are coplanar and at 120 degrees to each other, so only can only pick one of these for any given orthogonal coordinate basis, much as in the above K-coordinate example. 
The other components point in an E and an S direction (c.f. Fig \ref{Fig4}).
E and D are then the two main useful choices of principal axes, furnishing the \{E, D, S\} and \{D, E, S\}.   
Moreover the component pointing in the D direction has the form of a  pure relative angular momenta, 
${\cal S}_3 = {\cal J}$ of the \{23\} subsystem relative to the 1 subsystem.   
The other two ${\cal S}_{\sfA}$'s are mixed dilational and angular momenta with shape-valued coefficient [dropping (a) labels]: 
\beq
\mbox{sin}\,\Phi\, {\cal D}_{\triangle} + \mbox{cos}\,\Phi\,\mbox{cot}\,\Theta\,{\cal J}   \mbox{ } \mbox{ and } 
\mbox{ } \mbox{ }
-\mbox{cos}\,\Phi\, {\cal D}_{\triangle} + \mbox{sin}\,\Phi\,\mbox{cot}\,\Theta\,{\cal J}   \mbox{ } .
\eeq 
For triangleland, the {\it total shape momentum} counterpart of the total angular momentum is 

\noindent 
${\cal S}_{\sT\so\st} = \sum_{\sfA = 1}^3{\cal S}_{\sfA}\mbox{}^2 = {\cal D}_{\triangle}\mbox{}^2 + \mbox{sin}^{-2}\Theta\,{\cal J}^2  = 2\fT$. 
Finally, $\fH = {\cal D}_{\sT\so\st}/2 + \fV$.

\section{Quadrilateralland case}\label{Quad-Case}

Quadrilateralland's isometry group is Isom($\fS(4, 2)$) = Isom($\mathbb{CP}^2$) = $PSU(3)$ = $SU(3)/\mathbb{Z}_3$. 
This gives the same representation theory and mathematical form 
of conserved quantities as in the idealized flavour $SU$(3) or the colour $SU$(3) of Particle Physics [these {\sl also} have this quotienting].   
MacFarlane studied this and the difference between it and $SU(3)$ in \cite{Macfarlane68}; they share the same algebra, but there are some topological differences between them.
There are some parallels with the extent of the similarities between $SU(2)$ and $SU(2)/\mathbb{Z}_2 = SO(3)$ 
[which itself is relevant to RPM's via Isom($\fS(3, 2)$) = Isom($\mathbb{S}^1$) =  Isom($\mathbb{CP}^1$) = $PSU(2)$ = $SU(2)/\mathbb{Z}_2$].

\subsection{Particle Physics analogues}

Analogy 1) Flavour symmetry (constitution of hadrons in terms of up, down and strange quarks).
My use of 1, 2, 3, +, 

\noindent and -- is the standard one of $SU(2)$ mathematics.
$SU(3)$ contains three overlapping such ladders (in fact three overlapping 
$SU(2) \times U(1)$'s, with the 
$SU(2)$'s being isospins $I_+$, $I_-$, $I_3$, $V_+$, $V_-$, $V_3$ and $U_+$, $U_-$, $U_3$ and the 
$U(1)$'s being hypercharges $Y$, $Y_V$ and $Y_U$).  
The usual set of independent such objects, $I_3$, $I_+$, $I_-$, $V_+$, $V_-$, $U_+$, $U_-$ 
and ${Y}$, are then  represented by the Gell-Mann $\lambda$-matrices up to proportion. 
Then one can obtain ${V}_3$, ${U}_3$ ${Y}_{U}$,and ${Y}_{V}$ in terms of these, these other quantities being 
useful on grounds of even-handedness between the three $SU(2) \times U(1)$'s (see the next subsection).
One can then define ${I}_{\sT\so\st } = \sum_{{\cal A} = 1}^3{I}_{{\cal A}}\mbox{}^2$, 
                    ${U}_{\sT\so\st } = \sum_{{\cal A} = 1}^3{U}_{{\cal A}}\mbox{}^2$ and 
                    ${V}_{\sT\so\st } = \sum_{{\cal A} = 1}^3{V}_{{\cal A}}\mbox{}^2$.
In total, $SU(3)$ has 3 independent commuting quantities, which are usually taken to be ${I}_{\sT\so\st }$, ${I}_3$, ${Y}$.   


\noindent Note that flavour symmetry is broken by mass differences -- it is only an approximate symmetry.  
\noindent Also note that the word `hypercharge' is also not conceptually descriptive.  
In the Particle Physics of flavour, it is an `extra charge' that partly contains strangeness, unlike the isospin which is purely in terms of up and down quarks.  
Thus Serna and I prefer the more conceptually descriptive names `strange charge' for Particle Physics and `extra charge' for the quadrilateral itself. 
(Take due note that `charge' used in an unqualified way is taken to imply the presence of a very common $U(1)$ symmetry rather than some generalized non-abelian symmetry.)  
We also use {\sl angular charge} in place of isospin when describing the quadrilateral; here the qualification `angular' is taken to carry $SU(2)$ symmetry connotations.    

\mbox{ } 

\noindent Analogy 2) (MacFarlane) Colour symmetry.
This use of $SU(3)$ differs in being postulated to be exact, and in the red, green and blue labels being frivolous choices, so that one is really dealing with $SU(3)/\mathbb{Z}_3$.

\subsection{\underline{Key 9}: Quadrilateralland's conserved quantities}

I calligraphize all of the above symbols in the quadrilateralland case, to distinguish these quantities clearly from their Particle Physics analogues.  
This application is in fact more like colour physics than approximate flavour physics, in that the symmetry is exact. 
However, whilst for uninterpreted $\mathbb{CP}^2$ one can take the three types of ladder to be frivolous labels.  
Thus they involve $SU(3)/\mathbb{Z}_3$. 
The quadrilateralland interpretation pins distinction upon the three ladders. 
Thus one wishes for the whole $\mathbb{CP}^2$ with its three uniform states per hemi-$\mathbb{CP}^2$ rather than a folded-up version in which the three coincide.  


On the basis of the above discussion, Serna and I call ${\cal I}_3$ the {\it angular charge} and ${\cal Y}$ the {\it extra angular charge}    
due to its coming alongside the usual angle charge's $SU(2)$ but not within it, as a picked out $SU(2) \times U(1)$.
In the $\mathbb{CP}^2$ realization of $SU(3)$, this is  \cite{MF79, MacFarlane} not only picked out by the basis but also by the Gibbons--Pope type coordinates in use.  

\noindent
\beq
{\cal Y} = 2p_{\psi} = 2{\cal J}_{\psi} \mbox{ } \mbox{ } , \mbox{ }  \mbox{ } {\cal I}_3 = {p}_{\phi} = {\cal J}_{\phi} \mbox{ } .
\eeq
In terms of the quadrilateralland-significant inhomogeneous bipolar coordinates, these are then 
\beq
{\cal Y} = - 2\{ p_{\Phi_1} + p_{\Phi_2}  \} \mbox{ } , \mbox{ }  \mbox{ }
{\cal I}_3 = p_{\Phi_2} - p_{\Phi_1} \mbox{ } .  
\eeq
The meanings of ${\cal Y}$ and ${\cal I}$ are immediately inherited from those of $p_{\psi}$ and $p_{\phi}$.

\subsection{Generators of the isometries in $\mZ^{\barp}$ coordinates from Noether's theorem}

Conserved quantities in terms of $\mZ^{\barp}$ and $\Pi_{\barp}$ are presented below.  
MacFarlane \cite{MacFarlane} derived these from the Euler--Lagrange action.
The present paper uses instead the r-formulation's Jacobi-type action, the outcome from which is equivalent to MacFarlane's result by the following Lemma. 

\mbox{ } 

\noindent{\bf Lemma 3}.  
The quantities arising from Noether's theorem as applied to a Jacobi-type action are equivalent to those arising from the corresponding Euler--Lagrange type action.

\mbox{ } 

\noindent Thus MacFarlane's results carry over to the context of relational Jacobi-type actions, and provide the following conserved quantities.
\beq
2i{\cal I}_3 = \bPi \sigma_3 {\bZ} - \overline{\bPi}\sigma_3\overline{\bZ} \mbox{ } , \mbox{ } \mbox{ } 
 i{\cal Y} = \bPi \cdot {\bZ} - \overline{\bPi}\cdot\overline{\bZ} 
\eeq
\beq
i{\cal I}_+ = \Pi_1\mZ^2 - \overline{\Pi}_2\overline{\mZ}^1 \mbox{ } , \mbox{ } \mbox{ } 
i{\cal I}_- = \Pi_2\mZ^1 - \overline{\Pi}_1\overline{\mZ}^2 \mbox{ } ,
\eeq
\beq
i{\cal V}_+  = \Pi_1 + \overline{\bPi}\cdot\overline{\bZ}\,\overline{\mZ}^1 \mbox{ } , \mbox{ } \mbox{ } 
-i{\cal V}_- = \overline{\bPi}_1 + \bPi\cdot{\bZ}\,\mZ^1 \mbox{ } , 
\eeq
\beq
i{\cal U}_+  = \Pi_2 + \overline{\bPi}\cdot\overline{\bZ}\,\overline{\mZ}^2 \mbox{ } , \mbox{ } \mbox{ } 
-i{\cal U}_- = \overline{\bPi}_2 + \bPi\cdot{\bZ}\,\mZ^2 \mbox{ } . 
\eeq
\mbox{ } 
\noindent Note that the above constitutes a nonlinear realization of the $SU(3)$. 
\noindent Also note that for the triangleland counterpart,    
${\cal Y}_{\cal U} = -\frac{1}{4}\{{\cal I}_3 + 3{\cal Y}\}$, 
${\cal Y}_{\cal V} = -\frac{1}{4}\{{\cal I}_3 - 3{\cal Y}\}$, 
${\cal U}_3 =  \frac{1}{2}\{{\cal I}_3 - {\cal Y}\}$ and 
${\cal V}_3 =  \frac{1}{2}\{{\cal I}_3 + {\cal Y}\}$.  

\noindent The generators are of types
\beq
\Pi \mZ - \overline{\Pi}\overline{Z}          \mbox{ } , \mbox{ } \mbox{ } 
{\Pi} + \overline{\mZ}^2\overline{\Pi}        \mbox{ } , \mbox{ } \mbox{ } 
\overline{\Pi} + \mZ^2\Pi                      \mbox{ } ,
\eeq
i.e., respectively, what ${\cal I}_3$ and ${\cal Y}$, ${\cal U}_+$ and ${\cal V}_+$, and ${\cal U}_-$ and ${\cal V}_-$ pairwise collapse to; 
${\cal I}_{\pm}$ cease to exist at all.  
Quantities proportional to these generators are then a ${\cal J}_3$ and ${\cal J}_{\pm}$ for the triangle.  

\mbox{ } 

\noindent Finally, the three $SU(2)$ ladders correspond to the three triangles (or two triangles and a rhombus) of coarse-graining in Fig \ref{FigX}.
[Each of these, of course, is associated with a coarse-grained shape space sphere whose isometry group is the corresponding $SO(3)$.]
Furthermore, each ladder is paired with an `extra charge' to form three overlapping embedded $SU(2) \times U(1)$ 's.  
${\cal I}_1$, ${\cal I}_2$, ${\cal I}_3$, ${\cal Y}$ is one of the embedded $SU(2) \times U(1)$ groups within the $SU(3)$, the others 
being the ${\cal U}$ and ${\cal V}$ counterparts of this.

\subsection{Generators of the isometries in terms of Gibbons--Pope type momenta}

\beq
\mbox{As well as} \hspace{1in}
{\cal Y} = 2p_{\psi} = 2{\cal J}_{\psi} = - 2\{ p_{\Phi_1} + p_{\Phi_2}  \} \mbox{ } \mbox{ } , \mbox{ }  \mbox{ } 
{\cal I}_3 = {p}_{\phi} = {\cal J}_{\phi} = p_{\Phi_2} - p_{\Phi_1}\mbox{ } , \hspace{3in}  
\eeq
\beq
\mbox{one has} \hspace{1in}
{\cal I}_1 = 
- \mbox{sin}\,\phi\, p_{\beta} + \frac{\mbox{cos}\,\phi}{\mbox{sin}\,\beta}\{p_{\psi} - \mbox{cos}\,\beta\,p_{\phi}\} = 
- \mbox{sin}\,\phi\, {\cal D}_{\beta} + \frac{\mbox{cos}\,\phi}{\mbox{sin}\,\beta}\{{\cal J}_{\psi} - \mbox{cos}\,\beta\,{\cal J}_{\phi}\}                                                       \mbox{ } , \mbox{ } \hspace{3in}  
\eeq
\beq
{\cal I}_2 = 
\mbox{cos}\,\phi\,p_{\beta} + \frac{\mbox{sin}\,\phi}{\mbox{sin}\,\beta}\{p_{\psi} - \mbox{cos}\,\beta\,{p}_{\phi}\} =
\mbox{cos}\,\phi\,{\cal D}_{\beta} + \frac{\mbox{sin}\,\phi}{\mbox{sin}\,\beta}\{{\cal J}_{\psi} - \mbox{cos}\,\beta\,{\cal J}_{\phi}\} \mbox{ } .  
\eeq 
\beq
\mbox{Finally, } \hspace{0.3in} 
{\cal I}_{\sT\so\st} := {\cal I}^2 = 
p_{\beta}\mbox{}^2 + \frac{1}{\mbox{sin}^2\beta}
\{p_{\phi}\mbox{}^2 - 2\mbox{cos}\,\beta\, p_{\psi}{p}_{\phi} + p_{\psi}\mbox{}^2\} = 
{\cal D}_{\beta}^2 + \frac{1}{\mbox{sin}^2\beta}\{{\cal J}_{\phi}^2 - 2\mbox{cos}\,\beta\,{\cal J}_{\phi}{\cal J}_{\psi} + {\cal J}_{\psi}^2\} 
\mbox{ } .  \hspace{2in}  
\eeq
Thus, whether for H's or for K's there is also a pair of coordinates $\beta$ and $\chi$: additionally dependent on only one corresponding ratio of relative separations, 
i.e. the ${\cal I}_1$ and ${\cal I}_2$ depend on $\beta$ alone rather than on $\chi$.  
These are conjugate to quantities that involve relative distance momenta in addition to relative angular momenta.

\noindent Note that the other expressions (${\cal U}_{\pm}, {\cal V}_{\pm}$) are much more complex and less insightful in these particular 
${\cal I}$-adapted Gibbons--Pope type coordinates.
Of course, ${\cal U}$ and ${\cal V}$ adapted Gibbons--Pope type coordinates exist as well, via omitting in each case a different choice of Jacobi vector. 
E.g. in the Jacobi H case, ${\cal I}$ is tied to the collapse to the rhombus, with ${\cal U}$ and ${\cal V}$ corresponding to the two one-post collapse triangles.
In  terms of each of these coordinate systems, the corresponding sets of picked-out $SU(2) \times U(1)$ 
quantities (i.e. \{${\cal U}_{\pm}, {\cal U}_3, {\cal Y}_{\cal U}, {\cal U}_{\sT\so\st}$\} and  
                 \{${\cal V}_{\pm}, {\cal V}_3, {\cal Y}_{\cal V}, {\cal V}_{\sT\so\st}$\} have the same expressions as above 
(with ${\cal U}$, ${\cal V}$ labels, respectively, understood but dropped on the Gibbons--Pope type coordinates in use.)

\subsection{Interpretation: the collapse of the above to the usual $SU(2)$ operators for the ``${\cal I}$" coarse-graining}

For $\rho_3 = 0$, i.e. $\chi = \pi/2$, one recovers the usual ${\cal I}_{\Gamma}$, $\Gamma = 1$ to $3$, of $SU(2)$ with $\beta$ playing the role of $\theta$. 
\beq
{\cal I}_1 = - \mbox{sin}\,\Phi\, p_{\Theta} - \mbox{cos}\,\Phi\,\mbox{cot}\,\Theta\, p_{\Phi}   =  {\cal S}_2
\mbox{ } , \mbox{ } \mbox{ }
{\cal I}_2 =   \mbox{cos}\,\Phi\, p_{\Theta} - \mbox{sin}\,\Phi\,\mbox{cot}\,\Theta\, p_{\Phi}   =  {\cal S}_1
\mbox{ } , \mbox{ } \mbox{ }
{\cal I}_3 = p_{\Phi} = {\cal S}_3 \mbox{ } .  
\eeq
N.B. that this case is not a triangle; it is the rhombic coarse-graining of Fig \ref{FigX}g).

\subsection{Interpretation: quadrilateralland isometry generators}

I mention the parallel with ${\cal S}_3 = {\cal J} = p_{\Phi}$ pure relative angular momentum in in triangleland whilst ${\cal S}_1$ and ${\cal S}_2$ 
are mixtures of relative angular momentum and relative dilational momentum.   
There is a looser parallel with ${\cal D} = p_{\phi}$ in 4-stop metroland which has, however, a different meaning.   

\noindent
In H 
coordinates, the meaning of ${\cal I}_1$ and ${\cal I}_2$ coordinates is that of mixed relative angular momentum and relative  
dilation of the $\beta$ type. 
I.e. a rate of  change in the contents inhomogeneity (the ratio of the sizes of the two constituent subclusters).  
The meaning of ${\cal I}_{\sT\so\st}$ is the total angular momentum of the third, rhombic, coarse-graining triangle of the H in Fig \ref{FigX}g).  
In each case, changing which ratios one regards as primary gives similar presentations for the ${\cal U}$'s and ${\cal V}'$s.  

\noindent In K coordinates, the meaning of ${\cal I}_1$ and ${\cal I}_2$  is that of mixed relative angular momentum and relative dilation of the $\beta$ type. 
I.e. a rate of  change in the ratio of the face of the blade to the thickness of the blade: a sharpness/flatness shape quantity for the obvious triangle subsystem.         
The meaning of ${\cal I}_{\sT\so\st}$ is the total angular momentum of the second coarse-graining triangle of K in Fig \ref{FigX}.

\noindent In each case, changing which ratios one regards as primary gives similar presentations for the ${\cal U}$'s and ${\cal V}'$s.

\section{Classical PoT applications of quadrilateralland model}\label{Cl-PoT-Q} 

\subsection{Configurational Relationalism resolved by Best Matching for $N$-a-gonland}  

\underline{Key 10} Obtaining Sec 3's r-formulation down the reduced path in Sec 3 precisely amounts to this resolution of this PoT facet.

\subsection{Emergent Jacobi--Barbour--Bertotti time} 

This (\underline{Key 11}) is based on an ephemeris time type procedure, as inspired by the course of action in the practical modelling of the Solar System from the 
1890's through to 1970's period.
What was originally thought to be anomalous motion of the Moon was explained instead by e.g. Willem de Sitter \cite{deSitter} to be a fault in the assumption that the time read off by the 
rotation of the Earth was the same as Newton's absolute time to the requisite accuracy to explain the lunar observations in question. 
It was eventually decided \cite{Clemence} to read time off the motions of larger sets of celestial bodies; while cumbersome for everyday use, this can be tabulated 
and other types of clock can be kept calibrated with it by occasionally carrying out checks.
Whilst the astronomers involved in this `ephemeris time' program were in it for the practicalities, it has since been noted that their work implements `Mach's time principle' 
\cite{B94I, ARel2}, which I furthermore implement as `GLET is to be abstracted from a STLRC' (generalized local ephemeris time). 
This paper has a classical generalization of this chroniferous modelling and the next one has a semiclassical counterpart -- 
applying these insights of de Sitter and Astronomer Gerald Clemence \cite{Clemence} to Quantum Cosmology.

\noindent \underline{Key 12} For RPM models in 1- and 2-$d$, one can pass from the indirectly-formulated version of (\ref{G-tem}) to the below directly-formulated expressions 
due to the Best Matching Problem being resolved for these examples.  
In the pure-shape case, it is 
\beq
\lt^{\se\sm(\sJ\sB\sB)} = \int {\d s_{\sF\sS}(\bz)}\left/{\sqrt{2\{E - V(\bz)\}}}\right. = 
                          \int {\d s_{\sF\sS}(\chi, \beta, \phi, \psi)}\left/{\sqrt{2\{E - V(\chi, \beta, \phi, \psi)\}}}\right.  \mbox{ } .
\eeq
In the scaled case, it is  
\beq
\lt^{\se\sm(\sJ\sB\sB)} = \int {\sqrt{\d\rho^2 + \rho^2\d s_{\sF\sS}^2(\bz)}}\left/{\sqrt{2\{E - V(\rho, \bz)\}}}\right. = 
                          \int {\sqrt{\d\rho^2 + \rho^2\d s_{\sF\sS}^2(\chi, \beta, \phi, \psi)}}\left/{\sqrt{2\{E - V(\rho, \chi, \beta, \phi, \psi)\}}}\right.  \mbox{ } .
\label{Ricic}
\eeq
\mbox{ } \mbox{ } For the scaled case of most cosmological interest, the h-approximand is 
\beq
\lt^{\se\sm(\sJ\sB\sB)}_{(0)} = \int  {\d\rho}\left/{\sqrt{2\{E - V_{\rho}(\rho)\}}} \right. \mbox{ } .
\eeq
This is not a GLET from STLRC form of Mach's Time Principle because it does not give the l-changes an opportunity to contribute.  
Classical GLET implementations are, rather of the form 
\beq
\lt^{\se\sm(\sJ\sB\sB)}_{(1)} = {\cal F}[h, l, \d h, \d l] \mbox{ } .  
\label{callie}
\eeq
The first approximation that is satisfactory in this manner is 
\beq
\lt^{\se\sm(\sJ\sB\sB)}_{(1)} = \lt^{\se\sm(\sJ\sB\sB)}_{(0)} + \frac{1}{2\sqrt{2}}
\int\frac{J_{\rho S}\d \rho}{W_{\rho}^{3/2}} + 
\frac{1}{4}\int\frac{\d\rho}{\sqrt{W_{\rho}}} 
\left\{
\frac{\d S }{\d \, \mbox{ln} \,\rho}
\right\}^2 
+ O\left(\left\{\frac{J_{\rho S}}{W_{\rho}}\right\}^2\right) + O\left(\left\{\frac{\d S}{\d \,\ml\mn\,\rho}\right\}^4\right) 
\label{Cl-Expansion}
\eeq
$J_{\rho S}/W_{\rho}$ is a dimensionless measure of interaction strength.  
$\d \mS/\d \mbox{ln}\,\rho$ small is the `scale dominates shape' approximation \cite{Cones, FileR}, which is appropriate in cosmological modelling, where inhomogeneous dynamics effects 
are 5 orders of magnitude smaller than scalefactor dynamics.

\subsection{Classical \K beables for quadrilateralland}

\noindent Now, shape variables and shape momenta have the additional interpretation as regards \K beables.
The shape variables lucidly correspond to/are centred about geometrically significant configurations and their momenta lucidly correspond to changes of these acquires further significance.  
One gets the sense that these are practically interesting beables and, at least sometimes, correspond to localized clusters.

Now, for the pure-shape case, classical Dirac beables B = D = F[$\bQ, \bP$]  obey the Poisson brackets 
\beq
\{{\cal E}, \mB\}       =  0 \mbox{ } ,
\label{OE0}  
\eeq
\beq
\{{\cal L}_{\mu}, \mB\} = 0 \mbox{ } ,                     
\label{OLi0}
\eeq
\beq
\{{\cal D}, \mB\} = 0 \mbox{ } .  
\label{OD0}
\eeq
For the scaled case, (\ref{OE0}, \ref{OLi0}) hold.
Also classical \K beables B = K = F[$\bQ, \bP$] solve (\ref{OLi0}) for the scaled case, and (\ref{OLi0}, \ref{OD0}) for the pure-shape case.
Classical \K beables are a resolved issue in 1- and 2-$d$ RPM's because the Best Matching Problem is solved for these \cite{FORD, Cones, FileR}.  
For the pure-shape case, these are the set of all functionals of the shape variables and the shape momenta, $\mK = \mF[\bS, \bP_{\sS}]$.
For the scaled case, these are the set of all functionals of the scale and shape variables and the scale and shape momenta, $\mK = \mF[\sigma, \bS, \mP_{\sigma}, \bP_{\sS}]$.
The quantum counterpart of the above then `straightforwardly' involves some operator form for the canonical variables and commutators in place of Poisson brackets.

Note that here the Best Matching Problem is solved for 1- and 2-$d$ RPM's, whether pure-shape or scaled by results summarized in Sec \ref{Q-1}.  
And we have been able to straightforwardly construct and interpret a resolution of the Problem of Beables in the sense of \K.  
The corresponding Kucha\v{r} beables are those quantities whose brackets with the linear constraints vanish.  
This occurs in pure-shape RPM for precisely the set of all functions of the shape variables and the shape momenta. 
Likewise, the set of  Kucha\v{r} beables for pure-shape RPM is precisely the set of all functions of the scale and shape variables and the scale and shape momenta.

As per Sec \ref{Q-1}, the 1-$d$ pure-shape RPM r-configuration spaces are \cite{06II} $\mathbb{S}^{N - 2}$ and suitable shape variables thereupon are the (ultra)spherical angles \cite{AF}, 
interpreted as functions of ratios of relative separations, and the corresponding shape momenta are as per Sec \ref{Mom-Int}.
The same Secs give that 2-$d$ pure-shape RPM r-configuration spaces are $\mathbb{CP}^{N - 2}$ and suitable shape variables fore these are the inhomogenous coordinates ${\mZ}^{\barr}$.  
To interpret these complex coordinates in terms of the $N$-a-gons, it is useful to pass to their polar forms, ${\mZ}^{\barr} = {\cal R}^{\barr}\mbox{exp}(i\Phi^{\barr})$.  
Then the moduli are, again, ratios of relative separations, and the phases are now relative angles.  
In the specific case of the scalefree triangle, there is one of each, e.g. in coordinates based around the \{1,23\} clustering, these are 
\cite{TriCl} $\Theta = 2\,\mbox{arctan}(\rho_2/\rho_1)$ and $\mbox{arccos}\big( \brho_1 \cdot \brho_3 / \rho_1 \rho_3 \big)$.
The shape momenta for the $N$-a-gon are \cite{FileR} 
\beq
{\cal P}^{{\cal R}}_{\barp} = 
\left\{
\frac{\delta_{\barp\barq}}{1 + ||{\cal R}||^2}   - 
\frac{{\cal R}_{\barp}{\cal R}_{\barq}}{\{1 + ||{\cal R}||^2\}^2}
\right\}
{\cal R}_{\barq}^{*}
\mbox{ } \mbox{ } , \mbox{ } \mbox{ }  
{\cal P}^{\Theta}_{\widetilde{\sp}} = 
\left\{
\frac{\delta_{\overline{\sp}\overline{\sq}}}{1 + ||{\cal R}||^2} - 
\frac{{\cal R}_{\overline{\sp}}{\cal R}_{\overline{\sq}}}{\{1 + ||{\cal R}||^2\}^2}
\right\}
{\cal R}_{\overline{\sp}}{\cal R}_{\overline{\sq}}\Theta_{\widetilde{\sp}}^{*} \mbox{ } .  
\eeq
\noindent I gave triangleland in \cite{AHall} as a specific 2-$d$ example. 
This casts the mathematics of the example in \cite{H03} into a whole-universe, 
nontrivially linearly constrained context; in the present paper I give the quadrilateralland case as a larger and mathematically new example.  
\K beables for this problem are, in Gibbons--Pope type coordinates and using Secs \ref{Q-1} and \ref{Q-2} for shapes and Sec \ref{Mom-Int} for shape momenta, functionals of the form 
\beq
\mbox{K = F[$\chi, \beta, \phi, \psi, \pi_{\chi}, \pi_{\beta}, \pi_{\phi}, \pi_{\psi}$ alone] } .   
\eeq
For the corresponding scaled case, these are functionals of the form 
\beq
\mbox{K = F[$\rho$, $\chi, \beta, \phi, \psi, \pi_{\rho}, \pi_{\chi}, \pi_{\beta}, \pi_{\phi}, \pi_{\psi}$ alone] } . 
\eeq
\noindent These all make for geometrically (in space) meaningful propositions and some are sometimes locally determinable/locally beable.  
\noindent Actual propositions involve approximate values of quantities, and this then rests on configuration space regions (Sec \ref{Eregion}).  

\noindent As regards the use of conserved quantities in preference to/alongside the momenta (see Sec II.3 for more), 

\noindent 1) these, or functions thereof, commute also with the Hamiltonian constraint and are thus Dirac beables.  
They manage to be this way via not encountering an obstruction from the potential term in \{B, ${\cal E}$\}.    

\noindent 2) They feature in the kinematical quantization procedure, making them even more natural at the quantum level.  
For the sphere, these are the $SU(2)$ quantities ${\cal S}_{\sfA}$; for the quadrilateral, these are the $SU(3)$ quantities, especially the 
${\cal I}_{{\cal A}}$ and ${\cal Y}$ that remain conserved quantities for a wider range of potentials.
Here also e.g. for the sphere, $\Phi$ and $\Theta$ do not become good operators at the quantum level, it is the unit Cartesian vectors that do.  

\mbox{ } 

\noindent Halliwell's approach \cite{H03, H09-1, H09-2} allows for this construct to be uplifted \cite{AHall} to produce classical Dirac beables.

\mbox{ }

\noindent The formal counterpart for GR-as-geometrodynamics is the classical \K beables are, formally, coordinate-patch-wise defined functionals of the 3-geometries ${\capFrG}$ 
and their conjugates alone, 
\beq
\mK = \mF[{\capFrG}, \Pi^{\scapFrG} \mbox{ alone}] \mbox{ } .  
\eeq
`Formal' here refers to us not having an equivalent for the 3-geometries of the concrete Dragt coordinates for triangleland's \K beables).

\subsection{Three more PoT facets}

\underline{Key 13} The $N$-a-gonland r-presentation has only the one constraint --  the r-formulation's version of the quadratic energy constraint, ${\cal E}^{\sr}$, 
so there is no classical Constraint Closure Problem.
RPM indirectly formulated has in general a point-identification map, but no further GR-like spacetime structure. 
The r-presentation has the point-identification map already built into it.
Finally, RPM's have no foliation-dependence or further spacetime reconstruction issues in it by their nature as simple models.

\section{Equations of motion and conserved quantities for various potentials}\label{dyn}

\subsection{$N$-stop metrolands and triangleland}

\noindent For triangleland \cite{TriCl, 08I, 08III} [and suppressing (a)-labels]: $\Phi$-independence in the potential corresponds to there being no means for angular momentum to 
be exchanged between the subsystem composed of particles 2, 3 and that composed of particle 1.  
This corresponds to having an $SO(2)$ invariance (`special case')
If the potential is additionally $\Theta$-independent and so constant, one has the full $SO(3)$ invariance (`very special case') 
For $N$-stop metroland \cite{AF, ScaleQM}, there is likewise a sequence of special, very special, ... very$^{Y - 2}$ special potentials corresponding to $SO(2)$, $SO(3)$, ... $SO(Y)$ .
These above observations are useful as regards finding a nice range of analytic solutions of increasing complexity \cite{08I, 08II, +tri, AF, ScaleQM, 08III}. 
Moreover, e.g. for triangleland, there are in fact three particular $SO(2)$'s, corresponding to the three $\Phi$'s defined relative to the three DM axes present, 
albeit only one of these can be realized in any given model.
I will next consider the quadrilateralland counterparts of these statements.

\subsection{Equations of motion for quadrilateralland in Gibbons--Pope type coordinates}

The $\psi$-, $\phi$-, $\beta$- and $\chi$-equations are, respectively, 
\beq
\{\mbox{sin}^2\chi\,\mbox{cos}^2\chi\{\psi^* + \mbox{cos}\,\beta\,\phi^*\}/4\}^* = - {\pa \fV}/{\pa \psi}
\mbox{ } ,
\eeq
\beq
\{\mbox{sin}^2\chi\{\mbox{cos}^2\chi\{\phi^* + \mbox{cos}\,\beta\,\psi^*\} +  \mbox{sin}^2\chi\,\mbox{sin}^2\beta\,\phi^*\}/4\}^* 
= - {\pa \fV}/{\pa \phi}
\mbox{ } , 
\eeq
\beq
\{\mbox{sin}^2\chi\,\beta^*/4\}^* = 
\mbox{sin}^2\chi\,\mbox{sin}\,\beta\{\mbox{sin}^2\chi\mbox{cos}\,\beta\,\phi^* - \mbox{cos}^2\chi\,\psi^*\}\phi^*/4  - {\pa \fV}/{\pa \beta}
\mbox{ } , \mbox{ } \mbox{ and } 
\eeq
\beq
\chi^{**} = \mbox{sin}\,\chi\,\mbox{cos}\,\chi
\{\beta^{*\,2} + \mbox{cos}\,2\chi\,\{\phi^{*\,2} + \psi^{*\,2} + 2\,\phi^*\,\psi^*\mbox{cos}\,\beta\} + 2\mbox{sin}^2\chi\,\mbox{sin}^2\beta\,\phi^{*\,2} \}/4  - {\pa \fV}/{\pa \chi}
\mbox{ } .
\eeq
One of these can be supplanted by the energy first-integral,
\beq
{\chi^{* \, 2}}/{2} + {\mbox{sin}^2\chi}
\{\beta^{* \, 2} + \mbox{cos}^2\chi\{\phi^{*\,2} + \psi^{*\,2} + 2\,\phi^*\psi^*\mbox{cos}\,\beta\} + \mbox{sin}^2\chi\,\mbox{sin}^2\beta\,\phi^{*\, 2}\}\}/8 + \fV = \fE
\mbox{ } .
\eeq
(See \cite{AF} for the 4-stop metroland equations of motion, and \cite{08I} for the triangleland ones.)

\subsection{Which potentials realize which subgroups?}  

\noindent i)  For $\fV$ explicitly dependent on all of $\chi, \beta, \phi, \psi$, no isometry generator survives as a conserved quantity.  

\noindent ii) For $\fV$ $\psi$-independent, $\psi$ is a cyclic coordinate and yields one constant of the motion, 
\beq
\mbox{sin}^2\chi\,\mbox{cos}^2\chi\{\psi^* + \mbox{cos}\,\beta\,\phi^*\} =  C \mbox{ } .
\eeq
This corresponds to a $U(1)$ symmetry.  
I identify this constant $C$ as taking the value $2\,{\cal Y}$.

\noindent
iii) For $\fV$ $\phi$-independent, $\phi$ is a cyclic coordinate and yields another constant of the motion, 
\beq
\mbox{sin}^2\chi\{\mbox{cos}^2\chi\{\phi^* + \mbox{cos}\,\beta\,\psi^*\} +  \mbox{sin}^2\chi\,\mbox{sin}^2\beta\,\phi^*\} = K \mbox{ } .  
\eeq
This also corresponds to a $U(1)$ symmetry.  
I identify this constant $K$ as taking the value $4\,{\cal I}_3$.

\noindent 
iv) Potentials independent of both $\psi$ and $\phi$ yield both of these at once, corresponding to a $U(1) \times U(1)$ symmetry.  

\noindent 
Note that all of the symmetries considered so far may be viewed as phase factors in the complex representation, by which their $U(1)$ nature is rendered manifest.

\noindent v) Potentials independent of both $\beta$ and $\phi$ yield three conserved quantities, corresponding to $SU(2)$ symmetry.  

\noindent vi) Potentials independent of all of $\beta$, $\phi$ and $\psi$ yield four, corresponding to $SU(2) \times U(1)$ symmetry.  

\noindent vii) If the potential is constant, one has all eight conserved quantities corresponding to the full $SU(3)$ isometry group. 

\noindent There are also $U$ and $V$ counterparts of all of the above, so that there are three versions of all the partial symmetries.  

{            \begin{figure}[ht]
\centering
\includegraphics[width=0.55\textwidth]{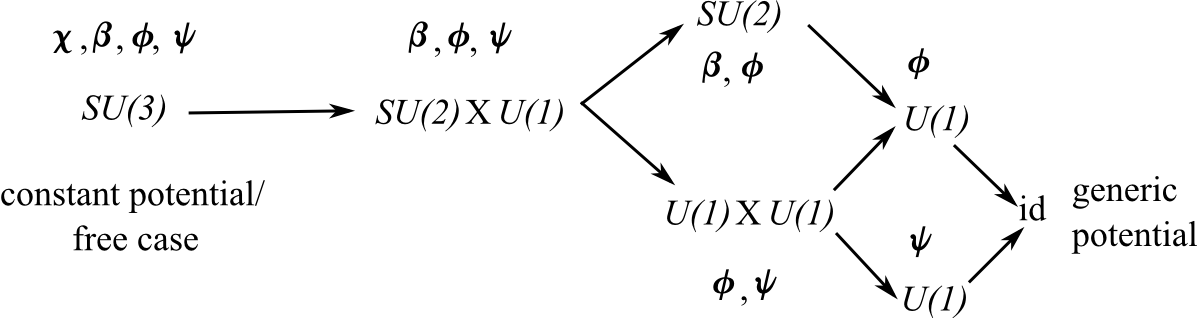}
\caption[Text der im Bilderverzeichnis auftaucht]{        \footnotesize{Flow diagram for the breakdown of the $SU(3)$ 
symmetry group due to various different potentials.  
The coordinates listed are those required to be absent from the potential. 
In fact, there are 3 different $SU(2) \times U(1)$ possibilities, each adapted to one of ${\cal I}$, ${\cal V}$ and ${\cal U}$.} }
\label{Fig1-II} \end{figure}          }

\subsection{Hamiltonians with conserved quantities substituted in}

For 3-stop metroland with $\varphi$-independent potential, $\fH = {\cal D}_{\sT\so\st} + \fV$, constant. 
For 4-stop metroland, with $\phi$-independent potentials, 

\noindent
\beq
\fH = {p_{\theta}^2}/{2} + {{\cal D}_{\phi}^2}/{2\mbox{sin}^2{\theta}} + \fV(\theta) \mbox{ } .
\eeq
For triangleland, with $\Phi$-independent potentials [$U(1)$-symmetric case], one has \cite{TriCl, 08I}
\beq
\fH = {p_{\Theta}^2}/{2} + {{\cal J}^2}/{2\mbox{sin}^2{\Theta}} + \fV(\Theta) \mbox{ } .
\eeq
Each of these two has a constant case for the full $SO(3)$ symmetry. 
The question then is what is the quadrilateralland counterpart of these simplified (partly) symmetric cases.

One can now use the conserved quantity equations to write down Hamiltonians with more constants of the motion inside instead of momenta.  
If there is a $U(1)$ symmetry of the $\psi$ type, 
\beq
\fH = \frac{p_{\chi}^2}{2} + \frac{2}{\mbox{sin}^2\chi}\left\{p_{\beta}^2 + \frac{1}{\mbox{sin}^2\beta}
\left\{p_{\phi}^2 + \frac{{\cal Y}^2}{4} - p_{\phi}{\cal Y}\,\mbox{cos}\,\beta\right\}\right\} + \frac{{\cal Y}^2}{2\mbox{cos}^2\chi}  + \fV(\chi, \beta, \phi) 
\mbox{ } , \mbox{ } \mbox{ ${\cal Y}$ constant } .
\eeq
If there is a $U(1)$ symmetry of the $\phi$ type, 
\beq
\fH = \frac{p_{\chi}^2}{2} + \frac{2}{\mbox{sin}^2\chi}\left\{p_{\beta}^2 + \frac{1}{\mbox{sin}^2\beta}
\{{\cal I}_{3}^2 + p_{\psi}^2 - 2 \, {\cal I}_{3}p_{\psi}\mbox{cos}\,\beta\}\right\} + \frac{2p_{\psi}}{2\mbox{cos}^2\chi}  + \fV(\chi, \beta, \psi) 
\mbox{ } , \mbox{ } \mbox{ ${\cal I}_3$ constant } .
\eeq
If there is a $U(1) \times U(1)$ symmetry, 
\beq
\fH = \frac{p_{\chi}^2}{2} + \frac{2}{\mbox{sin}^2\chi}\left\{p_{\beta}^2 + \frac{1}{\mbox{sin}^2\beta}
\left\{{\cal I}_{3}^2 + \frac{{\cal Y}^2}{4} - {\cal I}_{3}{\cal Y}\,\mbox{cos}\,\beta \right\}\right\} + \frac{{\cal Y}^2}{2\mbox{cos}^2\chi} 
+ \fV(\chi, \beta) \mbox{ } , \mbox{ } \mbox{ ${\cal I}_3$, ${\cal Y}$ constant } .
\eeq
If there is a $SU(2)$ symmetry, 
\beq
\fH = {p_{\chi}^2}/{2} + {2{\cal I}^2}/{\mbox{sin}^2\chi}  + {2p_{\psi}^2}/{\mbox{cos}^2\chi}  + \fV(\chi, \psi) 
\mbox{ } , \mbox{ } \mbox{ ${\cal I}$ constant } .
\eeq
If there is an $SU(2) \times U(1)$ symmetry, 
\beq
\fH = {p_{\chi}^2}/{2} + {2{\cal I}^2}/{\mbox{sin}^2\chi} + {{\cal Y}^2}/{2\mbox{cos}^2\chi}  + \fV(\chi) 
\mbox{ } , \mbox{ } \mbox{ ${\cal I}$, ${\cal Y}$ constant } . 
\eeq
The $SU(3)$-symmetric case has $\fH$ constant.   

The simplest nontrivial case of $SU(2) \times U(1)$ symmetry can be straightforwardly represented as (via $u = \mbox{cos}^2\chi$)      
\beq
\lt^{\se\sm(\sJ\sB\sB)} = - \frac{1}{2}\int {\md u}\left/{\sqrt{- {\cal Y}^2 + \{{\cal Y}^2 + 2\fW(u) - 4{\cal I}^2\}u - 2\fW(u) u^2}}\right. \mbox{ } .
\label{uint}
\eeq

\section{\underline{Key 14}: HO-type potentials}\label{HO-PoT}

These are motivated as the perturbations in the parallel with Halliwell--Hawking inhomogeneous quantum cosmology in Paper II.  
As explained in e.g. \cite{FileR} these are not HO's per se in the pure-shape case, since they have to be homogeneous of degree zero in order to be consistent.
This is attained by dividing the usual HO expression for the potential by the moment of inertia of the system (which subsequently turns out to be a constant).   
The scaled case has the usual HO potentials.  
For 3-stop metroland \cite{FileR},
$V = K_1 \mn_1^2/2 + K_2 \mn_2^2/2 + L \mn_1 \mn_2 = A + B\,\mbox{cos}\,2\varphi + C\mbox{sin}\,2\varphi$ for 
$A = \{K_1 + K_2\}/2 \mbox{ } , \mbox{ } \mbox{ } B = \{K_2 - K_1\}/2 \mbox{ } , \mbox{ } \mbox{ } C = L/2$.
There is a special case with $SO(2) = U(1)$ symmetry, for $B = 0 = C$ i.e. $L = 0$ and $K_1 = K_2$, so that cluster 1 and cluster 2 have the same `constitution'. 
I.e. the same Jacobi--Hooke coefficient per Jacobi cluster mass. 
Here the `constituent springs'\!' potential contributions balance out to produce the constant potential, $\fV = A$. 
This is a kind of `homogeneity requirement' on the `structure' of the model universe.
For 4-stop metroland \cite{AF, ScaleQM},   
\beq
\fV = \sum\mbox{}_{a = 1}^3\{K_{a}\mn^{a\, 2}/2 + L_{a}\mn^{b}\mn^{c}\}
    = \ttA + \ttB\,\mbox{cos}\,2\theta + \ttC\,\mbox{sin}^2\,\theta\,\mbox{cos}\,2\phi + 
\ttD\,\mbox{sin}^2\theta\,\mbox{sin}\,2\phi + \ttE\,\mbox{sin}\,2\theta\,\mbox{cos}\,\phi + \ttF\,\mbox{sin}\,2\theta\,\mbox{sin}\,\phi 
\label{Bro}
\eeq 
\beq
\mbox{ for } \mbox{ } \ttA = \frac{1}{4}\left\{K_3 + \frac{K_1 + K_2}{2}\right\} \mbox{ } , \mbox{ } 
\ttB = \frac{1}{4}\left\{K_3 - \frac{K_1 + K_2}{2}\right\} \mbox{ } , \mbox{ } 
\ttC = \frac{K_1 - K_2}{4} \mbox{ } .  
\eeq
This has a special case with $U(1)$ symmetry, for $L_a = 0$ and $K_1 = K_2$ [i.e. $\ttC, \ttD, \ttE, \ttF = 0$],
$\fV = \ttA + \ttB\, \mbox{cos}\,2\theta$.  
It also has a very special case with $SO(3)$ symmetry, for $\ttB, \ttC, \ttD, \ttE, \ttF = 0$, i.e. $L_a = 0$ and $K_1 = K_2 = K_3$, for which 
high-symmetry situation the various potential contributions balance out to produce the constant, $\fV = \ttA$.  
\beq
\mbox{ } \mbox{ } \mbox{For triangleland \cite{+tri, 08III}, } \hspace{0.2in} 
\fV = K_1\mn_1\mbox{}^2/2 + K_2\mn_2\mbox{}^2/2 + L \bn_1 \cdot \bn_2 = 
      A + B\,\mbox{cos}\,\Theta + C\,\mbox{sin}\,\Theta\,\mbox{cos}\,\Phi \mbox{ } . \hspace{3in}
\eeq 
This has a special case with $U(1)$ symmetry, for $L = 0$ (i.e. $C = 0$), $\fV =  A + B\,\mbox{cos}\,\Theta$.  
It also has a very special case with $SO(3)$ symmetry, for $L = 0$, $K_1 = K_2$, i.e. $B, C = 0$, $\fV =  A$.  

\noindent Then for quadrilateralland, a parametrization of the HO-type potential at level of Jacobi vectors is
\beq
\fV =  \{K_1\mn_1^2 + K_2\mn_2^2 + K_3\mn_3^2\}/2 + L_1\bn_2 \cdot \bn_3 + L_2\bn_3 \cdot \bn_1 + L_3\bn_1 \cdot \bn_2 \mbox{ } .  
\eeq
The Kuiper coordinates are {\sl very} HO-adapted:  
\beq
\fV = \{K_1N_1 + K_2N_2 + K_3N_3 + L_1\mbox{aniso}(23) + L_2\mbox{aniso}(31) + L_3\mbox{aniso}(12)\}/2 \mbox{ } .  
\eeq
However, redundancy and non-adaptation of the kinetic term limit the usefulness of this expression.  
\noindent I use $\omega = \sqrt{K} = \sqrt{2A}$.  


The first three terms of this in Gibbons--Pope type coordinates form the combination as the first three terms of (\ref{Bro}). 
This is because it involves solely the real parts for which the 2-$d$ 4 particle problem reduces to the 1-$d$ 4 particle one (in fact the mirror image identified version of this).
For the other 3 cross-terms, however, the analysis has specific 2-$d$ character in contrast to (\ref{Bro})'s 1-$d$ character. 
These are, for H 
coordinates, and using eqs (\ref{n1n2}-\ref{n3n1}),
\beq
\fV =  
\left.
\left\{
L_3\mbox{sin}\,\beta\, \mbox{sin}^2\chi \mbox{cos}\,f_3 +  
L_1 \mbox{sin}\mbox{$\frac{\beta}{2}$} \mbox{sin}\,2\chi\, \mbox{cos}\,f_1 +  
L_2 \mbox{cos}\mbox{$\frac{\beta}{2}$} \mbox{sin}\,2\chi\, \mbox{cos}\,f_2
\right\}
\right/
2   \mbox{ }  .
\eeq
On the other hand, for K
%
%
coordinates, one has the $2  \leftrightarrow 3$ of the above, and likewise by Sec \ref{Q-2}'s transpositions argument for the other choices 
of tree and of ratios.  
\noindent As regards HO-like potentials possessing particular symmetries, see Fig \ref{FlowD2}.
%
{            \begin{figure}[ht]
\centering
\includegraphics[width=0.95\textwidth]{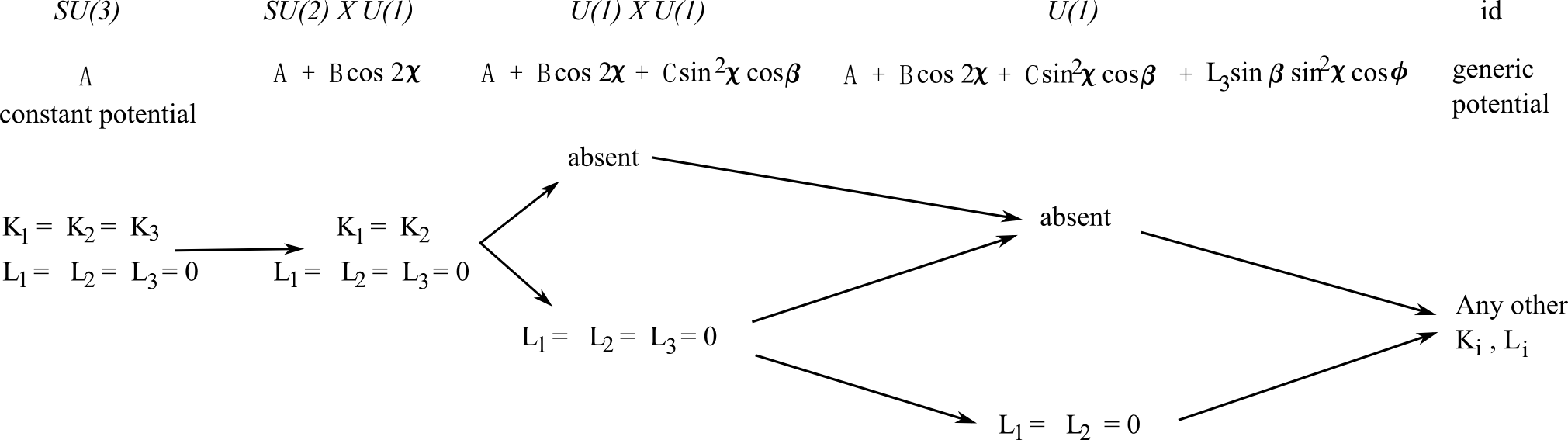}
\caption[Text der im Bilderverzeichnis auftaucht]{        \footnotesize{Flow diagram of the breakdown of the kinetic term's $SU(3)$ 
symmetry group due to various different HO potentials.
In the $SU(2) \times U(1)$ symmetric: if realized alongside at most the first term, it is the partial balance $K_1 = K_2$ of particular 
significance in the H-case as universe contents homogeneity, and in which case $\ttB = \{K_3 - K_1\}/4$. } }
\label{FlowD2} \end{figure}          }

\section{Classical solutions for quadrilateralland}\label{cl-soln}

\subsection{Geodesics of $\mathbb{CP}^{k}$ and their $N$-a-gonland interpretation}

For the general $\mathbb{CP}^k$ \cite{W82a}, geodesics through the origin are particularly simply expressed in complex form,  
\beq
Z^I = \tau C^I
\label{Warner}
\eeq
for $\tau$ a parameter and $C^I$ a constant vector.

\subsection{Triangleland geodesics}

This is new in the context of this program (insofar as \cite{08I} treated triangleland's $\mathbb{S}^2$ as a real manifold).    
Eliminating $\tau$ from (\ref{Warner}) in this case, and passing to the convenient spherical coordinates amounts to $\Phi$ being constant whilst $\Theta$ varies.   
The first of these conditions means that ${\cal J}$ = 0.  
These are the set of meridians with the origin being the D-pole corresponding to the underlying choice of clustering 
and the infinity being the M-pole antipodal to this.  
These are indeed a subset of the great circles that are well-known to be the geodesics in this case, and which 
were interpreted in terms of quadrilaterals in \cite{08I, +tri}.  
A particular such is the meridian of collinearity and another such is the meridian of isoscelesness. 
For later comparison, I furthermore note that these run between the clustering's two notions of collapse: 
$\rho_1 \longrightarrow 0$ and $\rho_2 \longrightarrow 0$.
I.e. from the arbitrarily sharp triangle to the arbitrarily flat one.  
The $N$-stop metroland spheres give (generalized) great circles but with complex formulations essentially absent (only present for $N$ = 3).

\subsection{\underline{Key 15}: Quadrilateralland geodesics}

\noindent In the quadrilateralland case, eliminating $\tau$ from (\ref{Warner}) and passing to the useful Gibbons--Pope coordinates 
amounts to $\psi$, $\phi$ and $\beta$ being constant whilst $\chi$ varies.  
The first three of these conditions imply that ${\cal Y}$ and ${\cal I}_{\sT\so\st}$ are both zero.
These also run between two collapsed cases, although now there is a diversity of such collapses available and of interpretations for these geodesics, according to the choices H or K 
and then of which common denominator to pick in making the subsequent two ratios (the usual five choices of Sec \ref{Q-2}).  
The 0 end is the 2 Jacobi distance collapse case and the $\infty$ end is the 1 Jacobi distance collapse complementary to it.  
Thus, these motions pick out the following.

\noindent For the usual H, this geodesic family corresponds to the posts to crossbar ratio increasing from 
`both posts collapsed to form a DD' of Fig \ref{FigX}.o) to crossbar collapsed to form the rhombus of Fig 8.g). 
[This requires a slight distortion away from the right-angles case displayed in the particular picture of a chopping board provided in Fig \ref{EndGame}.]  

\noindent For the usual K, this geodesic family corresponds to the face and thickness of the blade to handle ratio increasing from
`face and thickness of the blade collapsed to form a T' of Fig 8.q)  to `handle-collapsed particle 3 onto T' triangle of Fig \ref{FigX}.k).

\subsection{Time-traversal formulae}

\noindent Some simple time-traversal cases that have analytical integrals (in terms of the emergent time) are as follows [from (\ref{uint}].

\noindent 1) For constant potential, ${\cal I}_{\sT\so\st\sa\sll}$ and ${\cal Y}$ are both 0, and then 
\beq
\chi = \sqrt{2W}\lt^{\se\sm(\sJ\sB\sB)} \mbox{ } .  
\eeq  
The shapes here are just $\chi$ changes as per above, and the time traversal confirms that these do not turn around, so the $\chi$ runs from one extreme value to the other.

\noindent 2) Formulae such as (precise trig/hyp functions involved depend on the signs of the various coefficients involved) 
\beq
\chi = \mbox{arccos}\left(\sqrt{W -  2{\cal I}_{\sT\so\st\sa\sll}}\,\mbox{sin}\left(-\sqrt{2W}\lt^{\se\sm(\sJ\sB\sB)}\right)\right)
\eeq 
for nonzero ${\cal I}_{\sT\so\st\sa\sll}$ and zero ${\cal Y}$ and still having constant potential.  

\noindent 3) Formulae such as 
\beq
\chi = \mbox{arccos}\left(\sqrt{1 -  2{\cal Y}^2/2W}\mbox{sin}\left(\sqrt{2W}\lt^{\se\sm(\sJ\sB\sB)} \right)\right)
\eeq
for the ${\cal I}_{\sT\so\st\sa\sll} \longrightarrow {\cal Y}$ of the above.
\beq 
4) \hspace{1in} \chi = \mbox{arccos}
\left(
\sqrt{{\left\{\mbox{exp}\left(\sqrt{C}\lt^{\se\sm(\sJ\sB\sB)}\right) - B \right\}
       \left\{\mbox{exp}\left(\sqrt{C}\lt^{\se\sm(\sJ\sB\sB)}\right) - B - 2A\sqrt{C}   \right\}}/{2\sqrt{C}\left\{B + 2 \sqrt{C}\right\}}}
\right)   \hspace{3in} \mbox{ }  
\eeq
for both of these conserved quantities being nonzero, and where $A = - {\cal Y}^2$, $B = 2\fW - 4{\cal I}^2 + {\cal Y}^2$ and $C = - 2\fW$.
4-stop metroland and triangleland are analogous to zero extra charge cases of the above.

\subsection{Further solutions}

\noindent Unlike for 4-stop metroland and triangleland, where there is a range of further classically-tractable solutions \cite{AF, 08I, FileR}, 
further cases for quadrilateralland, and the simplest HO counterparts, give at best combinations including elliptic functions (checked with Maple).  
The QM of these additional problems is not analytically tractable either.  
However, the direction I will take is treating the HO potentials as small perturbations about the free case at the quantum level.

The following Figure provides useful qualitative analysis of ${\cal Y}$ = 0 and $\neq 0$ and ${\cal I}_{\sT\so\st\sa\sll} = 0$ and $\neq 0$ (\underline{Key 16}).  

{            \begin{figure}[ht]
\centering
\includegraphics[width=0.66\textwidth]{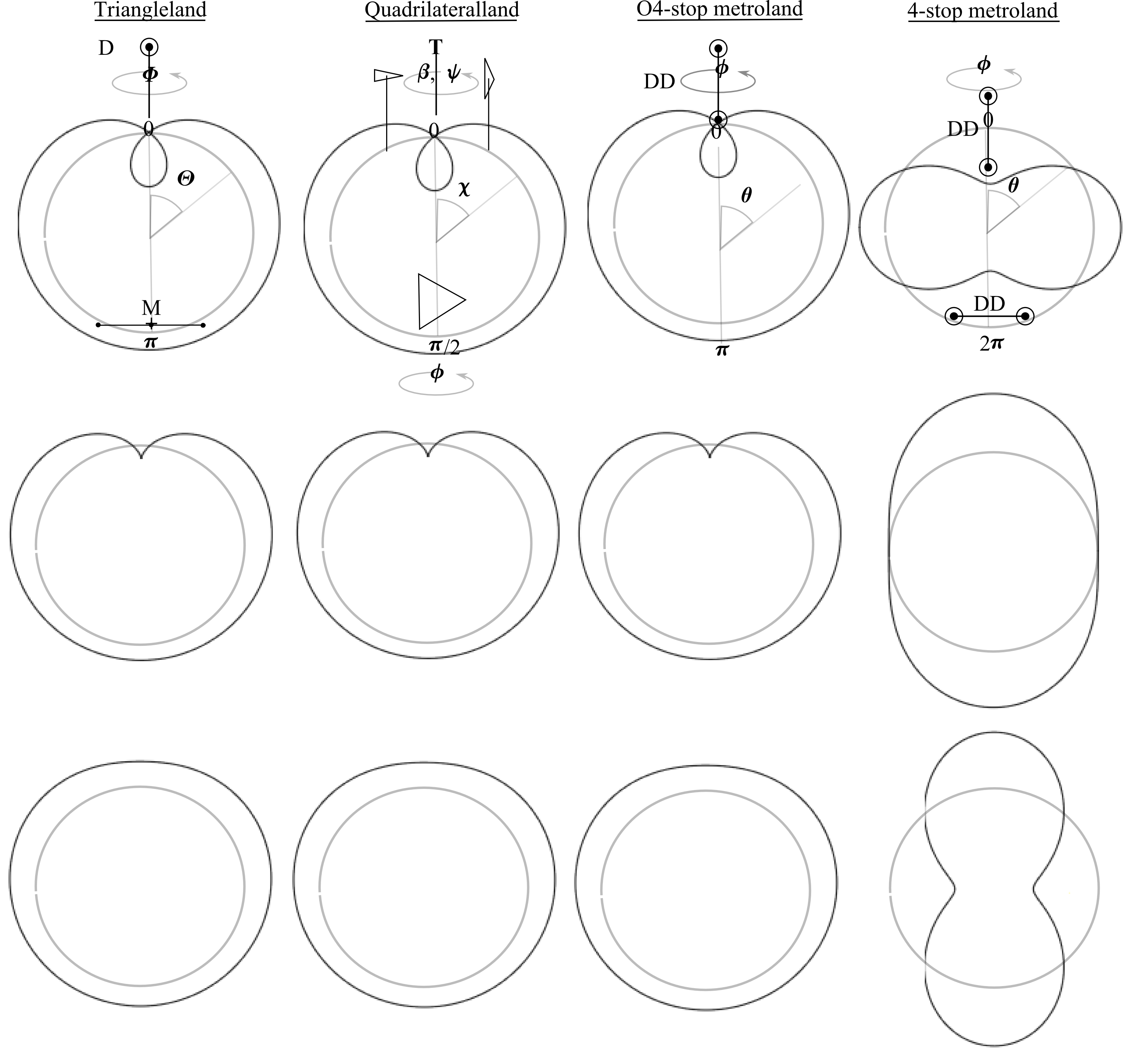}
\caption[Text der im Bilderverzeichnis auftaucht]{        \footnotesize{Qualitative behaviour of some of the simpler HO dynamics: 
for the potential $\ttA + \ttB\,\mbox{cos}\, 2\chi$ for various ratios of $\ttA$ and $\ttB$.  
In each case the back-cloth gives a qualitative indication of the typical and atypical dynamical configurations.  
The quadrilateralland case of this (column 2) is confined around the handle configuration (i.e. long-handled axes of various blade shapes, rather than axehead-like configurations. 
It is usefully compared with the corresponding potentials from two existing works and a further variant.  
Namely 1) triangleland (column 1 \cite{+tri}): confined around a double collision D as opposed to around a merger M.
2) 4-stop metroland (column 4, \cite{AF}): confined to each of 2 double--double collision DD configuration wells in the first row (tyre-shaped potential) or the opposite in 
the other two rows (egg and peanut shaped potentials). 
This mirror-image identified `O4-stop metroland' case is recovered in the study of the collinear submanifold of the quadrilateralland case.

Furthermore, a centrifugal spike as felt for $L_{\tT\to\ttt} \neq 0$ in planar mechanics or spherical mechanics, 
${\cal S}_{\tT\to\ttt} \neq 0$ triangleland and ${\cal D}_{\tT\to\ttt} \neq 0$ for 4-stop metroland.  
That the on-sphere cases of this have a spike at {\sl each} pole is standard.
This occurs for the triangleland case and the 4-stop metroland case.  
The O4-stop metroland case has just the North Pole spike.  
A feature that first occurs in this paper's quadrilateralland case is having {\sl independent} spikes at each `pole'.  
Namely, a spike felt for ${\cal I}_{\tT\to\ttt\ta\tl} \neq 0$ at the `North pole' $\chi = 0$ and a different spike felt by ${\cal Y} \neq 0$ at the `South pole' $\chi = \pi/2$.

In this paper's quadrilateralland case, the motions are qualitatively as follows. 
In the cross-section for conserved quantities being 0, and `round and round' in the other cases.  
That the $SU(2)$ round and round is some higher-$d$ $\mathbb{S}^2$ analogue of axisymmetry. 
The $U(1)$ of the $\psi$ is standard axisymmetry. 
\noindent [Note that each of these corresponds to going round-and-round in {\sl different} suppressed dimensions.] } }
\label{Temporary} \end{figure}          }

\section{Conclusion}\label{Concl}

\subsection{Classical Quadrilateralland Results}

\noindent Quadrilateralland is the smallest model that possesses both linear constraints and nontrivial subsystem structure.    
Applications of this include Records Theory, structure formation, investigating the extent to which one subsystem can be used as a clock for the other parts of the model universe 
\cite{PW83, Rovellibook}, Histories Theory and Combined Semiclassical--Histories--Records Schemes \cite{H03, H09-1, H09-2, AHall, FileR}.

\noindent Quadrilateralland in solved as follows at the classical level.  

\noindent 1) Regardless of an RPM's particle number $N$ or dimension, the classical and quantum $N$-body problem's Jacobi coordinates remain 
available to diagonalize relative interparticle cluster coordinates
For quadrilaterals, these are two qualitatively different choices of Jacobi tree of clusterings: H-shaped and K-shaped, 
adapted to 2 + 2 and \{2 + 1\} + 1 splits of the particles into subsystem clusters respectively.  

\noindent 2) For any $N$-a-gonland, the shape space is $\mathbb{CP}^{N - 2}$ with Fubini--Study metric, as first pointed out in Kendall's work in Shape Statistics.
In particular, for quadrilateralland, it is $\mathbb{CP}^2$.  

\noindent 3)  Inhomogeneous coordinates (from basic projective geometry) have some uses, but further uses require casting the Fubini--Study metric in `minimal block form' 
(i.e. as close to diagonal as possible).  
For quadrilateralland, this arises for Gibbons--Pope type coordinates.  
These are in a number of ways analogous to (ultra)spherical coordinates. 
This represents an increase in complexity from triangleland that parallels 
passing from diagonal Bianchi IX models to nondiagonal ones.
Gibbons--Pope type coordinates are useful in characterizing conserved quantities (Sec \ref{SSec: Cons})
and in separating the free-potential time-independent Schr\"{o}dinger equation as exposited in Paper II. 

\noindent 4) Any $N$-a-gon's relationalspace is the cone over the corresponding shape space.

\noindent 5)  Triangleland's useful Dragt-type shape quantities (ellipticity, anisoscelesness and four times the mass-weighted area per unit $I$) 
from Molecular Physics extend to quadrilateralland as follows.  
There is one ellipticity, one linear combination of ellipticities, three anisoscelesnesses and a quantity 
proportional to the square root of the sum of the squares of the three mass-weighted areas per unit $I$.
For Jacobi K-coordinates, these quantities refer to three cluster-dependent `coarse-graining triangles' [Fig \ref{FigX}.j)--l)].  
On the other hand, for Jacobi H-coordinates, these quantities refer 
to two of these and a `coarse-graining' rhombus [Fig \ref{FigX}.e)--g)]. 
In each case, these are obtained from the quadrilateral by striking out each Jacobi vector in turn.   
The last of these quantities, is furthermore like the triangle case's 4 $\times$ area [= demo($N$ = 3)] in being a  
democratic invariant (cluster-independent quantity -- a notion also from Molecular Physics), demo($N$ = 4).   

\noindent 6) Triangleland's parabolic coordinates, that pick out the base and apex Jacobi magnitudes as separate subsystems alongside the 
relative angle between the corresponding Jacobi vectors, generalize in a number of ways to the quadrilateralland context as the Kuiper coordinates \cite{Kuiper} from pure Geometry. 
There are 6 of these for quadrilateralland: three partial moments of inertia of the system and three constituent subsystem anisoscelesneses. 
These can be rearranged into total moment of inertia I and 5 Kuiper quantities. 
Together with the areas of the three constituent subsystems, these form a set of 8 Gell-Mann quadratic forms.
Casting these in terms of Gibbons--Pope type coordinates places them on a common footing with the conserved quantities for convenience of subsequent algebraic computation.  

\noindent 7) For the simpler models considered hitherto, the shape space dimension was small enough to allow for complete graphical presentation 
of the interpretation of the regions of the shape space in RPM terms.  
E.g. 4-stop metroland and triangleland were both represented as distinct tesselations of their configuration space spheres \cite{AF} and Fig \ref{Fig-4-Stop}, 
and \cite{Kendall89, +tri}, and Fig \ref{Fig4}.  
$\mathbb{CP}^2$ being 4-dimensional, one is more limited in visualization.  
Nevertheless, I provided a characterization of the RPM-significant submanifolds of this, and also the complex-projective chopping board visualization: the 
pure-ratio counterpart of Kendall's spherical blackboard.  
I used Kuiper coordinates to determine the quadrilateralland counterpart of the split of triangleland's sphere into two hemispheres of mirror-image configurations 
that join along an equator of collinear configurations (which is one of \cite{+tri}'s most important and useful results).  
One has now two topological $\mathbb{S}^4$ `hemi-$\mathbb{CP}^2$'s' that join along the `$\mathbb{CP}^2$-equator' of collinear configurations 
that form the mirror-image-identified 4-stop metroland and thus are a $\mathbb{RP}^2$ submanifold.  
Upon codimension-1 embedding into a surrounding Euclidean space, this join maps into the Veronese surface, in accord with Kuiper's theorem.  
Quadrilateralland's $\mathbb{CP}^2$ is additionally decorated by a net of 6 $\mathbb{S}^2$ trianglelands (in each case with one vertex being a double collision corresponding to the 
$ \mbox{\Large (} \stackrel{\mbox{\scriptsize 4}}{\mbox{\scriptsize 2}}\mbox{\Large )}$  pairs of particles coinciding). 
I also showed how $\mathbb{S}^2$'s corresponding to the coarse-graining `trianglelands' (and `rhombusland') occur within quadrilateralland, 
by applying restrictions on Gibbons--Pope type coordinates.

\noindent As regards uniform configurations, I also used Gibbons--Pope type and Kuiper coordinates to investigate the 6 possible labellings of square configurations 
and a weaker criterion of the cosmologically and quantum-cosmologically interesting topic of uniform states based on the extremization of demo($N$ = 4).

\noindent 8) Shape momenta conjugate to the above shape variables were considered.  
Physically, these are relative dilational momenta, relative angular momenta and mixtures of these.  
Conserved quantities for RPM's in 1-$d$ were also considered; these are particular combinations of the preceding.  
Functionals of the shape variables and shape momenta can be used to resolve the Problem of Observables in the sense of \K for these models.  
This paper consolidates the above for small RPM's and is the first place to extend as far as the explicit quadrilateralland examples of these things.  
This is a substantial extension due to how quadrilateralland is the smallest RPM to simultaneously possess nontrivial 
subsystem structure and linear constraints in the quantum cosmologically significant case. 
It also has more general and typical mathematics for an $N$-a-gonland than triangleland does (since its shape space is $\mathbb{CP}^2$ whilst 
triangleland's is $\mathbb{CP}^1$, which is atypical through also being $\mathbb{S}^2$).

\noindent 9) The isometry group of $\mathbb{CP}^2$ is $SU(3)/\mathbb{Z}_3$. 
$SU(3)$ is well-known to contain $SU(2) \times U(1)$ subgroups; a particular choice of Gibbons--Pope type coordinates then happens to be adapted to each such.
Thus quadrilateralland exhibits a number of parallels with the Particle Physics of the strong force.  
One new question addressed in this paper then is how to interpret the $SU(3)$ quantities in terms of the quadrilateral.   
Gibbons--Pope type coordinates are useful in this investigation - they have clear geometrical interpretation in quadrilateralland 
terms, and they are additionally cyclic coordinates for some of the simpler potentials
I used these to give the geometric interpretations of the momenta and conserved quantities.  
 
\noindent 10) I considered the multi HO-like potential for quadrilateralland in terms of the useful intrinsic Gibbons--Pope coordinates.  

\noindent 11) I interpreted a geodesics result for $\mathbb{CP}^2$  in terms of quadrilaterals, and link it to the various collapses of the quadrilateralland trees.

\noindent 12) In the qualitative dynamics Figure \ref{Temporary} for the simpler HO-like potential, I provide a nice extension of work in \cite{AF, +tri} with one additional case 
necessitated so as to parallel the quadrilateral better.  
I.e. the qualitative analysis of HO type potentials on O4-stop metroland (i.e. the mirror-image-identified version).  
The new feature in quadrilateralland is two repulsive spikes that are felt independently of each other according to whether the 
system in question has each of the two types of charge.  
I.e. {\it angular charge} (paralleling Particle Physics' isospin), and extra charge (paralleling Particle Physics' hypercharge).  
\noindent These qualitative dynamics figures also combine nicely with the tessellation interpretations of the 4-stop metroland and triangleland 
shape spheres and of the characterized submanifolds of $\mathbb{CP}^2$ as provided in Paper I.  

\mbox{ }

\noindent The Introduction listed which imports RPM models, the PoT and Quantum Cosmology obtain from these branches of physics.  
Which other subjects benefit? 
Classical Dynamics and Molecular Physics get a treatise of 2-$d$ 4-particle model kinematics and imports of shape-geometry and shape-statistics tools \cite{AStats}. 
MacFarlane's Particle Physics inspired work at the quantum level has a number of relevant generalizations built for it in Paper II. 
This work provides some major long-term avenues for Shape Statistics, namely the quantum mechanical, GR and quantum gravitational versions of the standing stones problem viewed 
as a Timeless Records problem.
It also shows how a mechanics, instead of statistics, can be built upon the common foundation of shape geometry.  
Moreover, doing so directly addresses the famous absolute versus relational motion debate.

\vspace{10in}

\subsection{$N$-a-gonland generalization of this paper}

Most of Key points 1) to 4) readily carry over to $N$-a-gonland.  

\noindent The numbers and complexities of the qualitatively different types of Jacobi trees increase.

{            \begin{figure}[ht]
\centering
\includegraphics[width=0.75\textwidth]{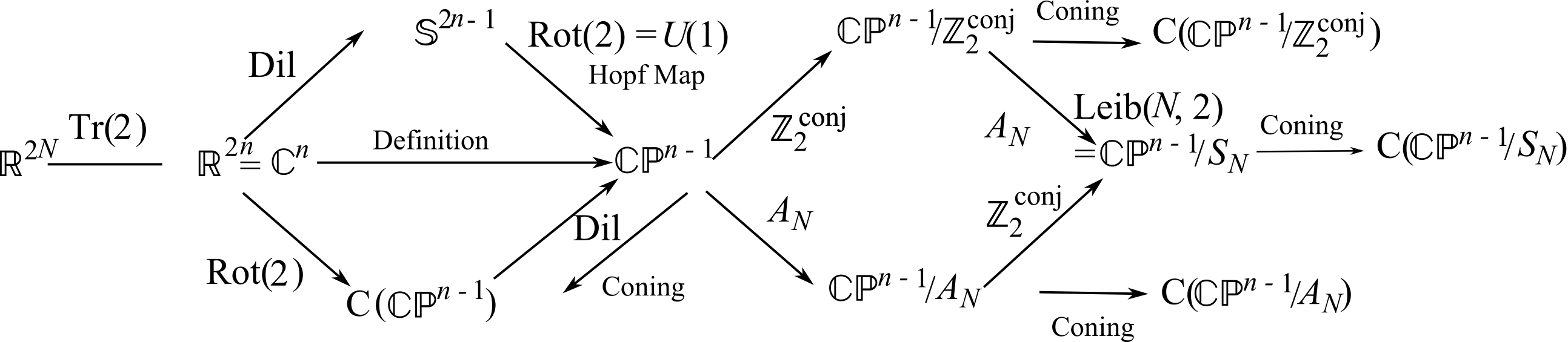}
\caption[Text der im Bilderverzeichnis auftaucht]{        \footnotesize{The general sequence of configuration 
spaces for $N$-a-gonland.} }
\label{Fig7} \end{figure}          }

\noindent The topological part of 2) is covered in Fig \ref{Fig7} for arbitrary $N$. 
The first part of 3)'s arbitrary-$N$ status should be contrasted with how the 3-$d$ models' shape spaces forming a much harder sequence \cite{Kendall}. 
This gives a strong practical 
reason to study the 2-$d$ models (particularly since the analogy with geometrodynamics does not require dimension 3 in order to work well, as exposited in \cite{FileR}).  
As regards 4), an arbitrary-dimensional generalization of Gibbons--Pope type coordinates is given in \cite{MF03b}, with explicit results for the $SU$(4) i.e. $\mathbb{CP}^3$ case 
that corresponds to pentagonland.
As regards 5) and 6), extending the Dragt/parabolic/Kuiper/shape type of redundant coordinates is straightforward via the arbitrary-p generalization of the Gell-Mann $\lambda$ 
matrices to $SU$(p).   
The truest name then is $SU(p)$-{\it adjoint-rep quadratic forms}, taken in what passes for a standard basis for each p.   

\noindent As regards 7), the $N$-a-gonland significance of two half-spaces of different orientation separated by an orientationless manifold of collinearities gives reason 
for double covers to the $N - 2 \geq 3$ $\mathbb{CP}^{N - 2}$ spaces to exist for all $N$.  
However, there is no known guarantee \cite{Kuroki} that these will involve geometrical entities as simple as or tractable as 
quadrilateralland's $\mathbb{S}^4$ for the half-spaces, or of the Veronese surface $V$ as the place of branching.  
It is clear that the manifold of collinearities within $N$-a-gonland's $\mathbb{CP}^{N - 2}$ itself is $\mathbb{RP}^{N - 2}$, so at least that is a known and geometrically-simple result 
for the structure of the general $N$-a-gonland.
Beyond the above, I leave $N$-a-gonland's significant submanifold structure as an open problem.      

\noindent MacFarlane \cite{MF03b} shows how subgroup-adapted coordinates carry over to $\mathbb{CP}^3$ and in a form that leaves in evidence the 
$\mathbb{CP}^{\sp}$ generalization in close parallel with arbitrary-$\mathbb{S}^{\sp}$ polar coordinates.  


\noindent As regards the Quantum Information Theory counterpart of this work (Sec 8), one possible usefulness of representing qutrit states as quadrilaterals is via the convenience of 
having a 2-$d$ graphical representation, which, moreover, remains 2-$d$ as one passes to the study of qu$n$its.  
In this picture, the relation between the 3 included $SU(2)$ ladders and the 3 coarse-graining triangles (or two triangles and one parallelogram) is that there are 3 constituent 
(overlapping) qubits in a qutrit, and this pattern continues for higher $n$.  

\noindent I finally comment that the large-$N$ limit of $N$-a-gonland should also be of interest, alongside the study of the statistical 
mechanics/entropy/notions of information/relative information/correlation that timeless approaches are concerned with.   

\mbox{ } 

\noindent The $N$-a-gonland interpretation of conserved quantities and of corresponding quantum numbers remains open.  
\noindent The geodesics result used \cite{W82a} is general-$N$. 
\noindent Some further results on the classical dynamics on $\mathbb{CP}^N$ can be found in e.g. \cite{CPNDyn1, CPNDyn2, CPNDyn3}.

\subsection{Applications to Papers II, III and IV}

\noindent Kinematical quantization and separability in Sec II.4 is based on this paper's work on shape quantities and conserved quantities. 
Gibbons--Pope type coordinates then serve to separate this problem's time-independent Schr\"{o}dinger equation.
The complex-projective chopping board is used as a back-cloth for the interpretation of quantum-mechanical wavefunctions in Sec II.6.  

\mbox{ } 

\noindent Papers I and II are then useful for subsequent investigations of Problem of Time in Quantum Gravity strategies and various other quantum-cosmological issues.  

\noindent In the present Paper I, we removed the Configurational Relationalism problem at the classical level by succeeding in performing Best Matching/Lagrangian-level reduction, 
WE also showed how this leads to an explicit form for the Machian emergent Jacobi--Barbour--Bertotti time as well as to classical \K beables. 
The classical constraint algebra also closing, and RPM's having no place for nontrivial GR-lie issues of foliation invariance and spacetime reconstruction, that covers 
all aspects needed for a local resolution of the PoT.

\noindent This resolution failing to unfreeze the quantum PoT, in Paper II Kneller and I need to re-start with Machian emergent semiclassical time in Sec II.10, 
supported by consideration of quantum \K beables in Sec II.2.
The Semiclassical Approach, however is unsatisfactory by itself at a deeper level (justifying the crucial chroniferous WKB ansatz). 
We get round this by combining it with the Histories Theory and Timeless Records approaches, which have their own classical and quantum forms to work out prior to attaining this 
combination (\cite{H03, H09-1, H09-2, AHall} and the subsequent Machian version in \cite{AHall, FileR}.   

\mbox{ }  

\noindent Configuration space is useful classically and as a place where QM unfolds; the study so far in this paper needs, however, 
to be supplemented with a study of regions as well as of the submanifolds considered so far.  
Classically, this corresponds to approximate knowledge of configuration.
One can pin a QM probability on that in some (part) timeless interpretations such as the \NSI or Halliwell's approach.   
Kneller and I provide the \NSI in Sec II.11; 
this combines the study of regions of $\mathbb{CP}^2$ configuration space in Appendix A below with the specific wavefunctions supplied in Paper II.  
The present paper's Gibbons--Pope type coordinates are useful both in considering regions of configuration space and in considering the wavefunctions themselves.
For now, the volume element in these is $\mbox{sin}^3\chi\,\mbox{cos}\,\chi\,\d\chi\,\mbox{sin}\,\beta\, \d\beta \, \d\phi \, \d\psi/8$, 
so that the quadrilateralland \NSI probability expression is 
\beq
\mbox{Prob}(\mbox{Region R}) \propto \int_{\sR}|\Psi(\chi,\beta,\phi,\psi)|^2  
\mbox{sin}^3\chi\,\mbox{cos}\,\chi\,\d\chi\, \mbox{sin}\,\beta\,\d\beta\, \d\phi \, \d\psi \mbox{ } . 
\eeq
The wavefunctions (at least in the free potential case \cite{MacFarlane, QuadIII}) also separate into (sums of) product of factors $C(\chi)D(\beta)E(\phi)F(\psi)$.
The characterization of the region of integration separates into individual integrals in each of these coordinates as well 
(at least for a number of physically-significant regions including the examples in Appendix A below). 
This is put to various Quantum Cosmology uses, such as the Peaking Interpretation and investigation of notions of uniformity, in Paper II.  

\mbox{ }

\noindent The Semiclassical Approach (also covered for quadrilateralland in Paper II) itself provides further reasons for the study of regions of configuration space: 
the many approximations used, including the crucial WKB and Born--Oppenheimer approximations, only hold in certain regions of configuration space.  
The Semiclassical Approach also constitutes a toy model for Quantum Cosmology as the originator of structure formation \cite{HallHaw}.  

\mbox{ }

\noindent The current Paper's above notions of closeness and Fubini--Study kinetic metric use additionally embody control over localization both in space and in configuration space. 
This allows for triangleland work toward establishing a Records Theory in \cite{FileR}; this concerns in particular notions of distance between shapes; 
it is to be extended to quadrilateralland (Paper III).  
Records are usually envisaged as localized subsystems. 
Thus having a certain richness of these available in one's model is conducive to a nontrivial study of this locality, 
associated notions of information and the emergence of a semblance of dynamics involving nontrivial structure formation.  
Moreover, as argued in the Introduction and in Sec \ref{Q-2}, quadrilateralland is rather richer as regards constituent subsystems than triangleland is, 
whilst retaining a nontrivial linear constraint (that parallels the GR momentum constraint) in the quantum-cosmologically relevant scaled case.  

\noindent As regards the Combined Approach itself, it requires development of Histories Theory in Paper III. 
An extra connection between these approaches is that \cite{H03, H09-1, H09-2} the Semiclassical Approach aids in the 
computation of timeless probabilities of histories entering given configuration space regions.
This, by the WKB assumption, gives a semiclassical flux into each region in terms of $S(\mh)$ and, in a simple case, the Wigner function (see e.g. \cite{H03}). 
This approach also promotes each of classical and semiclassical \K beables resolutions to Dirac beables resolutions.  
Halliwell studied this with a free particle, a working which has a direct, and yet more genuinely closed-universe, counterpart for scaled triangleland \cite{08I}.  
This is via the `Cartesian to Dragt coordinates analogy' that can be seen between Secs \ref{Demo-3} and \ref{Demo-4} allowing me to transcribe this working to a relational context.

\mbox{ } 

\noindent The quadrilateralland extension of this calculation is then desirable for its combination of further physical modelling features and mathematical novelty as a Histories Theory.  
This is because 

\noindent A) we are concerned with modelling quantum-cosmological structure formation, and it takes a quadrilateral to have splits into nontrivial subsystems and nested nontrivial 
subsystems. 
4-stop metroland had these too, but only at the cost of the model ceasing to involve linear constraints, which are the other key feature of midisuperspace that can be addressed by RPM's. 

\noindent B) Moreover, Records Theory's Shape Statistics quantification of correlations and Statistical Mechanics notions of information require $N > 3$ 
(and $N >> 3$ would further improve on that).  
E.g. in Shape Statistics, a common and relational diagnostic is to sample constituent triangles for approximate collinearity \cite{Kendall84, Kendall89, Kendall}. 
However, for $N = 3$ itself, the sampling size becomes 1, which is not tenable as regards the extraction of any statistical conclusions.

\noindent C) The triangleland case \cite{AHall, FileR} is underlied by similar $\mathbb{R}^3$ mathematics to Halliwell's own earlier non-Machian/non-relational examples. 
Mathematical novelty of the examples resulting from the RPM program therefore starts with the quadrilateral.  

\mbox{ }

\noindent {\bf Acknowledgements}: I thank those close to me for being supportive of me whilst this work was done. 
Mr Eduardo Serna for discussions and reading the manuscript and help with Secs \ref{Q-2}, \ref{Mom-Int}-\ref{cl-soln}.
Professors Don Page and Gary Gibbons for teaching me about $\mathbb{CP}^2$ in 2005 and 2007.  
Professor Julian Barbour for introducing me to RPM's in 2001.  
Professors Jonathan Halliwell, Chris Isham, Karel Kucha\v{r}, Julian Barbour and Christopher Small, Dr Julio Arce, Miss Sophie Kneller and one Anonymous Referee for discussions. 
Professors Belen Gavela, Marc Lachi\`{e}ze-Rey, Malcolm MacCallum, Don Page, Reza Tavakol, and Dr Jeremy Butterfield for support with my career.  
Fqxi grant RFP2-08-05 for travel money whilst part of this work was done in 2009-2010, and Universidad Autonoma de Madrid for funding in 2010--2011.  
The final drafting of this work was funded by a grant from the Foundational Questions Institute (FQXi) Fund, a donor-advised fund of the 
Silicon Valley Community Foundation on the basis of proposal FQXi-RFP3-1101 to the FQXi.  
I thank also Theiss Research and the CNRS for administering this grant, held at APC universit\'{e} Paris Diderot.  

\appendix
%
\section{Regions of configuration space}\label{Eregion}
%
{            \begin{figure}[ht]
\centering
\includegraphics[width=0.75\textwidth]{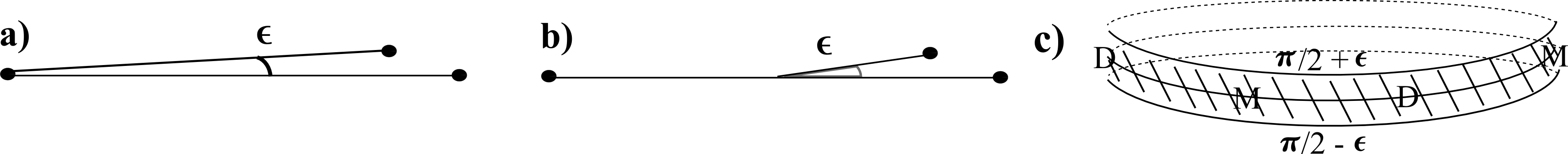}
\caption[Text der im Bilderverzeichnis auftaucht]{        \footnotesize{a) Kendall's $\epsilon$-collinearity in space for three points. 
b) A counterpart of this in variables useful for Mechanics. 
c) The corresponding configuration space belt of width 2$\epsilon$.} }
\label{Fig7a} \end{figure}          }

Here I characterize some physically-significant regions of the quadrilateralland configuration space.  
To complete the \NSI for these, we must wait for the wavefunctions themselves to be introduces in Paper II. 
4-stop metroland and triangleland counterparts of such regions are e.g. caps, belts and lunes that are simply-characterized in terms of spherical polar coordinates \cite{AF, +tri}. 
These are endowed with particular physical significance, e.g. the cap of $\epsilon$-equilaterality or Fig \ref{Fig7a}c)'s belt of $\epsilon$-collinearity. 
The present paper's regions are somewhat more involved due to $\mathbb{CP}^2$'s more complicated and higher-dimensional geometry.  
For the shape space spheres, I noted a correspondence between the size of the region in question and the size of the departure in space from the precise configuration.  
E.g. the width of the belt of collinearity on the shape sphere corresponds to a Kendall type \cite{Kendall, +tri} notion of $\epsilon$-collinearity of three points in space 
(Fig \ref{Fig7a}a).

Establishing such correspondences is then part of the study of physically-significant regions of the $\mathbb{CP}^2$ configuration space also.  
In particular, I consider 

\mbox{ }

\noindent I) approximately-collinear quadrilaterals.  
Exactly collinear was mathematically a $\mathbb{RP}^2$ corresponding to, using the H-coordinates most natural here, 
both angular coordinates $\phi$ and $\psi$ being 0 or an integer multiple of $\pi$.  
This condition is now $\epsilon$-relaxed.  
Thus the region of integration is all values of the ratio coordinates $\beta$ and $\chi$ whilst the angular coordinates are to live within the following union of products of intervals: 

\noindent 
$$
C_{\epsilon} := 
\big(
\{0 \leq \phi \leq \epsilon\} \cup \{\pi - \epsilon \leq \phi \leq \pi + \epsilon\} \cup \{2\pi - \epsilon \leq \phi \leq 2\pi\}
\big) 
\times
$$
\beq
\big(
\{0 \leq \psi \leq \epsilon\} \cup \{\pi - \epsilon \leq \psi \leq \pi + \epsilon\} \cup \{2\pi - \epsilon \leq \phi \leq 2\pi + \epsilon\}
\cup 
\{3\pi - \epsilon \leq \phi \leq 3\pi + \epsilon\} \cup \{4\pi - \epsilon \leq \phi \leq 4\pi\}
\big) \mbox{ } .  
\eeq
This is a collection of 2-lunes (in the `axial coordinates'; for the sphere, an interval of the axial coordinate for all azimuths is a lune).
From the spatial perspective, this notion of $\epsilon$-collinearity corresponds to the relative angles 
$\Phi_1$ and $\Phi_2$ each lying within $[0, \epsilon] \cup [\pi - \epsilon, \pi + \epsilon] \cup [2\pi - \epsilon, 2\pi]$.

\mbox{  }

\noindent II) Quadrilaterals that are approximately triangular or approximately one of the mergers depicted in the coarse-graining triangles or rhombi of Fig \ref{FigX}.     
The exact configuration in each of these cases is a sphere characterized by one ratio variable and one relative angle variable.  
In moving away from exact triangularity, a second $\epsilon$-sized ratio variable becomes involved, and, 
in doing so one is rendering the other relative angle meaningful, allowing it to take all values.  
Thus the region of integration here is all angles, one ratio variable taking all possible values also, 
and the other being confined to an $\epsilon$-interval about the value that corresponds to exact triangularity.    
To be more concrete, consider the \{+43\} triangle in Gibbons--Pope type coordinates that derive from H-coordinates.  
Here, $\beta = \pi$, so the region of integration is all $\phi$, all $\psi$, all $\chi$ and $\pi - \epsilon \leq \beta \leq \pi$; I take this to define a belt region $T_{\epsilon}$.  
Then, for example, the notion of $\epsilon$-\{+43\} triangular corresponds in space to the ratio (post 1)/crossbar  being of size $\epsilon/2$ or less.  

\mbox{ }  

\noindent III) One of the labellings of exact square configuration is at $\phi = 0$, $\psi = \pi$, $\beta = \pi/2$ and $\chi = \pi/4$.
Approximate squareness allows for all four of these quantities to be $\epsilon$-close. 
Thus the region of integration is a product of four intervals of width $\epsilon$ about these points. I denote this 4-box-like region by $S_{\epsilon}$.
That corresponds to the two sides of the Jacobi H being allowed to be $\epsilon$-close ($\beta$-variation), the quadrilateral to vary in height to length ratio ($\chi$-variation) and 
for each of the sides of the Jacobi H to become non-right with respect to the crossbar ($\psi$- and $\phi$-variation).  
This is an entirely intuitive parametrization of the possible small departures from exact squareness.
[If one is interested in all of the labellings of the square, one can straightforwardly characterize each with a similar construction and again take the union of these regions.]  
This example is clearly furthermore useful as a notion of approximate uniformity, which is of interest in Classical and Quantum Cosmology.  


\end{document}